\input{epsf}

\newcommand{\et}{\hspace{-0.08in}{\bf .}\hspace{0.1in}}
\newcommand{\BOX}{\hbox {$\sqcap$ \kern -1em $\sqcup$}}

\newcounter{letter}

\newcommand{\g}{{\frak g}}
\newcommand{\so}{{\frak so}}
\newcommand{\su}{{\frak su}}
\newcommand{\R}{{\Bbb R}}
\newcommand{\C}{{\Bbb C}}
\newcommand{\Z}{{\Bbb Z}}

\renewcommand{\to}{\rightarrow}
\newcommand{\tensor}{\otimes}
\newcommand{\maps}{\colon}

\newcommand{\iso}{\cong}
\newcommand{\we}{\wedge}
\newcommand{\ad}{{\rm ad}}
\renewcommand{\L}{{\cal L}}
\renewcommand{\H}{{\cal H}}
\newcommand{\F}{{\cal F}}
\newcommand{\E}{{\cal E}}
\newcommand{\V}{{\cal V}}
\newcommand{\A}{{\cal A}}
\newcommand{\T}{{\cal T}}

\newcommand{\G}{{\cal G}}
\newcommand{\SL}{{\rm SL}}
\newcommand{\SU}{{\rm SU}}
\newcommand{\U}{{\rm U}}

\newcommand{\SO}{{\rm SO}}

\newcommand{\Spin}{{\rm Spin}}

\newcommand{\Fun}{{\rm Fun}}

\newcommand{\Irrep}{\rm Irrep}

\newcommand{\tr}{{\rm tr}}

\newtheorem{thm}{Theorem}    

\newtheorem{defn}[thm]{Definition}

        \newcommand{\be}{\begin{equation}}
        \newcommand{\ee}{\end{equation}}
        \newcommand{\ba}{\begin{eqnarray}}
        \newcommand{\ea}{\end{eqnarray}}
        \newcommand{\ban}{\begin{eqnarray*}}
        \newcommand{\ean}{\end{eqnarray*}}
        \newcommand{\barr}{\begin{array}}
        \newcommand{\earr}{\end{array}}

\documentstyle[amsfonts,diagram]{article}

\begin{document}

      \begin{center}
      {\bf An Introduction to Spin Foam Models \\
      of $BF$ Theory and Quantum Gravity \\ }
      \vspace{0.5cm}
      {\em John C.\ Baez\\}
      \vspace{0.3cm}
      {\small Department of Mathematics, University of California\\ 
      Riverside, California 92521 \\
      USA\\ }
      \vspace{0.3cm}
      {\small email: baez@math.ucr.edu\\}
      \vspace{0.3cm}
      {\small May 20, 1999\\ }
      \end{center}

\begin{abstract}  
\noindent   
In loop quantum gravity we now have a clear picture of the quantum
geometry of {\it space}, thanks in part to the theory of spin networks.
The concept of `spin foam' is intended to serve as a similar picture
for the quantum geometry of {\it spacetime}.  In general, a spin network
is a graph with edges labelled by representations and vertices labelled
by intertwining operators.  Similarly, a spin foam is a 2-dimensional
complex with faces labelled by representations and edges labelled by
intertwining operators.  In a `spin foam model' we describe states as
linear combinations of spin networks and compute transition amplitudes 
as sums over spin foams.  This paper aims to provide a self-contained 
introduction to spin foam models of quantum gravity and a simpler field 
theory called $BF$ theory.
\end{abstract}

\section{Introduction}

Spin networks were first introduced by Penrose as a radical, purely
combinatorial description of the geometry of spacetime.  In their 
original form, they are trivalent graphs with edges labelled by 
spins:

\vskip 2em
\centerline{\epsfysize=2.0in\epsfbox{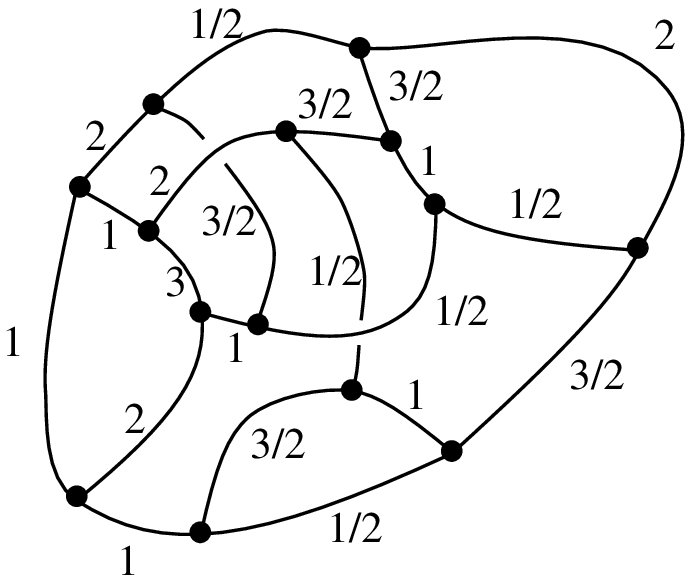}} \medskip

\noindent
In developing the theory of spin networks, Penrose seems to have been 
motivated more by the quantum mechanics of angular momentum than by the 
details of general relativity.  It thus came as a delightful surprise when 
Rovelli and Smolin discovered that spin networks can be used to describe 
states in loop quantum gravity.  

Fundamentally, loop quantum gravity is a very conservative approach to
quantum gravity.  It starts with the equations of general relativity and
attempts to apply the time-honored principles of quantization
to obtain a Hilbert space of states.  There are only two really new
ideas in loop quantum gravity.   The first is its insistence on a
background-free approach.  That is, unlike perturbative quantum gravity,
it makes no use of a fixed `background' metric on spacetime.  The second
is that it uses a formulation of Einstein's equations in which parallel
transport, rather than the metric, plays the main role.  It is
very interesting that starting from such ideas one is naturally led to
describe states using spin networks!

However, there is a problem.  While Penrose originally intended for spin
networks to describe the geometry of spacetime, they are really better
for describing the geometry of {\it space}.  In fact, this is how they are
used in loop quantum gravity.  Since loop quantum gravity is based on
canonical quantization, states in this formalism describe the geometry 
of space at a fixed time.  Dynamics enters the theory only in the form 
of a constraint called the Hamiltonian constraint.  Unfortunately this
constraint is still poorly understood.  Thus until recently, we had 
almost no idea what loop quantum gravity might say about the geometry of
{\it spacetime}.  
 
To remedy this problem, it is natural to try to supplement loop 
quantum gravity with an appropriate path-integral formalism.  In
ordinary quantum field theory we calculate path integrals using 
Feynman diagrams.   Copying this idea, in loop quantum gravity
we may try to calculate path integrals using `spin foams', which
are a 2-dimensional analogue of Feynman diagrams.   In general, 
spin networks are graphs with edges labelled by group representations 
and vertices labelled by intertwining operators.  These reduce to
Penrose's original spin networks when the group is $\SU(2)$ and the graph 
is trivalent.  Similarly, a spin foam is a 2-dimensional complex built
from vertices, edges and polygonal faces, with the faces labelled 
by group representations and the edges labelled by intertwining operators.
When the group is $\SU(2)$ and three faces meet at each edge, this
looks exactly like a bunch of soap suds with all the faces of the 
bubbles labelled by spins --- hence the name `spin foam'.

If we take a generic slice of a spin foam, we get a spin network.  Thus 
we can think of a spin foam as describing the geometry of spacetime, 
and any slice of it as describing the geometry of space at a given time.   
Ultimately we would like a `spin foam model' of quantum gravity, in
which we compute transition amplitudes between states by summming over 
spin foams going from one spin network to another:

\vskip 2em
\centerline{\epsfysize=2.0in\epsfbox{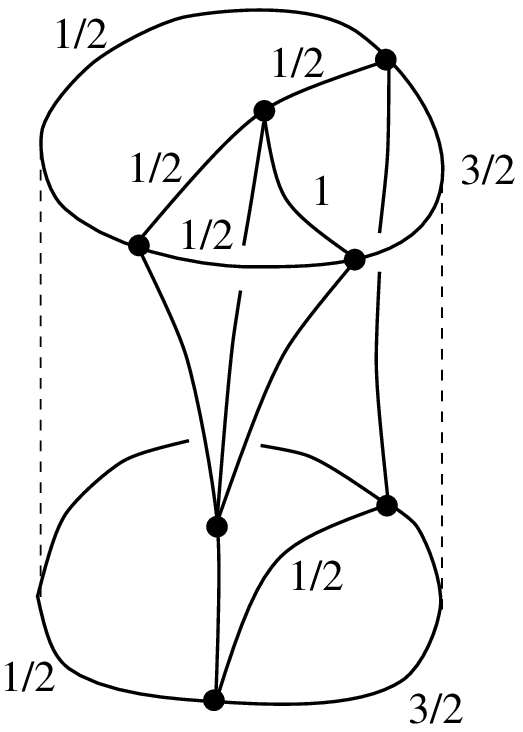}} \medskip
 
\noindent
At present this goal has been only partially attained.  For this reason
it seems best to start by discussing spin foam models of a simpler theory,
called $BF$ theory.  In a certain sense this the simplest possible 
gauge theory.  It can be defined on spacetimes of any dimension.  It 
is `background-free', meaning that to formulate it we do not need a 
pre-existing metric or any other such geometrical structure on spacetime.  
At the classical level, the theory has no local degrees of freedom: all 
the interesting observables are
global in nature.  This remains true upon quantization.  Thus $BF$
theory serves as a simple starting-point for the study of
background-free theories.  In particular, general relativity in 3
dimensions is a special case of $BF$ theory, while general relativity in
4 dimensions can be viewed as a $BF$ theory with extra constraints. 
Most work on spin foam models of 4-dimensional quantum gravity
seeks to exploit this fact.  

In what follows, we start by describing $BF$ theory at the classical level.  
Next we canonically quantize the theory and show the space of gauge-invariant
states is spanned by spin networks.   Then we use the path-integral
formalism to study the dynamics of the theory and show that the
transition amplitude from one spin network state to another is given as
a sum over spin foams.  When the dimension of spacetime is above 2, this
sum usually diverges.  However, in dimensions 3 and 4, we can render it
finite by adding an extra term to the Lagrangian of $BF$ theory.  In
applications to gravity, this extra term corresponds to the presence of
a cosmological constant.  Finally, we discuss spin foam models of 
4-dimensional quantum gravity. 

At present, work on spin foam models is spread throughout a large number
of technical papers in various fields of mathematics and physics.  
This has the unfortunate effect of making the subject seem more complicated 
and less beautiful than it really is.  As an attempt to correct this 
situation, I have tried to make this paper as self-contained as possible.   
For the sake of smooth exposition, I have relegated all references to the 
Notes, which form a kind of annotated bibliography of the subject.   The 
remarks at the end of each section contain information of a more technical 
nature that can safely be skipped.   

\section{$BF$ Theory: Classical Field Equations} \label{equations}

To set up $BF$ theory, we take as our gauge group any Lie group $G$
whose Lie algebra $\g$ is equipped with an invariant nondegenerate
bilinear form $\langle \cdot, \cdot \rangle$.  We take as our spacetime
any $n$-dimensional oriented smooth manifold $M$, and choose a principal
$G$-bundle $P$ over $M$.  The basic fields in the theory are then:
\begin{itemize}
\item a connection $A$ on $P$,
\item an $\ad(P)$-valued $(n-2)$-form $E$ on $M$.
\end{itemize}
Here $\ad(P)$ is the vector bundle associated to $P$ via the adjoint
action of $G$ on its Lie algebra.   The curvature of $A$ is an 
$\ad(P)$-valued 2-form $F$ on $M$.  If we pick a local trivialization we can
think of $A$ as a $\g$-valued 1-form on $M$, $F$ as a $\g$-valued
2-form, and $E$ as a $\g$-valued $(n-2)$-form.  

The Lagrangian for $BF$ theory is:
\[              \L =  \tr(E \we F)  .\]
Here $\tr(E \we F)$ is the $n$-form constructed by taking the wedge product 
of the differential form parts of $E$ and $F$ and using the bilinear 
form $\langle \cdot, \cdot \rangle$ to pair their $\g$-valued parts.  
The notation `$\tr$' refers to the fact that when $G$ is semisimple we
can take this bilinear form to be the Killing form $\langle x,y\rangle =
\tr(xy)$, where the trace is taken in the adjoint representation.

We obtain the field equations by setting the variation of the action to zero:
\ban 0 &=& \delta \int_M \L  \\
    &=& \int_M \tr(\delta E \we F + E \we \delta F)  \\
    &=& \int_M \tr(\delta E \we F + E \we d_A \delta A) \\
    &=& \int_M \tr(\delta E \we F + (-1)^{n-1} d_A E \we \delta A) \ean
where $d_A$ stands for the exterior covariant derivative.  Here in the
second step we used the identity $\delta F = d_A \delta A$, while in the
final step we did an integration by parts.   We see that the variation
of the action vanishes for all $\delta E$ and $\delta A$ if and only if
the following field equations hold: 
\[              F = 0 , \qquad d_A E = 0.\]

These equations are rather dull.  But this is exactly what we want,
since it suggests that $BF$ theory is a topological field theory!  In
fact, all solutions of these equations look the same locally, so $BF$
theory describes a world with no local degrees of freedom.  To see this,
first note that the equation $F = 0$ says the connection $A$ is flat.  
Indeed, all flat connections are locally the same up to gauge
transformations.  The equation $d_A E = 0$ is a bit subtler.  It is not
true that all solutions of this are locally the same up to a gauge
transformation in the usual sense.  However, $BF$ theory has another
sort of symmetry.  Suppose we define a transformation of the $A$ and $E$
fields by 
\[          A \mapsto A, \qquad   E \mapsto E + d_A \eta \]
for some $\ad(P)$-valued $(n-3)$-form $\eta$.  This transformation
leaves the action unchanged:
\ban \int_M \tr((E + d_A \eta) \we F) 
 &=& \int_M \tr(E \we F + d_A \eta \we F) \\
 &=& \int_M \tr(E \we F + (-1)^n \eta \we d_A F) \\
 &=& \int_M \tr(E \we F) \ean
where we used integration by parts and the Bianchi identity $d_A F = 0$.
In the next section we shall see that this transformation is a `gauge
symmetry' of $BF$ theory, in the more general sense of the term, meaning
that two solutions differing by this transformation should be counted as
physically equivalent.   Moreover, when $A$ is flat, any $E$ field with
$d_A E = 0$ can be written locally as $d_A \eta$ for some $\eta$; this
is an easy consequence of the fact that locally all closed forms are
exact.  Thus locally, all solutions of the $BF$ theory field equations
are equal modulo gauge transformations and transformations of the above
sort.

Why is general relativity in 3 dimensions a special case of $BF$ theory?
To see this, take $n = 3$, let $G = \SO(2,1)$, and let $\langle
\cdot,\cdot \rangle$ be minus the Killing form.  Suppose first  that $E
\maps TM \to \ad(P)$ is one-to-one.  Then we can use it to define  a
Lorentzian metric on $M$ as follows:
\[            g(v,w) = \langle Ev, Ew \rangle \]
for any tangent vectors $v, w \in T_x M$.   We can also use $E$ to pull
back the connection $A$ to a metric-preserving connection $\Gamma$ on
the tangent bundle of $M$.  The equation $d_A E = 0$ then says precisely
that $\Gamma$ is torsion-free, so that $\Gamma$ is the Levi-Civita
connection on $M$.  Similarly, the  equation $F = 0$ implies that
$\Gamma$ is flat.   Thus the metric $g$ is flat.  

In 3 dimensional spacetime, the vacuum Einstein equations simply say 
that the metric is flat.  Of course, many different $A$ and $E$ fields 
correspond to the same metric, but they all differ by gauge
transformations.   So in 3 dimensions, $BF$ theory with gauge group
$\SO(2,1)$ is really just an alternate formulation of Lorentzian general
relativity without matter fields --- at least when $E$ is one-to-one. 
When $E$ is not one-to-one, the metric $g$ defined above will be
degenerate, but the field equations of $BF$ theory still make perfect
sense.  Thus 3d $BF$ theory with gauge group $\SO(2,1)$ may be thought
of as an extension of the vacuum Einstein equations to the case of
degenerate metrics.   

If instead we take $G = \SO(3)$, all these remarks still hold except
that the metric $g$ is Riemannian rather than Lorentzian when $E$  is
one-to-one.   We call this theory `Riemannian general relativity'. We
study this theory extensively in what follows, because it is easier to
quantize than 3-dimensional Lorentzian general relativity.   However, it
is really just a warmup exercise for the Lorentzian case --- which in
turn is a warmup for 4-dimensional Lorentzian quantum gravity.

We conclude with a word about double covers.  We can also express
general relativity in 3 dimensions as a $BF$ theory by taking the double
cover $\Spin(2,1) \iso \SL(2,\R)$ or $\Spin(3) \iso \SU(2)$ as gauge
group and letting $P$ be the spin bundle.   This does not affect the
classical theory.   As we shall see, it does affect the quantum
theory.  Nonetheless, it is very popular to take these groups as gauge 
groups in 3-dimensional quantum gravity.  The question whether it
is `correct' to use these double covers as gauge groups seems to have 
no answer --- until we couple quantum gravity to spinors, at which 
point the double cover is necessary.   

\subsubsection*{Remarks}

1. In these calculations we have been ignoring the boundary terms that
arise when we integrate by parts on a manifold with boundary.  They are
valid if either $M$ is compact or if $M$ is compact with boundary and 
boundary conditions are imposed that make the boundary terms vanish.
$BF$ theory on manifolds with boundary is interesting both for its
applications to black hole physics --- where the event horizon may be
treated as a boundary --- and as an example of an `extended topological
field theory'.  

\section{Classical Phase Space} \label{phase}

To determine the classical phase space of $BF$ theory we assume spacetime 
has the form
\[                 M = \R \times S  \]
where the real line $\R$ represents time and $S$ is an oriented smooth
$(n-1)$-dimensional manifold representing space.  This is no real loss
of generality, since any oriented hypersurface in any oriented
$n$-dimensional manifold has a neighborhood of this form.  We can thus
use the results of canonical quantization to study the dynamics of $BF$
theory on quite general spacetimes.

If we work in temporal gauge, where the time component of the connection
$A$ vanishes, we see the momentum canonically conjugate to $A$ is
\[            {\partial \L \over \partial \dot A}  = E  .\]
This is reminiscent of the situation in electromagnetism, where the
electric field is canonically conjugate to the vector potential.   This
is why we use the notation `$E$'.  Originally people used the notation
`$B$' for this field, hence the term `$BF$ theory', which has
subsequently become ingrained.   But to understand the physical meaning
of the theory, it is better to call this field `$E$' and think of it as
analogous to the electric field.   Of course, the analogy is best when
$G = \U(1)$.

Let $P|_S$ be the restriction of the bundle $P$ to the `time-zero' slice
$\{0\} \times S$, which we identify with $S$.  Before we take into 
account the constraints imposed by the field equations, the
configuration  space of $BF$ theory is the space $\A$ of connections on
$P|_S$.  The corresponding classical phase space, which we call the
`kinematical phase space', is the cotangent bundle $T^\ast \A$.  A point
in this phase space consists of a connection $A$ on $P|_S$ and an
$\ad(P|_S)$-valued $(n-2)$-form $E$ on $S$.  The symplectic structure on
this phase space is given by
\[       \omega((\delta A,\delta E),(\delta A',\delta E')) = 
\int_S \tr(\delta A \we \delta E' - \delta A' \we \delta E) .\]
This reflects the fact that $A$ and $E$ are canonically conjugate
variables. However, the field equations of $BF$ theory put constraints
on  the initial data $A$ and $E$:
\[            B = 0 , \qquad d_A E = 0 \]
where $B$ is the curvature of the connection $A \in \A$, analogous to
the magnetic field in electromagnetism.  To deal with these constraints,
we should apply symplectic reduction to $T^\ast \A$ to obtain the
physical phase space.  

The constraint $d_A E = 0$, called the Gauss law, is analogous to the
equation in vacuum electromagnetism saying that the divergence of the
electric field vanishes.   This constraint generates the action of gauge
transformations on $T^\ast \A$.  Doing symplectic reduction with respect
to this constraint, we thus obtain the `gauge-invariant phase space'
$T^\ast(\A/\G)$, where $\G$ is the group of gauge transformations of the
bundle $P|_S$.    

The constraint $B = 0$ is analogous to an equation requiring 
the magnetic field to vanish.   Of course, no such equation exists
in electromagnetism; this constraint is special to $BF$ theory.  
It generates transformations of the form 
\[        A \mapsto A, \qquad E \mapsto E + d_A \eta, \]
so these transformations, discussed in the previous section, really are
gauge symmetries as claimed.  Doing symplectic reduction with respect
to this constraint, we obtain the `physical phase space' $T^\ast(\A_0/\G)$, 
where $\A_0$ is the space of flat connections on $P|_S$.  Points in this 
phase space correspond to physical states of classical $BF$ theory.  

\subsubsection*{Remarks}

1.  The space $\A$ is an infinite-dimensional vector space, and if we
give it an appropriate topology, an open dense set of $\A/\G$ becomes an
infinite-dimensional smooth manifold.   The simplest way to precisely 
define $T^\ast(\A/\G)$ is as the cotangent bundle of this open dense
set.  The remaining points correspond to connections with more symmetry
than the rest under gauge transformations.  These are called `reducible'
connections.   A more careful definition of the physical phase phase 
space would have to take these points into account.

\noindent
2.   The space $\A_0/\G$ is called the `moduli space of flat
connections on $P|_S$'.  We can understand it better as follows.   
Since the holonomy of a flat connection around a loop does not change 
when we apply a homotopy to the loop, a connection $A \in \A_0$ determines 
a homomorphism from the fundamental group $\pi_1(S)$ to $G$ after we
trivialize $P$ at the basepoint $p \in S$ that we use to define the
fundamental group.   If we apply a gauge transformation to $A$, this
homomorphism is conjugated by the value of this gauge transformation at
$p$.  This gives us a map from $\A_0/\G$ to $\hom(\pi_1(S),G)/G$, where
$\hom(\pi_1(S),G)$ is the space of homomorphisms from $\pi_1(S)$ to $G$, 
and $G$ acts on this space by conjugation.  When $S$ is connected this
map is one-to-one, so we have
\[        \A_0/\G \subseteq \hom(\pi_1(S),G)/G   .\]
The space $\hom(\pi_1(S),G)/G$ is called the `moduli space of flat
$G$-bundles over $S$'.  When $\pi_1(S)$ is finitely generated (e.g.\
when $S$ is  compact) this space is a real algebraic variety, and
$\A_0/\G$ is a subvariety.   Usually $\A_0/\G$ has singularities, but
each component has an open dense set that is a smooth manifold.   When
we speak of $T^\ast(\A_0/\G)$ above, we really mean the cotangent bundle
of this open dense set, though again a more careful treatment would deal
with the singularities.

We can describe $\A_0/\G$ much more explicitly in particular cases. 
For example, suppose that $S$ is a compact oriented surface of genus
$n$.  Then the group $\pi_1(S)$ has a presentation with $2n$ generators
$x_1,y_1, \dots, x_n,y_n$ satisfying the relation 
\[    R(x_i,y_i) := (x_1y_1x_1^{-1}y_1^{-1}) \cdots
(x_n y_n x_n^{-1} y_n^{-1}) = 1 .\] 
A point in $\hom(\pi_1(S),G)$ may thus be identified with 
a collection $g_1,h_1, \dots, g_n,h_n$ of elements of $G$ satisfying 
\[   R(g_i,h_i) = 1 ,\]
and a point in $\hom(\pi_1(S),G)/G$ is an equivalence class $[g_i, h_i]$
of such collections.   

The cases $G = \SU(2)$ and $G = \SO(3)$ are particularly interesting for
their applications to 3-dimensional Riemannian general relativity. 
When $G = \SU(2)$, all $G$-bundles  over a compact oriented surface $S$
are isomorphic, and $\A_0/\G = \hom(\pi_1(S),G)/G$.   When  $G =
\SO(3)$, there are two isomorphism classes of $G$-bundles over $S$,
distinguished by their second Stiefel-Whitney number $w_2 \in \Z_2$. 
For each of these bundles, the points $[g_i, h_i]$ that lie in 
$\A_0/\G$ can be described as follows.   Choose representatives  $g_i,
h_i \in \SO(3)$ and choose elements $\tilde g_i, \tilde h_i$  that map
down to these representatives via the double cover $\SU(2) \to \SO(3)$.
Then $[g_i, h_i]$ lies in $\A_0/\G$ if and only if 
\[ (-1)^{w_2} = R(\tilde g_i ,\tilde h_i) .\] 
For 3-dimensional Riemannian general relativity with gauge group
$\SO(3)$, the relevant bundle is the frame bundle of $S$, which has $w_2
= 0$.   For both $\SU(2)$ and $\SO(3)$, the space $\A_0/\G$ has
dimension $6n - 6$ for $n \ge 2$.  For the torus $\A_0/\G$ has
dimension 2, and for the sphere it is a single point.

\section{Canonical Quantization} \label{canonical.quantization}

In the previous section we described the kinematical, gauge-invariant
and physical phase spaces for $BF$ theory.  All of these are cotangent
bundles.   Naively, quantizing any one of them should give the Hilbert
space of square-integrable functions on the corresponding configuration
space.  We can summarize this hope with the following diagram:
\[
 \divide\dgARROWLENGTH by2
\begin{diagram}[L^2(\A_0/\G)]
\node{T^\ast(\A)} \arrow{s,l}{constrain} \arrow[2]{e,t}{quantize} 
\node[2]{L^2(\A)} \arrow{s,r}{constrain} \\
\node{T^\ast(\A/\G)}  \arrow{s,l}{constrain} \arrow[2]{e,t}{quantize}
\node[2]{L^2(\A/\G)} \arrow{s,r}{constrain} \\
\node{T^\ast(\A_0/\G)}  \arrow[2]{e,t}{quantize}
\node[2]{L^2(\A_0/\G)} 
\end{diagram}
\]
Traditionally it had been difficult to realize this hope with any degree
of rigor because the spaces $\A$ and $\A/\G$ are typically 
infinite-dimensional, making it difficult to define $L^2(\A)$ and
$L^2(\A/\G)$.  The great achievement of loop quantum gravity is that it
gives rigorous and background-free, hence diffeomorphism-invariant, 
definitions of these Hilbert spaces.   It does so by breaking away from
the traditional Fock space formalism and taking holonomies along paths
as the basic variables to be quantized.  The result is a picture in
which the basic excitations are not 0-dimensional particles but
1-dimensional `spin network edges'.  As we shall see, this eventually
leads us to a picture in which 1-dimensional Feynman diagrams are replaced
by 2-dimensional `spin foams'.  

In what follows we shall assume that the gauge group $G$ is compact and
connected and the manifold $S$ representing space is real-analytic.  The
case where $S$ merely smooth is considerably more complicated, but
people know how to handle it.   The case where $G$ is not connected 
would only require some slight modifications in our formalism.  However,
nobody really knows how to handle the case where $G$ is noncompact!
This is why, when we apply our results to quantum gravity, we consider
the quantization of the vacuum Einstein equations for Riemannian rather
than Lorentzian metrics: $\SO(n)$ is compact but $\SO(n,1)$ is not.  
The Lorentzian case is just beginning to receive the serious study that
it deserves.  

To define $L^2(\A)$, we start with the algebra $\Fun(\A)$ consisting of
all functions on $\A$ of the form
\[ \Psi(A) = f( Te^{\int_{\gamma_1}\! A}, \dots, Te^{\int_{\gamma_n}\! A}). \]
Here $\gamma_i$ is a real-analytic path in $S$, $T e^{\int_{\gamma_i}
A}$ is the holonomy of $A$ along this path, and $f$ is a continuous
complex-valued function of finitely many such holonomies.  Then we define
an inner product on $\Fun(\A)$ and complete it to obtain the Hilbert
space $L^2(\A)$.   To define this inner product, we need to think about
graphs embedded in space:

\begin{defn}\et A finite collection of real-analytic paths $\gamma_i \maps
[0,1] \to S$ form a {\rm graph in $S$} if they are embedded and
intersect, it at all, only at their endpoints.  We then call them {\rm
edges} and call their endpoints {\rm vertices}.  Given a vertex $v$,
we say an edge $\gamma_i$ is {\rm outgoing from $v$} if $\gamma_i(0) = 
v$, and we say $\gamma_i$ is {\rm incoming to $v$} if $\gamma_i(1) = v$.
\end{defn}

Suppose we fix a collection of paths $\gamma_1,\dots,\gamma_n$ that
form a graph in $S$.  We can think of the holonomies along 
these paths as elements of $G$.  Using this idea one can show that 
the functions of the form
\[  \Psi(A) = 
f( Te^{\int_{\gamma_1}\! A}, \dots, Te^{\int_{\gamma_n}\! A})  \]
for these particular paths $\gamma_i$ form a subalgebra of $\Fun(\A)$ 
that is isomorphic to the algebra of all continuous complex-valued 
functions on $G^n$.  Given two functions in this subalgebra, we can thus 
define their inner product by 
\[ \langle \Psi,\Phi \rangle = \int_{G^n} \overline{\Psi} \Phi  \]
where the integral is done using normalized Haar measure on $G^n$.  
Moreover, given any functions $\Psi,\Phi \in \Fun(\A)$ there is always
some subalgebra of this form that contains them.  Thus we can always
define their inner product this way.  Of course we have to check that 
this definition is independent of the choices involved, but this is not too
hard.  Completing the space $\Fun(\A)$ in the norm associated to this
inner product, we obtain the `kinematical Hilbert space' $L^2(\A)$.  

Similarly, we may define $\Fun(\A/\G)$ to be the space consisting of all 
functions in $\Fun(\A)$ that are invariant under gauge transformations,
and complete it in the above norm to obtain the `gauge-invariant 
Hilbert space' $L^2(\A/\G)$.   This space can be described in a very 
concrete way: it is spanned by `spin network states'.

\begin{defn}\et A {\rm spin network in $S$} is a triple $\Psi = (\gamma,
\rho,\iota)$ consisting of:
\begin{enumerate}
\item a graph $\gamma$ in $S$, 
\item for each edge $e$ of $\gamma$, an irreducible representation
$\rho_e$ of $G$,
\item for each vertex $v$ of $\gamma$, an intertwining operator 
\[  \iota_v \maps \rho_{e_1} \tensor \cdots \tensor \rho_{e_n}
\to \rho_{e'_1} \tensor \rho_{e'_m} \]
where $e_1,\dots, e_n$ are the edges incoming to $v$ and $e'_1,
\dots e'_m$ are the edges outgoing from $v$.  
\end{enumerate}
In what follows we call an intertwining operator an {\rm intertwiner}.
\end{defn}

There is an easy way to get a function in $\Fun(\A/\G)$ from a spin
network in $S$.  To explain how it works, it is easiest to give an
example.  Suppose we have a spin network $\Psi$ in $S$ with three edges
$e_1,e_2,e_3$ and two vertices $v_1,v_2$ as follows:

\vskip 2em
\centerline{\epsfysize=2.5in\epsfbox{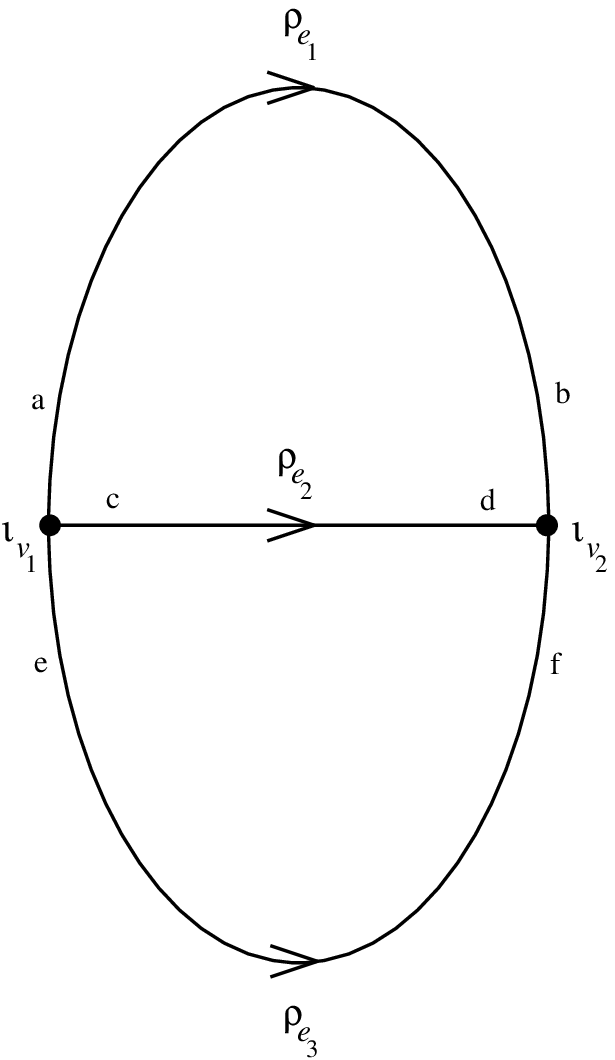}} \medskip

\noindent
We draw arrows on the edges to indicate their orientation, and 
write little letters near the beginning and end of each edge.  
Then for any connection $A \in \A$ we define
\[ \Psi(A) =   \rho_{e_1}(T e^{\int_{e_1} A})^a_b  \;
               \rho_{e_2}(T e^{\int_{e_2} A})^c_d  \;
               \rho_{e_3}(T e^{\int_{e_3} A})^e_f \;
               ({\iota_{v_1}})_{ace} \; ({\iota_{v_2}})^{bdf}  \]
In other words, we take the holonomy along each edge of $\Psi$, think of
it as a group element, and put it into the representation labelling that
edge.  Picking a basis for this representation we think of the result as
a matrix with one superscript and one subscript.   We use the little
letter near the beginning of the edge for the superscript and the little
letter near the end of the edge for the subscript.   In addition, we
write the intertwining operator for each vertex as a tensor.  This
tensor has one superscript for each edge incoming to the vertex and one
subscript for each edge outgoing from the vertex.  Note that this recipe
ensures that each letter appears once as a superscript and once as a
subscript!  Finally, using the Einstein summation convention we sum
over all repeated indices and get a number, which of course depends on
the connection $A$.  This is $\Psi(A)$.  

Since $\Psi \maps \A \to \C$ is a continuous function of finitely many
holonomies, it lies in $\Fun(\A)$.   Using the fact that the $\iota_v$
are intertwiners, one can show that this function is gauge-invariant.
We thus have $\Psi \in \Fun(\A/\G)$.   We call $\Psi$ a `spin network
state'.  The only hard part is to prove that spin network states span
$L^2(\A/\G)$.   We give some references to the proof in the Notes.

The constraint $F = 0$ is a bit more troublesome.   If we impose this
constraint at the classical level, symplectic reduction takes us from
$T^\ast(\A/\G)$ to the physical phase space $T^\ast(\A_0/\G)$.  
Heuristically, quantizing this should give the `physical Hilbert space'
$L^2(\A_0/\G)$.  However, for this to make sense, we need to choose
a measure on $\A_0/\G$.  This turns out to be problematic. 

The space $\A_0/\G$ is called the `moduli space of flat connections'. 
As explained in Remark 3 below, it has a natural measure when $S$ is
compact and of dimension 2 or less.   It also has a natural measure when
$S$ is simply connected, since then it is a single point, and we can use
the Dirac delta measure at that point.  In these cases the physical
Hilbert space is well-defined.  In most other cases, there seems to be no
natural measure on the moduli space of flat connections, so we cannot
unambiguously define the physical Hilbert space.  

If we are willing to settle for a mere vector space instead of a Hilbert
space, there is something that works quite generally.  Every function in
$\Fun(\A/\G)$ restricts to a gauge-invariant function on the space of
flat connections, or in other words, a function on $\A_0/\G$.    We
denote the space of such functions by $\Fun(\A_0/\G)$.   In the cases 
listed above where there is a natural measure on $\A_0/\G$, the space
$\Fun(\A_0/\G)$ is dense in $L^2(\A_0/\G)$.   In what follows, we abuse 
language by  calling elements of $\Fun(\A_0/\G)$ `physical states' even 
when there is no best measure on $\A_0/\G$.   Of course, a space of 
physical states without an inner product is of limited use.
Nonetheless the mathematics turns out to be very important for other 
things, so we proceed to study this space anyway.

We can understand $\Fun(\A_0/\G)$ quite explicitly using the fact that
every spin network in $S$ gives a function in this space.  In fact, if we
give $\Fun(\A_0/\G)$ a reasonable topology, like the sup norm topology,
finite linear combinations of spin network states are dense in this
space. Moreover, one can work out quite explicitly when two linear
combinations of spin networks define the same physical state.  For
example, two spin networks in $S$ differing by a homotopy define the
same physical state, because the holonomy of a flat connection along a
path does not change when we apply a homotopy to the path. There are
also other relations, called `skein relations', coming from the
representation theory of the group $G$.
 
For example, suppose $\rho$ is any irreducible representation of 
the group $G$.   Then the following skein relation holds:

\vskip 2em
\centerline{\epsfysize=1.5in\epsfbox{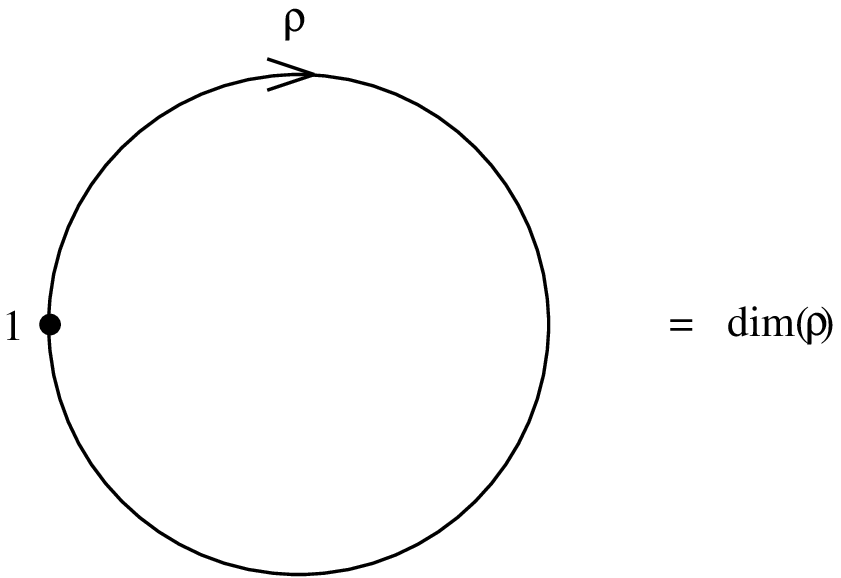}} \medskip

\noindent Here the left-hand side is a spin network with one edge $e$
labelled by the representation $\rho$ and one vertex labelled by the
identity intertwiner.   The edge is a contractible loop in 
$S$.  The corresponding spin network state $\Psi$ is given by
\[        \Psi(A) = \tr(\rho(T e^{\oint_e A})). \]
The skein relation above means that $\Psi(A) = \dim(\rho)$ when $A$
is flat.  The reason is that the holonomy of a flat connection around a
contractible loop is the identity, so its trace in the representation
$\rho$ is $\dim(\rho)$.   As a consequence, whenever a spin network has
a piece that looks like the above picture, if we eliminate that piece
and multiply the remaining spin network state by $\dim(\rho)$, we 
obtain the same physical state.   

People usually do not bother to draw vertices that are labelled by
identity intertwiners.  From now on we shall follow this custom.  Thus
instead of the above skein relation, we write:

\vskip 2em
\centerline{\epsfysize=1.5in\epsfbox{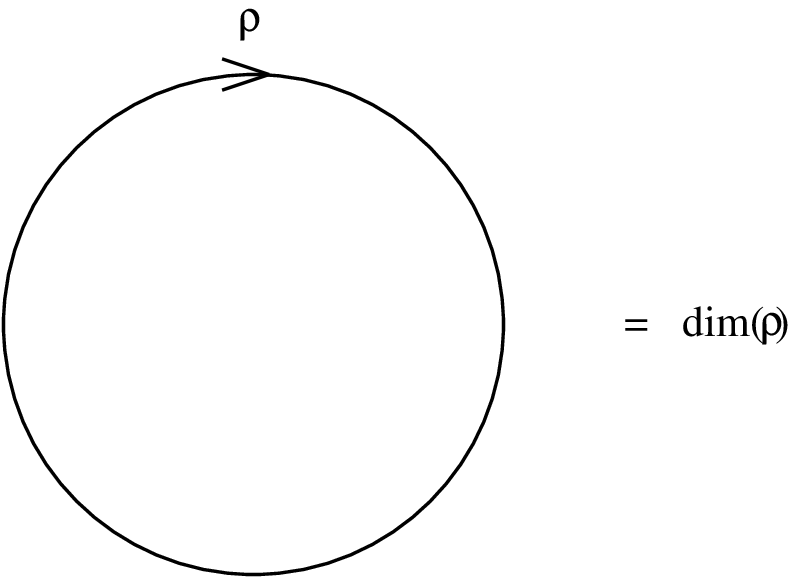}} \medskip

Moving on to something a bit more complicated, let us consider spin
networks with trivalent vertices.  Given any pair of irreducible
representations $\rho_1, \rho_2$ of $G$, their tensor product can be
written as a direct sum of irreducible representations.  Picking one of
these and calling it $\rho_3$, the projection from $\rho_1 \tensor
\rho_2$ to $\rho_3$ is an intertwiner that we can use to label a
trivalent vertex.  However, it is convenient to multiply this projection
by a constant so as to obtain an intertwiner $\iota \maps \rho_1
\tensor \rho_2 \to \rho_3$ with $\tr(\iota \iota^\ast)  = 1$.   We then
have the skein relation

\vskip 2em
\centerline{\epsfysize=2.5in\epsfbox{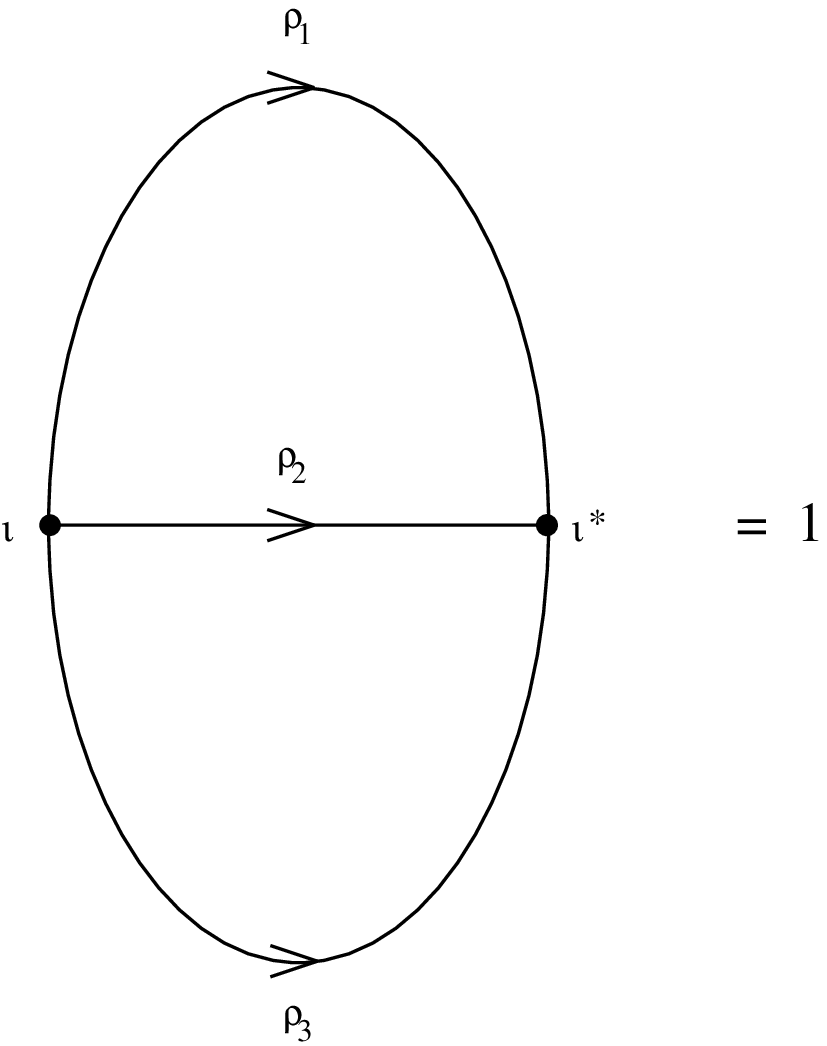}} \medskip

\noindent whenever this graph sits in $S$ in a contractible way.  Again,
this skein relation means that the spin network on the left side of the
equation defines a function $\Psi \in \Fun(\A/\G)$ that equals 1 on all
flat connections.    Whenever a spin network in $S$ has a piece that
looks like this, we can eliminate that piece without changing the
physical state it defines.

Of course, if the irreducible representation $\rho_3$ appears more than
once in the direct sum decomposition of $\rho_1 \tensor \rho_2$ there
will be more than one intertwiner of the above form.   We can always
pick a basis of such intertwiners such that $\iota_1 \iota_2^* = 0$ for
any two distinct intertwiners $\iota_1,\iota_2$ in the basis.  We then
have the following skein relation:

\vskip 2em
\centerline{\epsfysize=2.5in\epsfbox{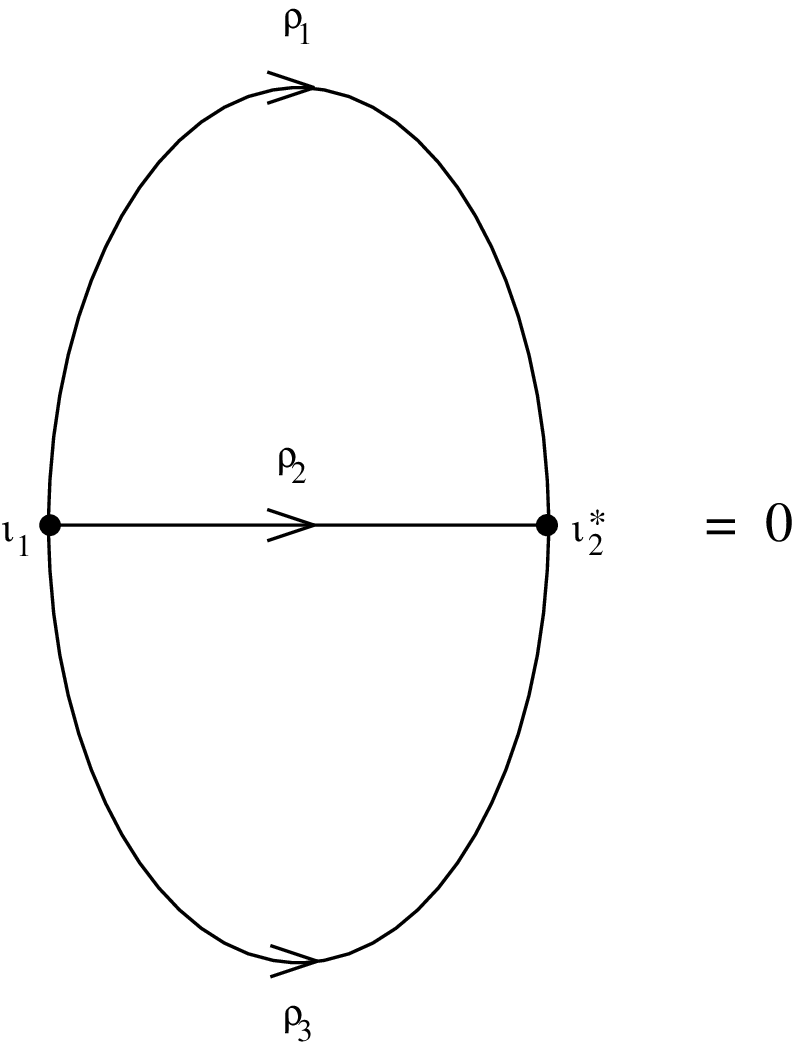}} \medskip

Let us pick such a basis of intertwiners for each triple of irreducible
representations of $G$.   To get enough states to span $\Fun(\A_0/\G)$,
it suffices to use these special intertwiners --- appropriately dualized
when necessary --- to label trivalent vertices.  What about vertices of
higher valence?  We can break any 4-valent vertex into two trivalent
ones using the following sort of skein relation:

\vskip 2em
\centerline{\epsfysize=1.5in\epsfbox{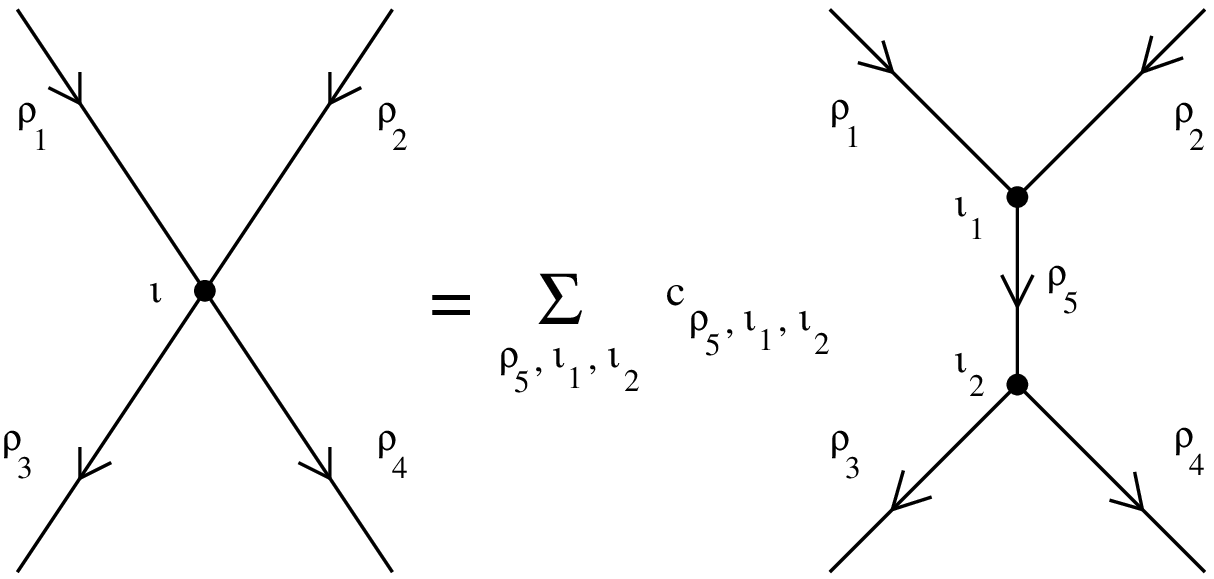}} \medskip

Here the sum is over irreducibles $\rho_5$ and intertwiners $\iota_1,
\iota_2$ in the chosen bases.  The coefficient depend on the details of
the intertwiners in question.  Both sides of this relation are to be
interpreted as part of a larger spin network.  The rest of the spin
network, not shown in the figure, is arbitrary but the same for both
sides.   Similar skein relations hold for vertices of valence 5 or more.
Using these skein relations and the tricks discussed in Remark 2 below,
we can write any physical state as a linear combination of states coming
from trivalent spin networks.

Philosophically, skein relations are intriguing because they can be
interpreted in two different ways: either as facts about $BF$ theory, 
or as facts about group representation theory.  In the first
interpretation, which we have emphasized here, the spin network edges
represent actual curves embedded in space.  In the second
interpretation, they are merely an abstract notation for 
representations of $G$.  The fact that both interpretations are possible
shows that in some sense $BF$ theory is nothing but a clever way to 
encode the representation theory of $G$ in a quantum field theory.   
Ultimately, this is the real reason why $BF$ theory is so interesting.

\subsubsection*{Remarks}

1.  The reason for assuming $S$ is real-analytic is that given a finite
collection of real-analytic paths $\gamma_i$ in $S$, there is always
some graph in $S$ such that each path $\gamma_i$ is a product of
finitely many edges of this graph.  This is not true in the smooth
context: for example, two smoothly embedded paths can intersect in a
Cantor set.  One can generalize the construction of $L^2(\A)$ and
$L^2(\A/\G)$ to the smooth context, but one needs a generalization of
graphs known as `webs'.  The smooth and real-analytic categories are
related as nicely as one could hope: a paracompact smooth manifold of
any dimension admits a real-analytic structure, and this structure is
unique up to a smooth diffeomorphism.  

\noindent
2.  There are various ways to modify a spin network in $S$ without changing
the state it defines: 
\begin{itemize}
\item  We can reparametrize an edge by any orientation-preserving
diffeomorphism of the unit interval.  
\item  We can reverse the orientation of an edge while simultaneously
replacing the representation labelling that edge by its dual and
appropriately dualizing the intertwiners labelling the endpoints of
that edge.
\item  We can subdivide an edge into two edges labelled with the
same representation by inserting a vertex labelled with the identity 
intertwiner.   
\item  We can eliminate an edge labelled by the trivial representation.
\end{itemize}
In fact, two spin networks in $S$ define the same state in $L^2(\A/\G)$
if and only if they differ by a sequence of these moves and their
inverses.  It is usually best to treat two such spin networks as `the
same'.

\noindent
3.  When $S$ is a circle, $\A_0/\G$ is just the space of conjugacy
classes of $G$.  The normalized Haar measure on $G$ can be pushed down
to this space.  We can easily extend this idea to put a measure on
$\A_0/\G$ whenever $S$ is a compact and 1-dimensional.   When $S$ is
compact and 2-dimensional the space $\A_0/\G$ is an algebraic variety
described as in Remark 2 of the previous section.  There is a natural
symplectic structure on the smooth part of this variety, given by
\[        \omega([\delta A],[\delta A']) = \int_S \tr(\delta A \we
\delta A')  \]
where $\delta A, \delta' A$ are tangent vectors to $\A_0$, i.e., 
$\ad(P)$-valued 1-forms.  Raising $\omega$ to a suitable power we
obtain a volume form, and thus a measure, on $\A_0/\G$.

\noindent 4.  The theory of Reidemeister torsion helps to explain why
there is typically a natural measure on the moduli space of flat
connections only in dimensions 2 or less.  The Reidemeister torsion
is a natural section of a certain bundle on the moduli space of flat
connections.  In dimensions 2 or less we can think of this section as
a volume form, but in most other cases we cannot.

\section{Observables}  \label{observables}

The true physical observables in $BF$ theory are self-adjoint operators
on the physical Hilbert space, when this space is well-defined. 
Nonetheless it is interesting to consider operators on the
gauge-invariant Hilbert space $L^2(\A/\G)$.    These are relevant not
only to $BF$ theory but also other gauge theories, such as 4-dimensional
Lorentzian general relativity in terms of real Ashtekar variables,
where the gauge group is $\SU(2)$.  In what follows we shall use the
term `observables' to refer to operators on the gauge-invariant Hilbert
space.  We consider observables of two kinds: functions of $A$ and
functions of $E$.   

Since $A$ is analogous to the `position' operator in elementary quantum
mechanics while $E$ is analogous to the `momentum', we expect that
functions of $A$ act as multiplication operators while functions of $E$
act by differentiation.  As usual in quantum field theory, we need to
smear these fields --- i.e., integrate them over some region of space
--- to obtain operators instead of operator-valued distributions.  Since
$A$ is like a 1-form, it is tempting to smear it by integrating it over
a path.  Similarly, since $E$ is like an $(n-2)$-form, it is tempting to
integrate it over an $(n-2)$-dimensional submanifold.  This is
essentially what we shall do.  However, to obtain operators on the
gauge-invariant Hilbert space $L^2(\A/\G)$, we need to quantize
gauge-invariant functions of $A$ and $E$.  

The simplest gauge-invariant function of the $A$ field is a `Wilson
loop': a function of the form
\[         \tr(\rho(T e^{\oint_\gamma A})) \]  
for some loop $\gamma$ in $S$ and some representation $\rho$ of $G$.
In the simplest case, when $G = \U(1)$ and the loop $\gamma$ bounds
a disk, we can use Stokes' theorem to rewrite $\oint_\gamma A$ 
as the flux of the magnetic field through this disk.   In general,
a Wilson loop captures gauge-invariant information about the holonomy 
of the $A$ field around the loop.  

A Wilson loop is just a special case of a spin network, and we can get
an operator on $L^2(\A/\G)$ from any other spin network in a similar
way.  As we have seen, any spin network in $S$ defines a function $\Psi
\in \Fun(\A/\G)$.  Since $\Fun(\A/\G)$ is an algebra, multiplication by
$\Psi$ defines an operator on $\Fun(\A/\G)$.  Since $\Psi$ is a bounded
function, this operator extends to a bounded operator on $L^2(\A/\G)$.  
We call this operator a `spin network observable'.   Note that since
$\Fun(\A/\G)$ is an algebra, any product of Wilson loop observables can
be written as a finite linear combination of spin network observables. 
Thus spin network observables give a way to measure correlations among
the holonomies of $A$ around a collection of loops.  

When $G = \U(1)$ it is also easy to construct gauge-invariant functions
of $E$.  We simply take any compact oriented $(n-2)$-dimensional submanifold
$\Sigma$ in $S$, possibly with boundary, and do the integral
\[          \int_\Sigma E  .\]
This measures the flux of the electric field through $\Sigma$. 
Unfortunately, this integral is not gauge-invariant when $G$ is nonabelian,
so we need to modify the construction slightly to handle the nonabelian
case.  Write
\[        E|_\Sigma = e \, d^{n-2}x \]
for some $\g$-valued function $e$ on $\Sigma$ and some $(n-2)$-form
$d^{n-2} x$ on $\Sigma$ that is compatible with the orientation
of $\Sigma$.   Then 
\[    \int_\Sigma \sqrt{\langle e,e\rangle} \, d^{n-2}x \]
is a gauge-invariant function of $E$.  One can check that it does not
depend on how we write $E$ as $e \, d^{n-2}x$.   We can think of it
as a precise way to define the quantity 
\[                 \int_\Sigma |E|  .\]

Recall that 3-dimensional $BF$ theory with gauge group $\SU(2)$ or
$\SO(3)$ is a formulation of Riemannian general relativity in 3
dimensions. In this case $\Sigma$ is a curve, and the above quantity has
a simple interpretation: it is the {\it length} of this curve. 
Similarly, in 4-dimensional $BF$ theory with either of these gauge
groups, $\Sigma$ is a surface, and the above quantity can be interpreted 
as the {\it area} of this surface.  The same is true for 4-dimensional
Lorentzian general relativity formulated in terms of the real Ashtekar
variables.

Quantizing the above function of $E$ we obtain a self-adjoint operator
${\cal E}(\Sigma)$ on $L^2(\A/\G)$, at least when $\Sigma$ is 
real-analytically embedded in $S$.  We shall not present the
quantization procedure here, but only the final result.  Suppose $\Psi$
is a spin network in $S$.  Generically, $\Psi$ will intersect $\Sigma$
transversely at finitely many points, and these points will not be
vertices of $\Psi$:

\vskip 2em
\centerline{\epsfysize=2.5in\epsfbox{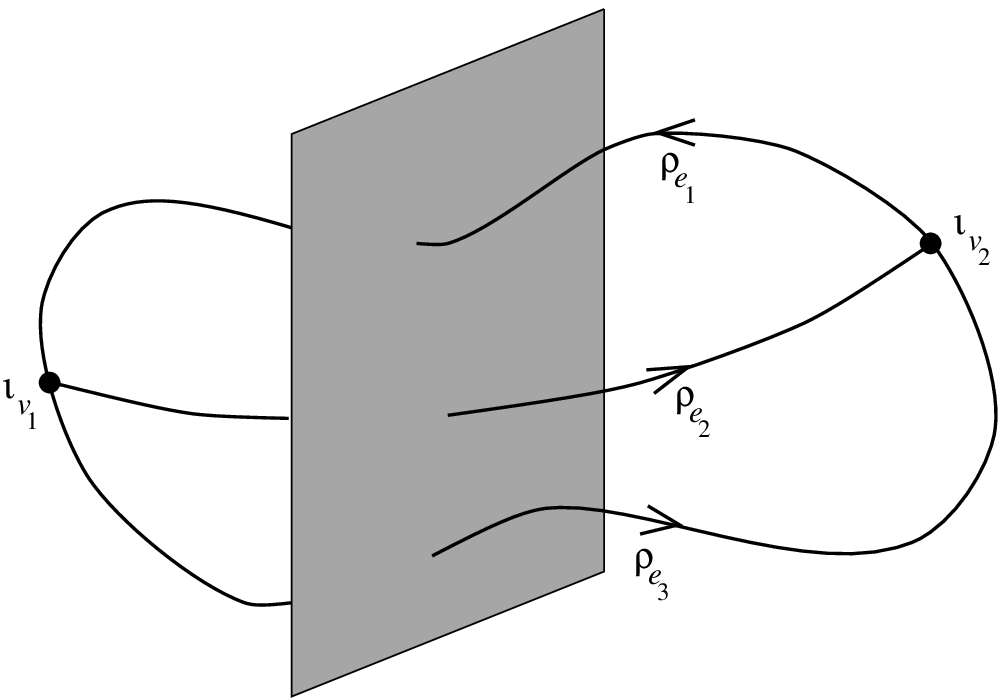}} \medskip

\noindent In this case we have
\[ {\cal E}(\Sigma) \Psi = 
\big( \sum_i C(\rho_i)^{1/2} \big) \Psi \]
Here the sum is taken over all points $p_i$ where an edge intersects the
surface $\Sigma$, and $C(\rho_i)$ denotes the Casimir of the
representation labelling that edge.   Note that the same edge may
intersect $\Sigma$ in several points; if so, we count each point
separately.   

This result clarifies the physical significance of spin network edges:
they represent {\it quantized flux lines of the $E$ field}.   In the
case of 3-dimensional Riemannian quantum gravity they have a particularly
simple geometrical meaning.  Here the observable ${\cal E}(\Sigma)$
measures the length of the curve $\Sigma$.  The irreducible
representations of $\SU(2)$ correspond to spins $j = 0, {1\over 2}, 1
\dots,$ and the Casimir equals $j(j+1)$ in the spin-$j$ representation. 
Thus a spin network edge labelled by the spin $j$ contributes a length
$\sqrt{j(j+1)}$ to any curve it crosses transversely.

As an immediate consequence, we see that the length of a curve is not a
continuously variable quantity in 3d Riemannian quantum gravity.  
Instead, it has a discrete spectrum of possible values!   We also see
here the difference between using $\SU(2)$ and $\SO(3)$ as our gauge
group: only integer spins correspond to irreducible representations of
$\SO(3)$, so the spectrum of allowed lengths for curves is sparser if we
use $\SO(3)$.   Of course, in a careful treatment we should also
consider spin networks intersecting $\Sigma$ nongenerically.  As
explained in Remark 1 below, these give the operator ${\cal E}(\Sigma)$
additional eigenvalues.  However, our basic qualitative conclusions here
remain unchanged.  

Similar remarks apply to 4-dimensional $BF$ theory with gauge group
$\SU(2)$, as well as quantum gravity in the real Ashtekar formulation. 
Here ${\cal E}(\Sigma)$ measures the area of the surface $\Sigma$, area
is quantized, and spin network edges give area to the surfaces they
intersect!   This is particularly intriguing given the
Bekenstein-Hawking formula saying that the entropy of a black hole is
proportional to its area.   It it natural to try to explain this result
by associating degrees of freedom of the event horizon to points where
spin network edges intersect it.  Attempts along these lines have been
made, and the results look promising.  Unfortunately, it is too much of
a digression to describe these here, so we refer the reader to the Notes
for more details.

\subsubsection*{Remarks}

1. The formula for ${\cal E}(\Sigma) \Psi$ is slightly more complicated
when the underlying graph $\gamma$ of $\Psi$ intersects $\Sigma$
nongenerically.  By subdividing its edges if necessary we may assume
this graph has the following properties: 
\begin{itemize}
\item If an edge of $\gamma$ contains a segment lying in $\Sigma$, 
it lies entirely in $\Sigma$.
\item Each isolated intersection point of $\gamma$ and $\Sigma$ is a vertex.
\item Each edge of $\gamma$ intersects $\Sigma$ at most once.  
\end{itemize}
For each vertex $v$ of $\gamma$ lying in $\Sigma$, we can divide the
edges incident to $v$ into three classes, which we call `upwards',
`downwards', and `horizonal'.  The `horizontal' edges are those lying in
$\Sigma$; the other edges are separated into two classes according to
which side of $\Sigma$ they lie on; using the orientation of $\Sigma$ 
we call these classes `upwards' and `downwards'.   Reversing
orientations of edges if necessary, we may assume all the upwards and
downwards edges are incoming to $v$ while the horizontal ones are
outgoing.   We can then write any intertwiner labelling $v$ as a linear
combination of intertwiners of the following special form:
\[    \iota_v \maps \rho_v^u \tensor \rho_v^d \to \rho_v^h  \]
where $\rho_v^u$ (resp.\ $\rho_v^d$, $\rho_v^h$) is an irreducible
summand of the tensor product of all the representations labelling
upwards (resp.\ downwards, horizontal) edges.   This lets us write any
spin network state with $\gamma$ as its underlying graph as a finite
linear combination of spin network states with intertwiners of  his
special form.   Now suppose $\Psi$ is a spin network state with 
intertwiners of this form.  Then we have
\[ {\cal E}(\Sigma) \Psi =  {1\over 2} 
\big( \sum_v  [2C(\rho_v^u) + 2C(\rho_v^d) - C(\rho_v^h)]^{1\over 2} \big) 
\Psi \]
where the sum is over all vertices at which $\Sigma$ intersects $\gamma$.
In the generic case $C(\rho_v^u) = C(\rho_v^d)$ and $C(\rho_v^h) = 0$, so 
this formula reduces to the previous one.  

\noindent
2.  When $G = \U(1)$ we can also quantize the observable $\int_\Sigma E$
when $\Sigma$ is real-analytically embedded in $S$, obtaining an
operator that measures the flux of the electric field through $\Sigma$.    
For any irreducible representation $\rho$ of $\U(1)$ there is an integer 
$Q(\rho)$ such that
\[     \rho(e^{i\theta}) = e^{iQ(\rho)\theta} ,\]
and using the notation of the previous remark this operator is given by
\[  \big(\int_\Sigma \hat{E}\big) \Psi = 
{1\over 2}\big( \sum_v Q(\rho_v^u) - Q(\rho_v^d) \big) \Psi. \]

\noindent
3.  As noted, the true physical observables in $BF$ theory are
self-adjoint operators on the physical Hilbert space $L^2(\A_0/\G)$. 
Examples include spin network observables: any spin network $\Psi$ in
$S$ defines a bounded function on $\A_0/\G$, and multiplication by  this
function defines a bounded operator on the physical Hilbert space.
Unlike the spin network observables on the gauge-invariant Hilbert
space, these operators remain unchanged when we apply any homotopy to
the underlying graph of the spin network, and they satisfy skein
relations. 

In the case of 3d $BF$ theory with gauge group $\SU(2)$ or $\SO(3)$, 
a maximal commuting algebra of operators on $L^2(\A_0/\G)$ is generated 
by Wilson loops corresponding to any set of generators of the fundamental 
group $\pi_1(S)$.  For example, we can use the generators described in 
Remark 2 of Section 3.  It suffices to use Wilson loops labelled by the
fundamental representation of the gauge group.    

\section{Canonical Quantization via Triangulations}  \label{triangulations}

Starting from classical $BF$ theory, canonical quantization has led us
to a picture in which states are described using spin networks embedded
in the manifold representing space.  But our discussion of skein
relations has shown that spin networks may also be regarded as abstract
diagrams arising naturally from the representation theory of the gauge
group $G$.  This is very appealing to those who cherish the hope that
someday quantum gravity will replace the differential-geometric
conception of spacetime by something more algebraic or combinatorial in
nature.   If something like this is true, spin networks may ultimately
be seen as more important than the manifold containing them!  To study
this possibility, we may isolate the following `abstract' notion of spin
network: 

\begin{defn}\et  A {\rm spin network} is a triple $\Psi = (\gamma,
\rho,\iota)$ consisting of:
\begin{enumerate}
\item a {\rm graph} $\gamma$: i.e., a finite set $\E$ of edges, a finite set
$\V$ of vertices, and {\rm source} and {\rm target} maps $s,t \maps \E \to \V$
assigning to each edge its two endpoints,
\item for each edge $e$ of $\gamma$, an irreducible representation
$\rho_e$ of $G$,
\item for each vertex $v$ of $\gamma$, an intertwiner
\[  \iota_v \maps \rho_{e_1} \tensor \cdots \tensor \rho_{e_n}
\to \rho_{e'_1} \tensor \cdots \tensor \rho_{e'_m} \]
where $e_1,\dots, e_n$ are the edges incoming to $v$ and $e'_1,
\dots, e'_m$ are the edges outgoing from $v$.  
\end{enumerate}
Here we say an edge is {\rm incoming to $v$} if its target
is $v$, and {\rm outgoing from $v$} if its source is $v$.
\end{defn}

People have already begun formulating physical theories in which such
abstract spin networks, not embedded in any manifold, describe the
geometry of space.  However it is still a bit difficult to relate such
theories to more traditional physics.  Thus it is useful to consider a
kind of halfway house: namely, spin networks in the dual 1-skeleton of a
triangulated manifold.  While purely combinatorial, these objects still
have a clear link to field theory as formulated on a pre-existing
manifold.

In this case of $BF$ theory this halfway house works as follows.   As
before, let us start with an $(n-1)$-dimensional real-analytic  manifold
$S$ representing space.  Given any triangulation of $S$ we can choose a
graph in $S$ called the `dual 1-skeleton', having one vertex at the
center of each $(n-1)$-simplex and one edge intersecting each
$(n-2)$-simplex.  Using homotopies and skein relations, we can express
any state in $\Fun(\A_0/\G)$ as a linear combination of states coming
from spin networks whose underlying graph is this dual 1-skeleton.  So
at least for $BF$ theory, there is no loss in working with spin networks
of this special form.  

It turns out that the working with a triangulation this way sheds new
light on the observables discussed in the previous section. Moreover,
the dynamics of $BF$ theory is easiest to describe using triangulations.   
Thus it pays to formalize the setup a bit more.  To do so, we 
borrow some ideas from lattice gauge theory.

Given a graph $\gamma$, define a `connection' on $\gamma$ to be an
assignment of an element of $G$ to each edge of $\gamma$, and denote
the space of such connections by $\A_\gamma$.  As in lattice gauge
theory, these group elements represent the holonomies along the
edges of the graph.  Similarly, define a `gauge transformation' on
$\gamma$ to be an assignment of a group element to each vertex, and
denote the group of gauge transformations by $\G_\gamma$.  This group
acts on $\A_\gamma$ in a natural way that mimics the usual action of
gauge transformations on holonomies.  Since $\A_\gamma$ is just a
product of copies of $G$, we can use normalized Haar measure on $G$ to
put a measure on $\A_\gamma$, and this in turn pushes down to a measure
on the quotient space $\A_\gamma/\G_\gamma$. Using these we can define
Hilbert spaces $L^2(\A_\gamma)$ and $L^2(\A_\gamma/\G_\gamma)$.   

In Section \ref{canonical.quantization} we saw how to extract a 
gauge-invariant function on the space of connections from any  spin
network embedded in space.  The same trick works in the present context:
any spin network $\Psi$ with $\gamma$ as its underlying graph defines a
function $\Psi \in L^2(\A_\gamma/\G_\gamma)$.  For example, if $\Psi$ is 
this spin network:

\vskip 2em
\centerline{\epsfysize=2.5in\epsfbox{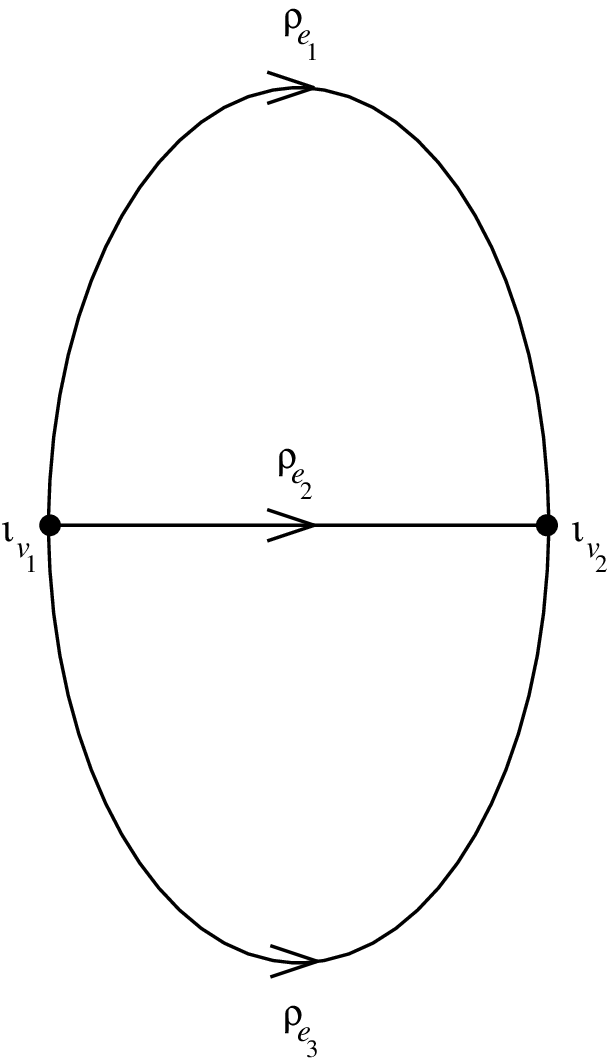}} \medskip

\noindent 
and the connection $A$ assigns the group elements $g_1,g_2,g_3$ to the
three edges of $\Psi$, we have 
\[ \Psi(A) =
\rho_{e_1}(g_1)_a^b \; \rho_{e_2}(g_2)^c_d \; \rho_{e_3}(g_3)^e_f \; 
({\iota_{v_1}})^f_{ac} \; ({\iota_{v_2}})^e_{db}  .  \] 
We again call such functions `spin network states'.  Not only do these
span $L^2(\A_\gamma/\G_\gamma)$, it is easy to choose an
orthonormal basis of spin network states.  Let $\Irrep(G)$ be a complete
set of irreducible unitary representations of $G$.  To obtain spin
networks $\Psi = (\gamma,\rho,\iota)$ giving an orthonormal basis of
$L^2(\A_\gamma/\G_\gamma)$, let $\rho$ range over all labellings of the
edges of $\gamma$ by representations in $\Irrep(G)$, and for each $\rho$
and each vertex $v$, let the intertwiners $\iota_v$ range over an 
orthonormal basis of the space of intertwiners
\[  \iota \maps \rho_{e_1} \tensor \cdots \tensor \rho_{e_n}
\to \rho_{e'_1} \tensor \cdots \tensor \rho_{e'_m} \]
where the $e_i$ are incoming to $v$ and the $e'_i$ are outgoing from $v$.

How do these purely combinatorial constructions relate to our previous
setup where space is described by a real-analyic manifold $S$ equipped
with a principal $G$-bundle?  Quite simply: whenever $\gamma$ is a graph
in $S$, trivializing the bundle at the vertices of this graph gives a
map from $\A$ onto $\A_\gamma$, and also a homomorphism from $\G$ onto
$\G_\gamma$.  Thus we have inclusions 
\[  L^2(\A_\gamma) \hookrightarrow L^2(\A) \] 
and  
\[   L^2(\A_\gamma/\G_\gamma) \hookrightarrow L^2(\A/\G) .\]

These constructions are particularly nice when $\gamma$ is the dual
1-skeleton of a triangulation of $S$.  Consider 3-dimensional 
Riemannian quantum gravity, for example.  In this case $\gamma$ is
always trivalent:

\vskip 2em
\centerline{\epsfysize=2.5in\epsfbox{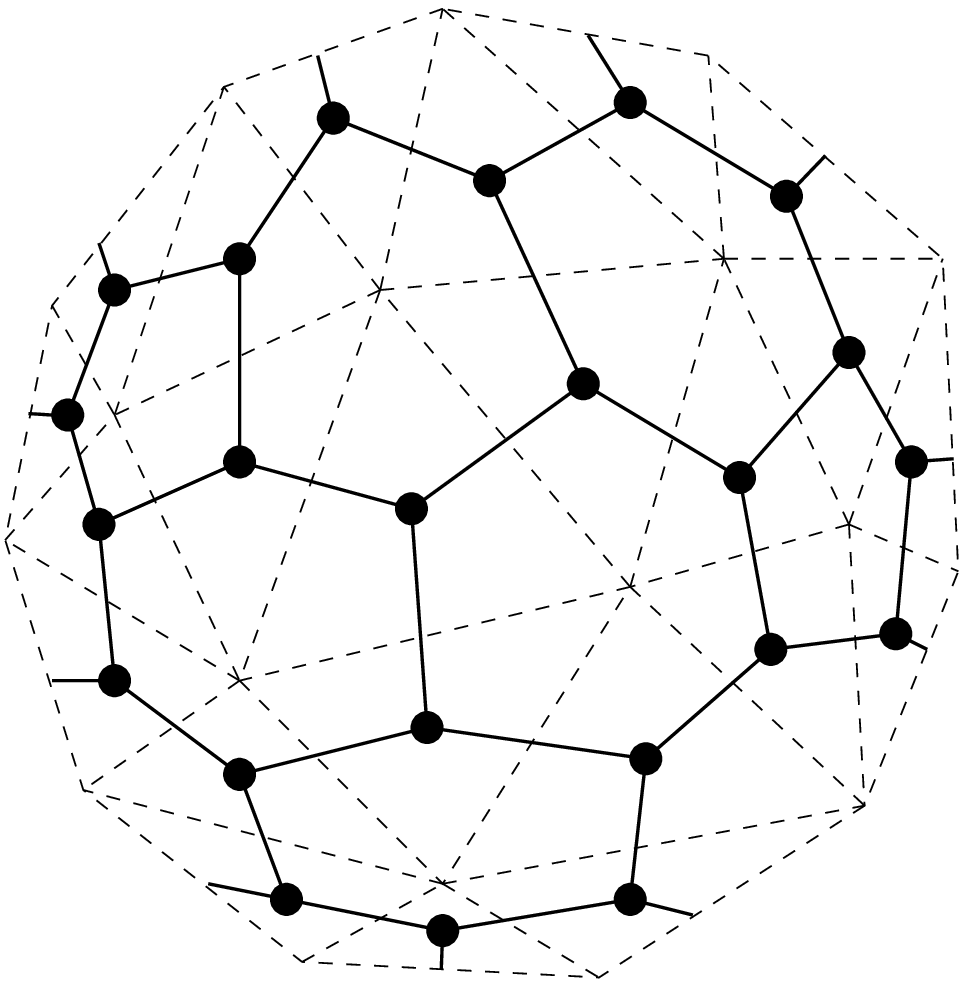}} \medskip

\noindent  Since the representations of $\SU(2)$ satisfy
\[    j_1 \tensor j_2 \iso |j_1 - j_2| \oplus \cdots
\oplus (j_1 + j_2) ,\]
each basis of intertwiners $\iota \maps j_1 \tensor j_2 \to j_3$
contains at most one element.   Thus we do not need to explicitly label
the vertices of trivalent $\SU(2)$ spin networks with intertwiners; we
only need to label the edges with spins.   We can dually think of these 
spins as labelling the edges of the original triangulation.  For example, 
the following spin network state:

\vskip 2em
\centerline{\epsfysize=2.5in\epsfbox{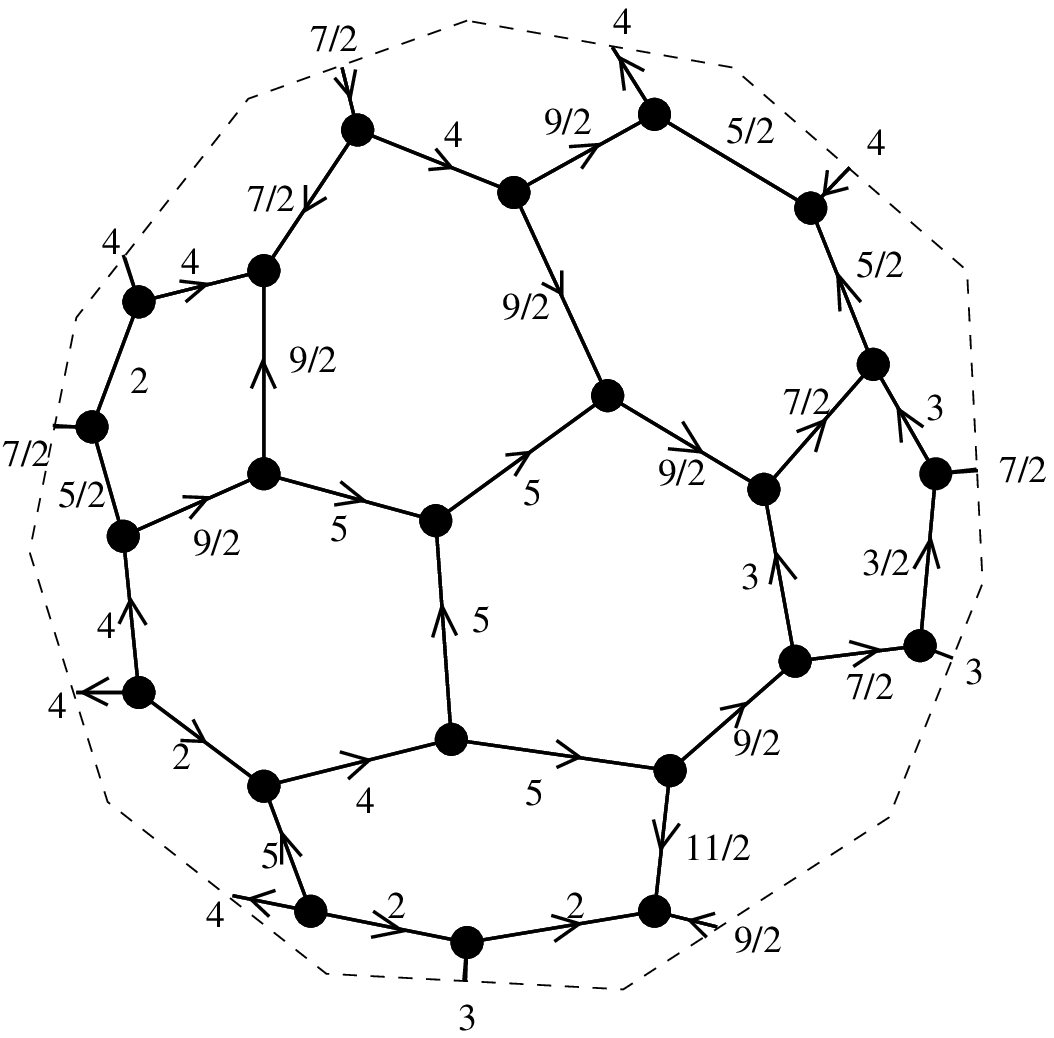}} \medskip

\noindent corresponds to a triangulation with edges labelled by
spins as follows:

\vskip 2em
\centerline{\epsfysize=2.5in\epsfbox{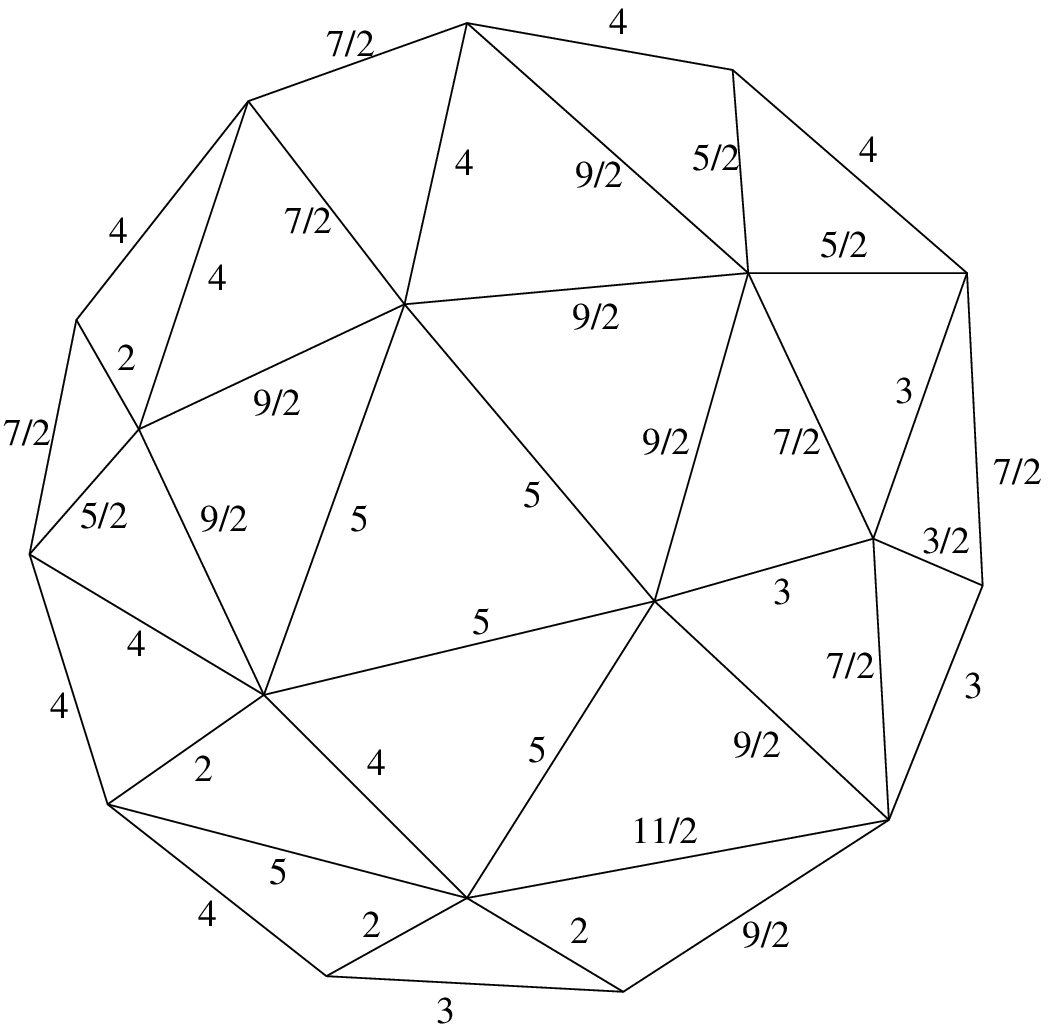}} \medskip

\noindent By the results of the previous section, these
spins specify the {\it lengths} of the edges, with spin $j$
corresponding to length $\sqrt{j(j+1)}$.   Note that for there
to be an intertwiner $\iota \maps j_1 \tensor j_2 \to j_3$, the
spins $j_1,j_2,j_3$ labelling the three edges of a given triangle 
must satisfy two constraints.  First, the triangle inequality must hold:
\[     |j_1 - j_2| \le j_3 \le  j_1 + j_2 . \]
This has an obvious geometrical interpretation.  Second, the spins
must sum to an integer.  This rather peculiar constraint would hold
automatically if we had used the gauge group $\SO(3)$ instead of
$\SU(2)$.  If we consider all labellings satisfying these constraints, we 
obtain spin network states forming a basis of $L^2(\A_\gamma/\G_\gamma)$.   

The situation is similar but a bit more complicated for 4-dimensional
$BF$ theory with gauge group $\SU(2)$.   Let $S$ be a triangulated
3-dimensional manifold and let $\gamma$ be its dual 1-skeleton.  Now
$\gamma$ is a 4-valent graph with one vertex in the center of each
tetrahedron and one edge intersecting each triangle.   To specify a 
spin network state in $L^2(\A_\gamma/\G_\gamma)$, we need to label 
each edge of $\gamma$ with a spin and each vertex with an intertwiner:

\vskip 2em
\centerline{\epsfysize=2.5in\epsfbox{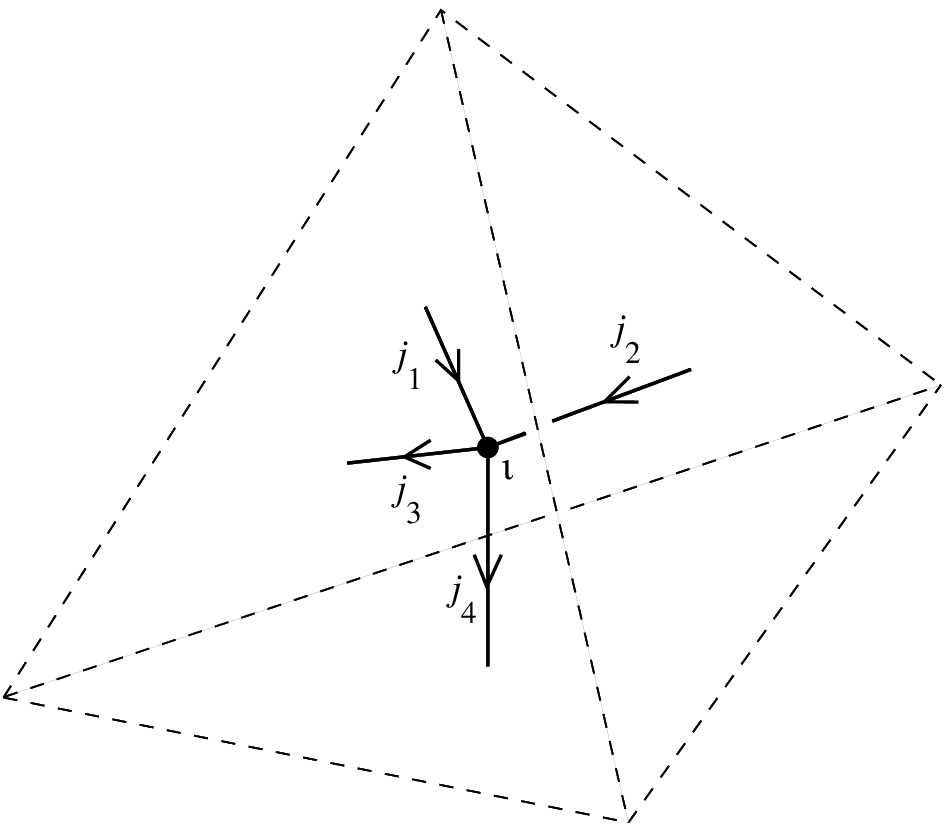}} \medskip

\noindent For each vertex there is a basis of intertwiners $\iota \maps j_1
\tensor j_2 \to j_3 \tensor j_4$ as described at the end of Section 
\ref{canonical.quantization}.  We can draw such an intertwiner by
formally `splitting' the vertex into two trivalent ones and labelling
the new edge with a spin $j_5$:

\vskip 2em
\centerline{\epsfysize=2.5in\epsfbox{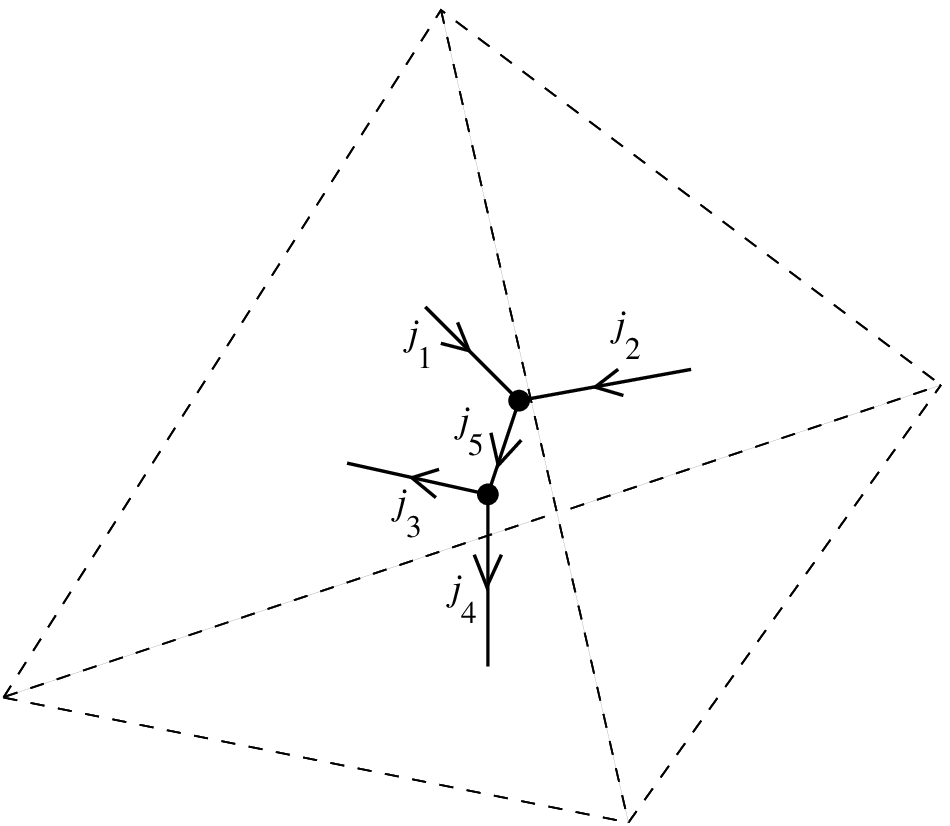}} \medskip

\noindent 
In the triangulation picture, this splitting corresponds to chopping the
tetrahedron in half along a parallelogram: 

\vskip 2em
\centerline{\epsfysize=2.5in\epsfbox{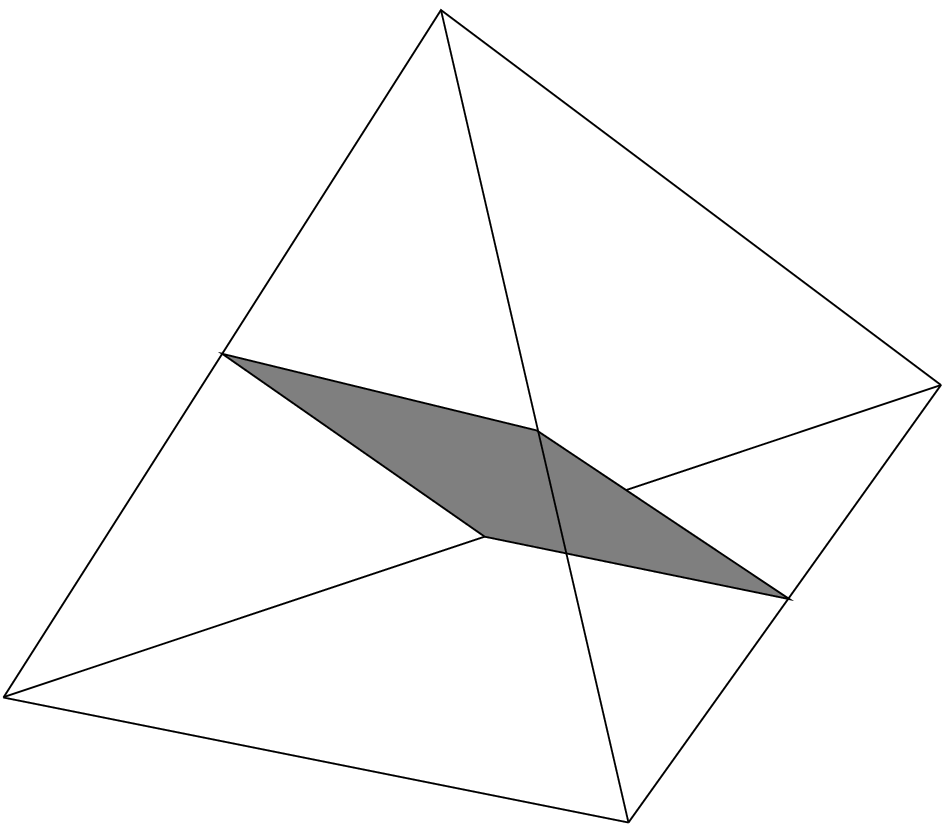}} \medskip

\noindent 
We can thus describe a spin network state in $L^2(\A_\gamma/\G_\gamma)$ 
by chopping each tetrahedron in half and labelling all the
resulting parallelograms, along with all the triangles, by spins.  These 
spins specify the {\it areas} of the parallelograms and triangles.  

It may seem odd that in this picture the geometry of each tetrahedron is
described by 5 spins, since classically it takes 6 numbers to specify
the geometry of a tetrahedron.  In fact, this is a consequence of the
uncertainty principle.  The area operators for surfaces
do not commute when the surfaces intersect.  There are three ways to 
chop a tetrahedron in half using a parallelogram, but we cannot 
simultaneously diagonalize the areas of these parallelograms, 
since they intersect.   We can describe a basis of states for the quantum 
tetrahedron using 5 numbers: the areas of its 4 faces and any
{\it one} of these parallelograms.  Different ways of chopping
tetrahedron in half gives us different bases of this sort, and the
matrix relating these bases goes by the name of the `$6j$ symbols':

\begin{center}

\setlength{\unitlength}{0.00083300in}%
\begingroup\makeatletter\ifx\SetFigFont\undefined
\def\x#1#2#3#4#5#6#7\relax{\def\x{#1#2#3#4#5#6}}%
\expandafter\x\fmtname xxxxxx\relax \def\y{splain}%
\ifx\x\y   
\gdef\SetFigFont#1#2#3{%
  \ifnum #1<17\tiny\else \ifnum #1<20\small\else
  \ifnum #1<24\normalsize\else \ifnum #1<29\large\else
  \ifnum #1<34\Large\else \ifnum #1<41\LARGE\else
     \huge\fi\fi\fi\fi\fi\fi
  \csname #3\endcsname}%
\else
\gdef\SetFigFont#1#2#3{\begingroup
  \count@#1\relax \ifnum 25<\count@\count@25\fi
  \def\x{\endgroup\@setsize\SetFigFont{#2pt}}%
  \expandafter\x
    \csname \romannumeral\the\count@ pt\expandafter\endcsname
    \csname @\romannumeral\the\count@ pt\endcsname
  \csname #3\endcsname}%
\fi
\fi\endgroup
\begin{picture}(5144,3044)(4779,-4583)
\thicklines
\put(5701,-2461){\circle*{88}}
\put(5701,-3661){\circle*{88}}
\put(8701,-3061){\circle*{88}}
\put(9301,-3061){\circle*{88}}
\put(6601,-1561){\line(-1,-1){900}}
\put(4801,-1561){\line( 1,-1){900}}
\put(5701,-3661){\line( 0, 1){1200}}
\put(4801,-4561){\line( 1, 1){900}}
\put(5701,-3661){\line( 1,-1){900}}
\put(8101,-1561){\line( 2,-5){600}}
\put(9901,-1561){\line(-2,-5){600}}
\put(8101,-4561){\line( 2, 5){600}}
\put(9301,-3061){\line( 2,-5){600}}
\put(8701,-3061){\line( 1, 0){600}}
\multiput(5634,-3061)(3.92222,-7.84444){19}{\makebox(6.6667,10.0000){\SetFigFont{7}{8.4}{rm}.}}
\multiput(5701,-3204)(3.75000,7.50000){21}{\makebox(6.6667,10.0000){\SetFigFont{7}{8.4}{rm}.}}
\multiput(5214,-4044)(-2.75000,-8.25000){19}{\makebox(6.6667,10.0000){\SetFigFont{7}{8.4}{rm}.}}
\multiput(5169,-4194)(8.38754,2.09689){18}{\makebox(6.6667,10.0000){\SetFigFont{7}{8.4}{rm}.}}
\multiput(5266,-1922)(2.75000,-8.25000){19}{\makebox(6.6667,10.0000){\SetFigFont{7}{8.4}{rm}.}}
\multiput(5311,-2072)(-8.38754,2.09689){18}{\makebox(6.6667,10.0000){\SetFigFont{7}{8.4}{rm}.}}
\multiput(6173,-4029)(2.75000,-8.25000){19}{\makebox(6.6667,10.0000){\SetFigFont{7}{8.4}{rm}.}}
\multiput(6218,-4179)(-8.38754,2.09689){18}{\makebox(6.6667,10.0000){\SetFigFont{7}{8.4}{rm}.}}
\multiput(6144,-1922)(-2.75000,-8.25000){19}{\makebox(6.6667,10.0000){\SetFigFont{7}{8.4}{rm}.}}
\multiput(6099,-2072)(8.38754,2.09689){18}{\makebox(6.6667,10.0000){\SetFigFont{7}{8.4}{rm}.}}
\multiput(8326,-3759)(2.06811,-8.27244){20}{\makebox(6.6667,10.0000){\SetFigFont{7}{8.4}{rm}.}}
\multiput(8366,-3916)(6.05263,6.05263){20}{\makebox(6.6667,10.0000){\SetFigFont{7}{8.4}{rm}.}}
\multiput(9503,-2311)(2.06811,-8.27244){20}{\makebox(6.6667,10.0000){\SetFigFont{7}{8.4}{rm}.}}
\multiput(9543,-2468)(6.05263,6.05263){20}{\makebox(6.6667,10.0000){\SetFigFont{7}{8.4}{rm}.}}
\multiput(9669,-3736)(-2.06811,-8.27244){20}{\makebox(6.6667,10.0000){\SetFigFont{7}{8.4}{rm}.}}
\multiput(9629,-3893)(-6.05263,6.05263){20}{\makebox(6.6667,10.0000){\SetFigFont{7}{8.4}{rm}.}}
\multiput(8499,-2311)(-2.06811,-8.27244){20}{\makebox(6.6667,10.0000){\SetFigFont{7}{8.4}{rm}.}}
\multiput(8459,-2468)(-6.05263,6.05263){20}{\makebox(6.6667,10.0000){\SetFigFont{7}{8.4}{rm}.}}
\multiput(8926,-3001)(7.99808,-3.19923){19}{\makebox(6.6667,10.0000){\SetFigFont{7}{8.4}{rm}.}}
\multiput(9069,-3061)(-7.86398,-3.14559){19}{\makebox(6.6667,10.0000){\SetFigFont{7}{8.4}{rm}.}}
\put(9788,-2431){\makebox(0,0)[lb]{\smash{$j_2$}}}
\put(9774,-3759){\makebox(0,0)[lb]{\smash{$j_4$}}}
\put(6294,-3174){\makebox(0,0)[lb]{\smash{$\displaystyle =\quad \sum_{j_5}
\left( \matrix{ j_1& j_2& j_6\cr
               j_4& j_3& j_5\cr}\right)$ }}}
\put(6361,-2101){\makebox(0,0)[lb]{\smash{$j_2$}}}
\put(6308,-4074){\makebox(0,0)[lb]{\smash{$j_4$}}}
\put(5003,-4051){\makebox(0,0)[lb]{\smash{$j_3$}}}
\put(4944,-2101){\makebox(0,0)[lb]{\smash{$j_1$}}}
\put(5408,-3130){\makebox(0,0)[lb]{\smash{$j_5$}}}
\put(8130,-3804){\makebox(0,0)[lb]{\smash{$j_3$}}}
\put(8115,-2439){\makebox(0,0)[lb]{\smash{$j_1$}}}
\put(8919,-2919){\makebox(0,0)[lb]{\smash{$j_6$}}}
\end{picture}
\end{center}

\subsubsection*{Remarks}

1.  For a deeper understanding of $BF$ theory with gauge group $\SU(2)$,
it is helpful to start with a classical phase space describing
tetrahedron geometries and apply geometric quantization to obtain a
Hilbert space of quantum states.  We can describe a tetrahedron in
$\R^3$ by specifying vectors $E_1,\dots,E_4$ normal to its faces, with
lengths equal to the faces' areas.  We can think of these vectors as
elements of $\so(3)^\ast$, which has a Poisson structure familiar from
the quantum mechanics of angular momentum:
\[       \{J^a,J^b\} = \epsilon^{abc} J^c  .\]
The space of 4-tuples $(E_1,\dots,E_4)$ thus becomes a Poisson manifold.
However, a 4-tuple coming from a tetrahedron must satisfy the constraint
$E_1 + \cdots + E_4 = 0$.  This constraint is the discrete analogue of
the Gauss law $d_A E = 0$.   In particular, it generates rotations, so
if we take $(\so(3)^\ast)^4$ and do Poisson reduction with respect to 
this constraint, we obtain a phase space whose points correspond to
tetrahedron geometries modulo rotations.   If we geometrically quantize
this phase space, we obtain the `Hilbert space of the quantum tetrahedron'.

We can describe this Hilbert space quite explicitly as follows.   If we
geometrically quantize $\so(3)^\ast$, we obtain the direct sum of all
the irreducible representations of $\SU(2)$:
\[        \H \iso \bigoplus_{j = 0,{1\over 2},1,\dots} j  . \]
Since this Hilbert space is a representation of $\SU(2)$, it has
operators $\hat J^a$ on it satisfying the usual angular momentum
commutation relations:
\[       [\hat J^a,\hat J^b] = i\epsilon^{abc} \hat J^c .  \]
We can think of $\H$ as the `Hilbert space of a quantum vector' and the
operators $\hat J^a$ as measuring the components of this vector.  If we
geometrically quantize $(\so(3)^\ast)^{\tensor 4}$, we obtain
$\H^{\tensor 4}$, which is the Hilbert space for 4 quantum vectors.
There are operators on this Hilbert space corresponding to the
components of these vectors:
\ban \hat E_1^a &=& \hat J^a \tensor 1 \tensor 1 \tensor 1  \\
     \hat E_2^a &=& 1 \tensor \hat J^a \tensor 1 \tensor 1  \\
     \hat E_3^a &=& 1 \tensor 1 \tensor \hat J^a \tensor 1  \\
     \hat E_4^a &=& 1 \tensor 1 \tensor 1 \tensor \hat J^a . \ean
One can show that the Hilbert space of the quantum tetrahedron is
isomorphic to 
\[  \T = \{ \psi \in \H^{\tensor 4} \colon \; 
(\hat E_1 + \hat E_2 + \hat E_3 + \hat E_4) \psi = 0 \}  .\]

On the Hilbert space of the quantum tetrahedron there are operators
\[             \hat A_i = (\hat E_i \cdot \hat E_i)^{1\over 2} \]
corresponding to the areas of the 4 faces of the tetrahedron, and
also operators
\[  \hat A_{ij} = 
((\hat E_i + \hat E_j)\cdot (\hat E_i + \hat E_j))^{1\over 2} \]
corresponding to the areas of the parallelograms.   Since $\hat A_{ij} = 
\hat A_{kl}$ whenever $(ijkl)$ is some permutation of the numbers $(1234)$,
there are really just 3 different parallelogram area operators.
The face area operators commute with each other and with the
parallelogram area operators, but the parallelogram areas do not commute
with each other.  There is a basis of $\T$ consisting of states that
are eigenvectors of all the face area operators together with any
one of the parallelogram area operators.  If for example we pick $\hat A_{12}$
as our preferred parallelogram area operator, any basis vector $\psi$ 
is determined by 5 spins:
\ban   \hat A_i \psi &=& \sqrt{j_i(j_i + 1)}   \qquad \qquad 1 \le i \le 4 , \\
       \hat A_{12} \psi &=& \sqrt{j_5(j_5 + 1)}  . 
\ean
This basis vector corresponds to the intertwiner $\iota_j \maps j_1 \tensor
j_2 \to j_3 \tensor j_4$ that factors through the representation $j_5$. 

In 4d $BF$ theory with gauge group $\SU(2)$, the Hilbert space
$L^2(\A_\gamma/\G_\gamma)$ described by taking the tensor product of
copies of $\T$, one for each tetrahedron in the 3-manifold $S$, and
imposing constraints saying that when two tetrahedra share a face their
face areas must agree.   This gives a clearer picture of the `quantum 
geometry of space' in this theory.  For example, we can define
observables corresponding to the volumes of tetrahedra.   The results
nicely match those of loop quantum gravity, where it has been shown that
spin network vertices give volume to the regions of space in which they
lie.  In loop quantum gravity these results were derived not from 
$BF$ theory, but from Lorentzian quantum gravity formulated in terms of
the real Ashtekar variables.  However, these theories differ only in 
their dynamics.  

\section{Dynamics} \label{dynamics}

We now turn from the spin network description of the kinematics of $BF$
theory to the spin foam description of its dynamics.  Our experience with
quantum field theory suggests that we can compute transition amplitudes
in $BF$ theory using path integrals.  To keep life simple, consider
the most basic example: the partition function of a closed manifold
representing spacetime.  Heuristically, if $M$ is a compact oriented
$n$-manifold we expect that
\ban  Z(M) &=&  \int\int {\cal D}A\, {\cal D}E\; e^{i \int_M \tr(E \we F)} \\
&=& \int {\cal D}A \; \delta(F), \ean
where formally integrating out the $E$ field gives a Dirac delta measure 
on the space of flat connections on the $G$-bundle $P$ over $M$.
The final result should be the `volume of the space of flat connections',
but of course this is ill-defined without some choice of measure.  

To try to make this calculation more precise, we can {\it discretize} it
by choosing a triangulation for $M$ and working, not with flat
connections on $P$, but instead with flat connections on the dual
2-skeleton.  By definition, the `dual 2-skeleton' of a triangulation has
one vertex in the center of each $n$-simplex, one edge intersecting each
$(n-1)$-simplex, and one polygonal face intersecting each
$(n-2)$-simplex.   We call these `dual vertices', `dual edges', and
`dual faces', respectively.  For example, when $M$ is 3-dimensional, the
intersection of the dual 2-skeleton with any tetrahedron looks like
this:

\vskip 2em
\centerline{\epsfysize=2.5in\epsfbox{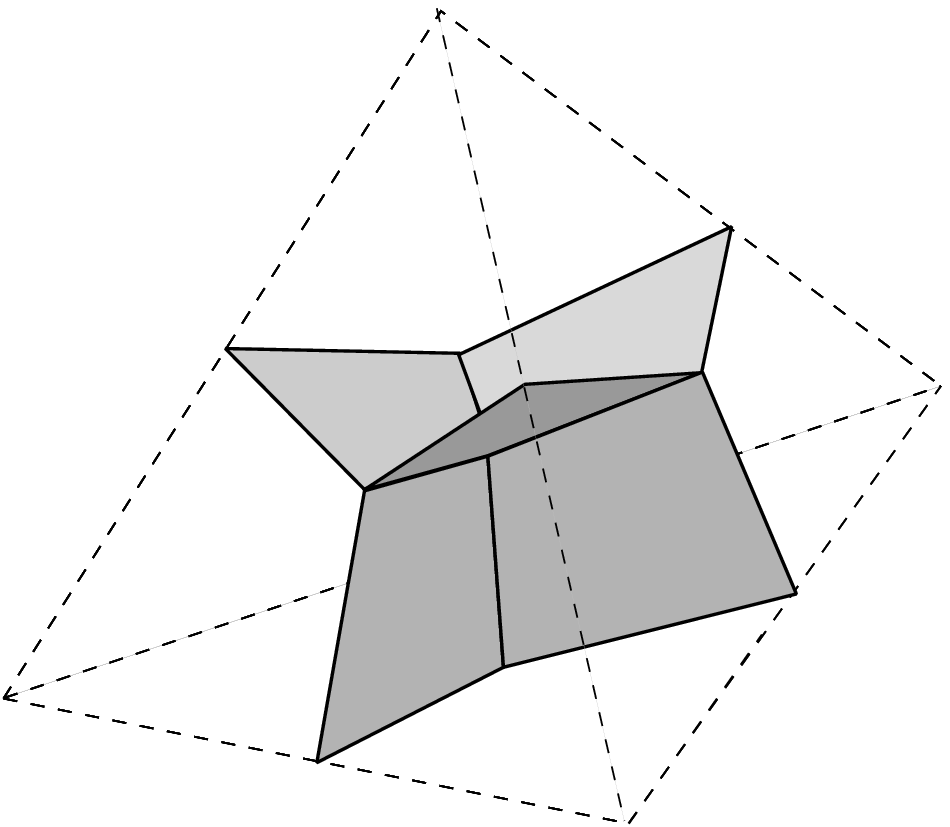}} \medskip

\noindent
while a typical dual face looks like this:

\vskip 2em
\centerline{\epsfysize=2.5in\epsfbox{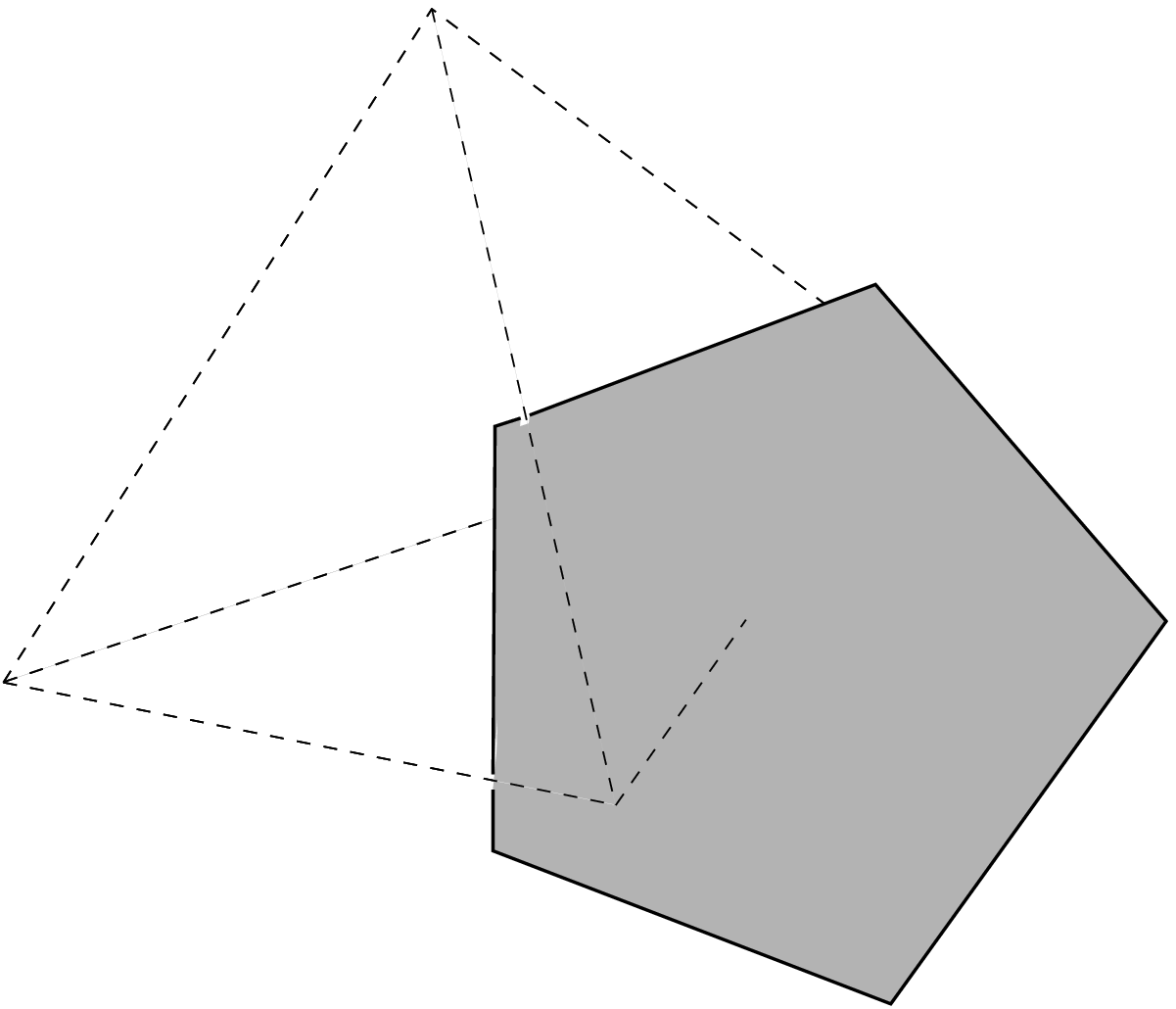}} \medskip

\noindent
Note that the dual faces can have any number of edges.   To keep track
of these edges, we fix an orientation and distinguished vertex for each
face $f$ and call its edges $e_1 f, \dots, e_N f$, taken in cyclic order
starting from the distinguished vertex.  Similarly, we call its verices
$v_1 f, \dots, v_N f$:

\vskip 2em
\centerline{\epsfysize=2.5in\epsfbox{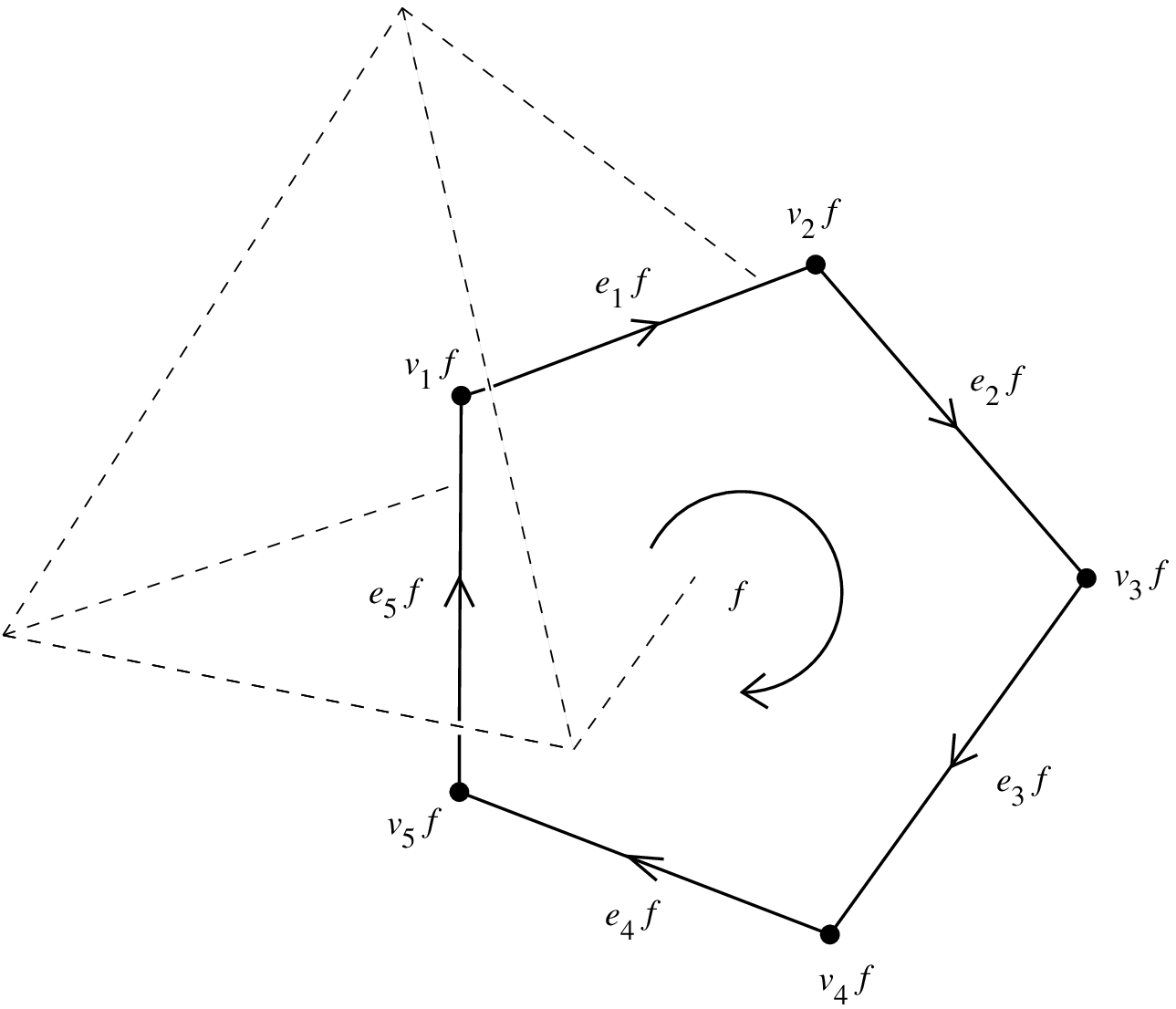}} \medskip

A `connection' on the dual 2-skeleton is an object assigning a group
element $g_e$ to each dual edge $e$.  For this to make sense we should
fix an orientation for each dual edge.  However, we can safely
reverse our choice of the orientation as long as we remember to replace 
$g_e$ by $g_e^{-1}$ when we do so.  We say that a connection on the dual 
2-skeleton is `flat' if that the holonomy around each dual face $f$ is 
the identity:
\[        g_{e_1 f} \cdots g_{e_N f} = 1  \]
where we use the orientation of $f$ to induce orientations of its
edges.  

To make sense of our earlier formula for the partition function of
$BF$ theory, we can try defining
\[        Z(M) = \int \prod_{e \in \E} dg_e\;
\prod_{f \in \F}\; \delta(g_{e_1 f} \cdots g_{e_N f}) , \]
where $\V$ is the set of dual vertices, $\E$ is the set of dual
edges, $\F$ is the set of dual vertices, and the integrals are done
using normalized Haar measure on $G$.  Of course, since we are taking a
product of Dirac deltas here, we run the danger that this expression
will not make sense.  Nonetheless we proceed and see what happens!

We begin by using the identity
\[        \delta(g) = \sum_{\rho \in \Irrep(G)} \dim(\rho) \tr(\rho(g)),  \]
obtaining
\[        Z(M) = \sum_{\rho \maps \F \to \Irrep(G)}
 \int \prod_{e \in \E} dg_e\; \prod_{f \in \F}\; 
\dim(\rho_f) \tr(\rho_f(g_{e_1 f} \cdots g_{e_N f})).  \]
This formula is really a discretized version of
\[ Z(M) =  \int\int {\cal D}A\, {\cal D}E\; e^{i \int_M \tr(E \we F)} .\]
The analogue of $A$ is the labelling of dual edges by group elements.
The analogue of $F$ is the labelling of dual faces by holonomies around
these faces.   These analogies make geometrical sense because $A$ is
like a 1-form and $F$ is like a 2-form.  What is the analogue of $E$? 
It is the labelling of dual faces by representations!  Since each dual
face intersects one $(n-2)$-simplex in the triangulation, we may dually
think of these representations as labelling $(n-2)$-simplices.    This
is nice because $E$ is an $(n-2)$-form.  The analogue of the pairing
$\tr(E \we F)$ is the pairing of a representation $\rho_f$ and the
holonomy around the face $f$ to obtain the number 
$\dim(\rho_f) \tr(\rho_f(g_{e_1 f} \cdots g_{e_N f}))$.  

Next we do the integrals over group elements in the formula for $Z(M)$. 
The details depend on the dimension of spacetime, and it is easiest to
understand them with the aid of some graphical notation.  In the
previous section we saw how an abstract spin network $\Psi$ together
with a connection $A$ on the underlying graph of $\Psi$ give a number
$\Psi(A)$.   Since the connection $A$ assigns a group element $g_e$ to
each edge of $\Psi$, our notation for the number $\Psi(A)$ will be a
picture of $\Psi$ together with a little circle containing the group
element $g_e$ on each edge $e$.  When the $g_e$ is the identity we  will
not bother drawing it.  Also, when two or more parallel edges  share the
same group element $g$ we use one little circle for both edges.   For
example, we define:

\vskip 2em
\centerline{\epsfysize=2.5in\epsfbox{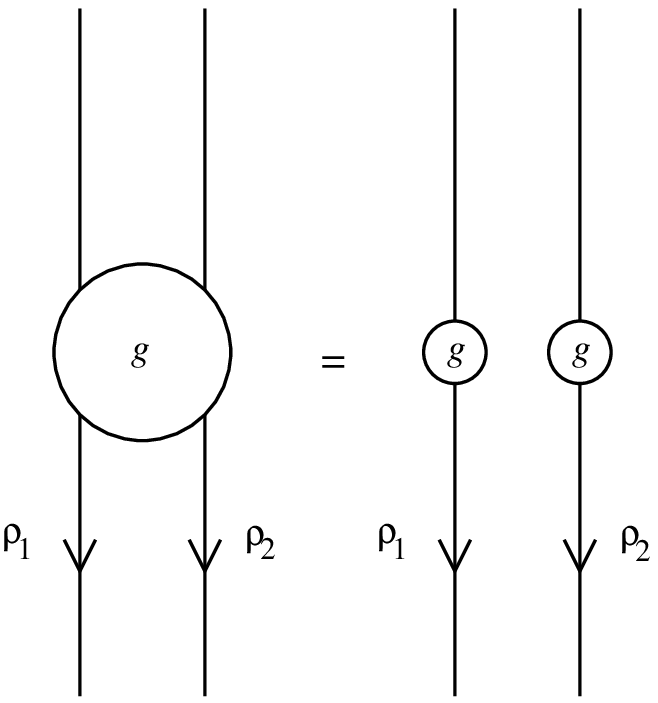}} \medskip

\noindent
This is just the graphical analogue of the equation $(\rho_1 \tensor
\rho_2)(g) = \rho_1(g) \tensor \rho_2(g)$.  

Now suppose $M$ is 2-dimensional.  Since each dual edge is the edge of
two dual faces, each group element appears twice in the expression 
\[   \prod_{f \in \F}
\tr(\rho_f (g_{e_1 f} \cdots g_{e_N f})) . \]
In our graphical notation, this expresssion corresponds to a spin
network with one loop running around each dual face:

\vskip 2em
\centerline{\epsfysize=2.5in\epsfbox{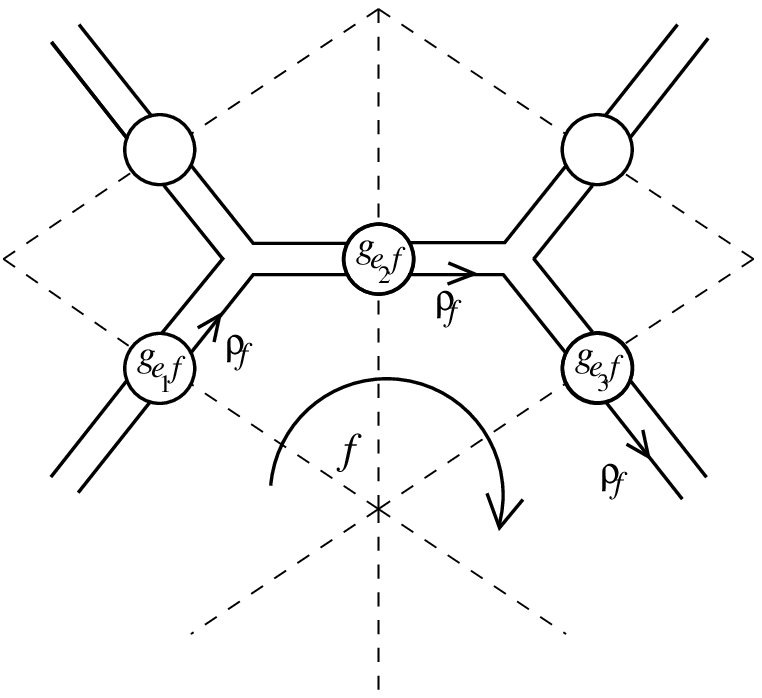}} \medskip

\noindent 
Here we have only drawn a small portion of the spin network.   We can
do the integral 
\[      \int \prod_{e \in E} dg_e\; \prod_{f \in F}\; 
\dim(\rho_f) \tr(\rho_f(g_{e_1 f} \cdots g_{e_N f}))  \]
by repeatedly using the formula
\[     \int dg \; \rho_1(g) \tensor \rho_2(g) =
 \cases {{\iota\iota^\ast \over \dim(\rho_1)}
  & if $\rho_1 \iso \rho_2^\ast$  \cr
         0 & otherwise \cr} \]
where $\iota \maps \rho_1 \tensor \rho_2 \to \C$ is the dual pairing
when $\rho_1$ is the dual of $\rho_2$.  This formula holds because
both sides describe the projection from $\rho_1 \tensor \rho_2$ onto
the subspace of vectors transforming in the trivial representation.
Graphically, this formula can be written as the following skein relation:

\vskip 2em

\begin{center}
\setlength{\unitlength}{0.00083300in}%
\begingroup\makeatletter\ifx\SetFigFont\undefined
\def\x#1#2#3#4#5#6#7\relax{\def\x{#1#2#3#4#5#6}}%
\expandafter\x\fmtname xxxxxx\relax \def\y{splain}%
\ifx\x\y   
\gdef\SetFigFont#1#2#3{%
  \ifnum #1<17\tiny\else \ifnum #1<20\small\else
  \ifnum #1<24\normalsize\else \ifnum #1<29\large\else
  \ifnum #1<34\Large\else \ifnum #1<41\LARGE\else
     \huge\fi\fi\fi\fi\fi\fi
  \csname #3\endcsname}%
\else
\gdef\SetFigFont#1#2#3{\begingroup
  \count@#1\relax \ifnum 25<\count@\count@25\fi
  \def\x{\endgroup\@setsize\SetFigFont{#2pt}}%
  \expandafter\x
    \csname \romannumeral\the\count@ pt\expandafter\endcsname
    \csname @\romannumeral\the\count@ pt\endcsname
  \csname #3\endcsname}%
\fi
\fi\endgroup
\begin{picture}(3870,3344)(4576,-4883)
\thicklines
\put(5701,-3211){\circle{848}}
\put(8101,-3361){\circle*{88}}
\put(8101,-3061){\circle*{88}}
\multiput(5926,-4111)(3.75000,-7.50000){21}{\makebox(6.6667,10.0000){\SetFigFont{7}{8.4}{rm}.}}
\multiput(6001,-4261)(3.75000,7.50000){21}{\makebox(6.6667,10.0000){\SetFigFont{7}{8.4}{rm}.}}
\multiput(5326,-4111)(3.75000,-7.50000){21}{\makebox(6.6667,10.0000){\SetFigFont{7}{8.4}{rm}.}}
\multiput(5401,-4261)(3.75000,7.50000){21}{\makebox(6.6667,10.0000){\SetFigFont{7}{8.4}{rm}.}}
\put(7801,-1561){\line( 1,-5){300}}
\put(8101,-3361){\line(-1,-5){300}}
\put(8401,-1561){\line(-1,-5){300}}
\put(8101,-3361){\line( 1,-5){300}}
\multiput(7851,-4206)(3.12159,-7.80399){20}{\makebox(6.6667,10.0000){\SetFigFont{7}{8.4}{rm}.}}
\multiput(7906,-4356)(5.50043,6.60052){20}{\makebox(6.6667,10.0000){\SetFigFont{7}{8.4}{rm}.}}
\multiput(8351,-4211)(-3.12159,-7.80399){20}{\makebox(6.6667,10.0000){\SetFigFont{7}{8.4}{rm}.}}
\multiput(8296,-4361)(-5.50043,6.60052){20}{\makebox(6.6667,10.0000){\SetFigFont{7}{8.4}{rm}.}}
\multiput(8016,-2191)(-3.12159,-7.80399){20}{\makebox(6.6667,10.0000){\SetFigFont{7}{8.4}{rm}.}}
\multiput(7961,-2341)(-5.50043,6.60052){20}{\makebox(6.6667,10.0000){\SetFigFont{7}{8.4}{rm}.}}
\multiput(8191,-2196)(3.12159,-7.80399){20}{\makebox(6.6667,10.0000){\SetFigFont{7}{8.4}{rm}.}}
\multiput(8246,-2346)(5.50043,6.60052){20}{\makebox(6.6667,10.0000){\SetFigFont{7}{8.4}{rm}.}}
\put(6001,-3361){\line( 0,-1){1425}}
\put(5401,-3361){\line( 0,-1){1425}}
\put(5401,-1636){\line( 0,-1){1425}}
\put(6001,-1636){\line( 0,-1){1425}}
\put(5648,-3248){\makebox(0,0)[lb]{\smash{\SetFigFont{12}{14.4}{it}
g
}}}
\put(6559,-3286){\makebox(0,0)[lb]
{\smash{\SetFigFont{14}{16.8}{rm}
=
}}}
\put(4576,-3286){\makebox(0,0)[lb]{\smash{\large $\int dg$}}}
\put(6976,-3286){\makebox(0,0)[lb]{\smash{\large 
${\delta_{\rho_1 \, \rho_2^*}\over {\rm dim}(\rho_1)}$}}}
\put(8400,-2184){\makebox(0,0)[lb]{\smash{$\rho_2$}}}
\put(8400,-4126){\makebox(0,0)[lb]{\smash{$\rho_2$}}}
\put(5101,-4126){\makebox(0,0)[lb]{\smash{$\rho_1$}}}
\put(6196,-4126){\makebox(0,0)[lb]{\smash{$\rho_2$}}}
\put(7651,-4126){\makebox(0,0)[lb]{\smash{$\rho_1$}}}
\put(7651,-2176){\makebox(0,0)[lb]{\smash{$\rho_1$}}}
\put(7876,-3136){\makebox(0,0)[lb]{\smash{$\iota$}}}
\put(7876,-3436){\makebox(0,0)[lb]{\smash{$\iota^*$}}}
\end{picture}
\end{center}
\medskip 

\noindent
Applying this to every dual edge, we see that when $M$ is 
connected the integral
\[      \int \prod_{e \in E} dg_e\; \prod_{f \in F}\; 
\dim(\rho_f) \tr(\rho_f(g_{e_1 f} \cdots g_{e_N f}))  \]
vanishes unless all the representations $\rho_f$ are the same representation
$\rho$, in which case it equals $\dim(\rho)^{|\V| - |\E| + |\F|}$.
The quantity $|\V| - |\E| + |\F|$ is a topological invariant of $M$,
namely the Euler characteristic $\chi(M)$.  
Summing over all labellings of dual faces, we thus obtain
\[   Z(M) = \sum_{\rho \in \Irrep(G)} \dim(\rho)^{\chi(M)}  \]
The Euler characteristic of a compact oriented surface of genus $n$ is
$2 - 2n$.  When $\chi(M) < 0$, the sum converges for any compact
Lie group $G$, and we see that the partition function of our
discretized $BF$ theory is well-defined and {\it independent of the
triangulation!}  This is precisely what we would expect in a topological 
quantum field theory.  For $\chi(M) \ge 0$, that is, for the sphere
and torus, the partition function typically does not converge.  

In the 3-dimensional case each group element shows up in 3 factors of
the product over dual faces, since 3 dual faces share each dual edge:

\vskip 2em
\centerline{\epsfysize=2.5in\epsfbox{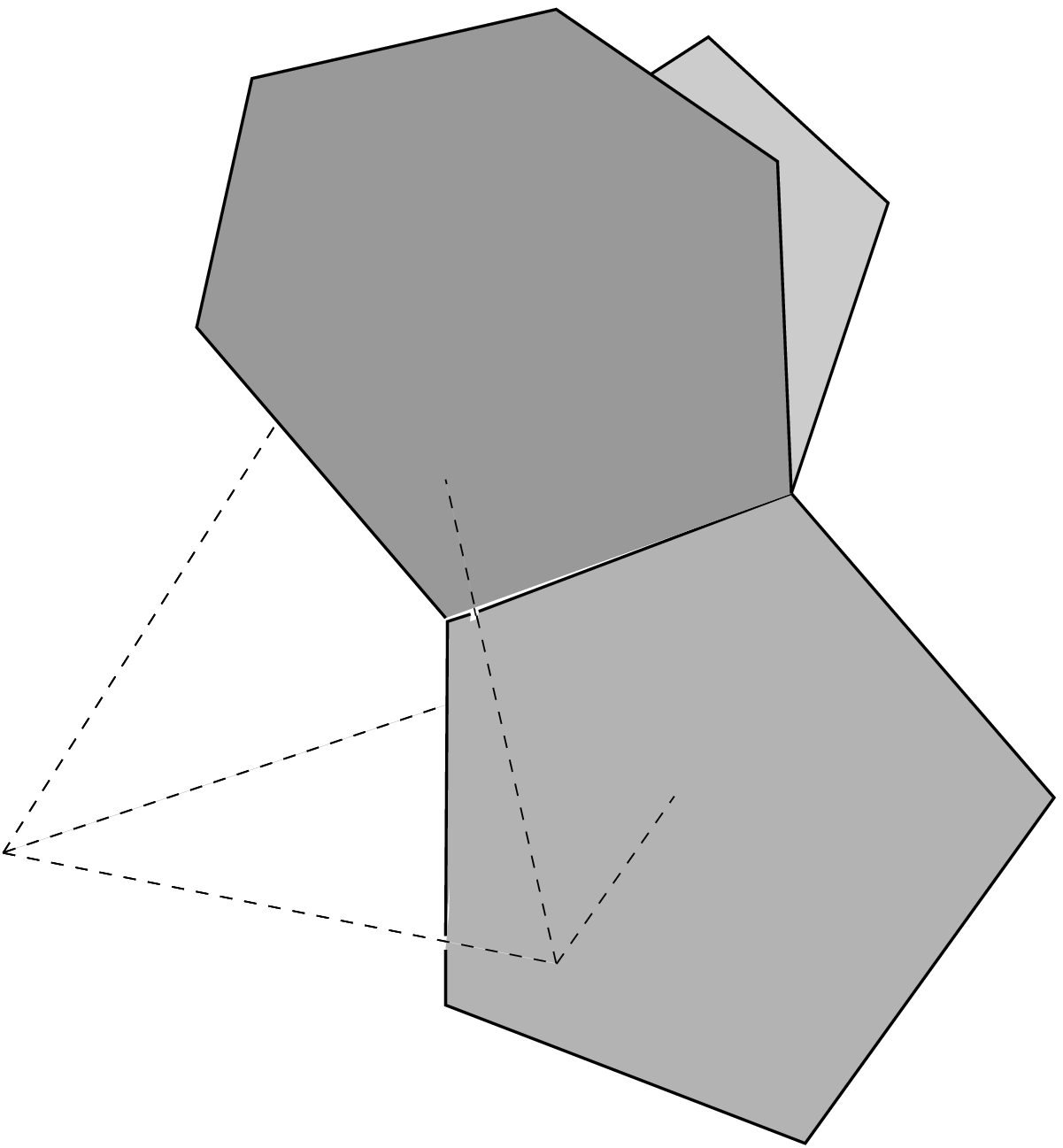}} \medskip

\noindent  
We can do the integral over each group element using the formula 
\[     \int dg \; \rho_1(g) \tensor \rho_2(g) \tensor \rho_3(g) =
\sum_\iota \iota \iota^\ast  \]
where the sum ranges over a basis of intertwiners 
$\iota \maps \rho_1 \tensor \rho_2 \tensor \rho_3 \to \C$, 
normalized as in Section \ref{canonical.quantization}, so that
$\tr(\iota_1 \iota_2^*) = \delta_{\iota_1 \iota_2}$ 
for any two intertwiners $\iota_1, \iota_2$ in the basis.  In 
our graphical notation this formula is written as:

\vskip 2em

\begin{center}
\setlength{\unitlength}{0.00083300in}%
\begingroup\makeatletter\ifx\SetFigFont\undefined
\def\x#1#2#3#4#5#6#7\relax{\def\x{#1#2#3#4#5#6}}%
\expandafter\x\fmtname xxxxxx\relax \def\y{splain}%
\ifx\x\y   
\gdef\SetFigFont#1#2#3{%
  \ifnum #1<17\tiny\else \ifnum #1<20\small\else
  \ifnum #1<24\normalsize\else \ifnum #1<29\large\else
  \ifnum #1<34\Large\else \ifnum #1<41\LARGE\else
     \huge\fi\fi\fi\fi\fi\fi
  \csname #3\endcsname}%
\else
\gdef\SetFigFont#1#2#3{\begingroup
  \count@#1\relax \ifnum 25<\count@\count@25\fi
  \def\x{\endgroup\@setsize\SetFigFont{#2pt}}%
  \expandafter\x
    \csname \romannumeral\the\count@ pt\expandafter\endcsname
    \csname @\romannumeral\the\count@ pt\endcsname
  \csname #3\endcsname}%
\fi
\fi\endgroup
\begin{picture}(4005,3765)(4576,-5119)
\thicklines
\put(5701,-3211){\circle{848}}
\put(8101,-3361){\circle*{88}}
\put(8101,-3061){\circle*{88}}
\put(7801,-1561){\line( 1,-5){300}}
\put(8101,-3361){\line(-1,-5){300}}
\put(8401,-1561){\line(-1,-5){300}}
\put(8101,-3361){\line( 1,-5){300}}
\put(6001,-3361){\line( 0,-1){1500}}
\put(5401,-1561){\line( 0,-1){1500}}
\put(6001,-1561){\line( 0,-1){1500}}
\put(5701,-1561){\line( 0,-1){1330}}
\put(5701,-3528){\line( 0,-1){1322}}
\put(8101,-3061){\line( 0, 1){1500}}
\put(8101,-3361){\line( 0,-1){1500}}
\multiput(7784,-4554)(3.12159,-7.80399){20}{\makebox(6.6667,10.0000){\SetFigFont{7}{8.4}{rm}.}}
\multiput(7839,-4704)(5.50043,6.60052){20}{\makebox(6.6667,10.0000){\SetFigFont{7}{8.4}{rm}.}}
\multiput(8434,-4561)(-3.12159,-7.80399){20}{\makebox(6.6667,10.0000){\SetFigFont{7}{8.4}{rm}.}}
\multiput(8379,-4711)(-5.50043,6.60052){20}{\makebox(6.6667,10.0000){\SetFigFont{7}{8.4}{rm}.}}
\multiput(8019,-4591)(3.75000,-7.50000){21}{\makebox(6.6667,10.0000){\SetFigFont{7}{8.4}{rm}.}}
\multiput(8094,-4741)(3.75000,7.50000){21}{\makebox(6.6667,10.0000){\SetFigFont{7}{8.4}{rm}.}}
\multiput(7931,-1801)(-3.12159,-7.80399){20}{\makebox(6.6667,10.0000){\SetFigFont{7}{8.4}{rm}.}}
\multiput(7876,-1951)(-5.50043,6.60052){20}{\makebox(6.6667,10.0000){\SetFigFont{7}{8.4}{rm}.}}
\multiput(8271,-1801)(3.12159,-7.80399){20}{\makebox(6.6667,10.0000){\SetFigFont{7}{8.4}{rm}.}}
\multiput(8326,-1951)(5.50043,6.60052){20}{\makebox(6.6667,10.0000){\SetFigFont{7}{8.4}{rm}.}}
\multiput(8026,-1824)(3.75000,-7.50000){21}{\makebox(6.6667,10.0000){\SetFigFont{7}{8.4}{rm}.}}
\multiput(8101,-1974)(3.75000,7.50000){21}{\makebox(6.6667,10.0000){\SetFigFont{7}{8.4}{rm}.}}
\put(5401,-3361){\line( 0,-1){1500}}
\multiput(5326,-4561)(3.75000,-7.50000){21}{\makebox(6.6667,10.0000){\SetFigFont{7}{8.4}{rm}.}}
\multiput(5401,-4711)(3.75000,7.50000){21}{\makebox(6.6667,10.0000){\SetFigFont{7}{8.4}{rm}.}}
\multiput(5626,-4561)(3.75000,-7.50000){21}{\makebox(6.6667,10.0000){\SetFigFont{7}{8.4}{rm}.}}
\multiput(5701,-4711)(3.75000,7.50000){21}{\makebox(6.6667,10.0000){\SetFigFont{7}{8.4}{rm}.}}
\multiput(5926,-4561)(3.75000,-7.50000){21}{\makebox(6.6667,10.0000){\SetFigFont{7}{8.4}{rm}.}}
\multiput(6001,-4711)(3.75000,7.50000){21}{\makebox(6.6667,10.0000){\SetFigFont{7}{8.4}{rm}.}}
\put(5648,-3248){\makebox(0,0)[lb]{\smash{\SetFigFont{12}{14.4}{it}
g
}}}
\put(6559,-3286){\makebox(0,0)[lb]{\smash{\SetFigFont{14}{16.8}{rm}
=
}}}
\put(4576,-3286){\makebox(0,0)[lb]{\smash{\large $\int dg$}}}
\put(7050,-3286){\makebox(0,0)[lb]{\smash{\large 
$\displaystyle \sum_{\iota} $}}}
\put(7876,-3136){\makebox(0,0)[lb]{\smash{$\iota$}}}
\put(7876,-3436){\makebox(0,0)[lb]{\smash{$\iota^*$}}}
\put(8071,-5078){\makebox(0,0)[lb]{\smash{$\rho_2$}}}
\put(8071,-1426){\makebox(0,0)[lb]{\smash{$\rho_2$}}}
\put(7598,-4666){\makebox(0,0)[lb]{\smash{$\rho_1$}}}
\put(7598,-1854){\makebox(0,0)[lb]{\smash{$\rho_1$}}}
\put(8500,-4675){\makebox(0,0)[lb]{\smash{$\rho_3$}}}
\put(8500,-1862){\makebox(0,0)[lb]{\smash{$\rho_3$}}}
\put(5176,-4651){\makebox(0,0)[lb]{\smash{$\rho_1$}}}
\put(5701,-5101){\makebox(0,0)[lb]{\smash{$\rho_2$}}}
\put(6170,-4651){\makebox(0,0)[lb]{\smash{$\rho_3$}}}
\end{picture}
\end{center}
\medskip

\noindent
Both sides represent intertwiners from $\rho_1 \tensor \rho_2 \tensor
\rho_3$ to itself.  Again, the formula is true because both sides are
different ways of describing the projection from $\rho_1 \tensor \rho_2
\tensor \rho_3$ onto the subspace of vectors that transform trivially
under $G$.   Using this formula once for each dual edge --- or equivalently, 
once for each triangle in the triangulation --- we can integrate out all the
group elements $g_e$.  Graphically, each time we do this, an 
integral over expressions like this:

\vskip 2em
\centerline{\epsfysize=2in\epsfbox{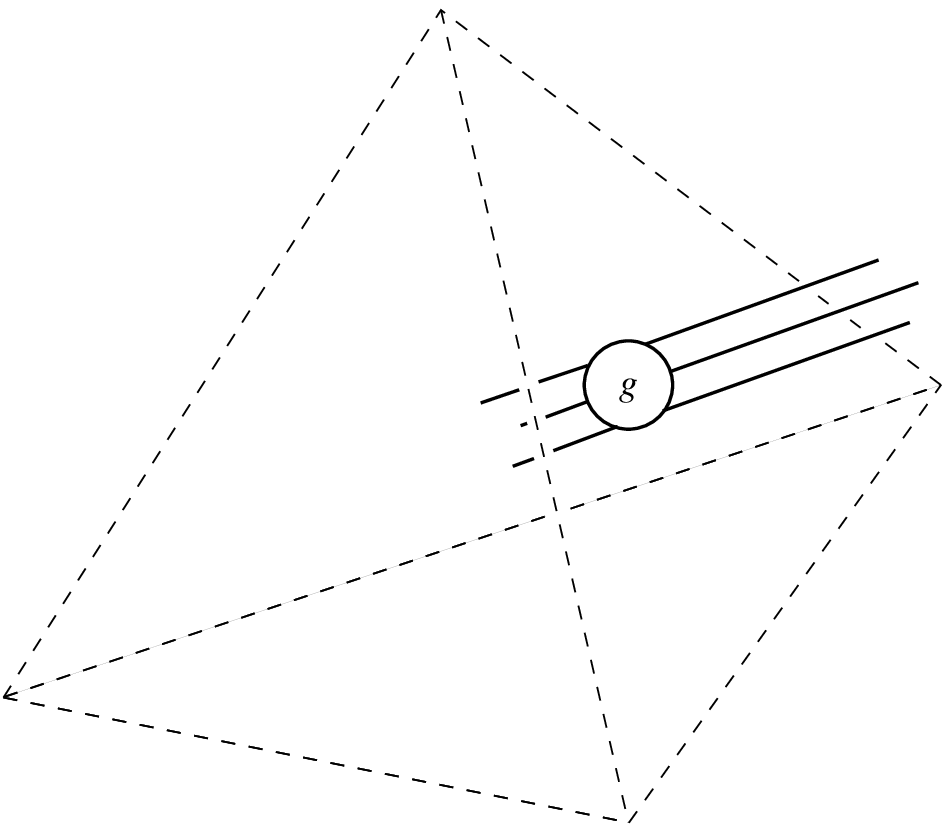}} \medskip

\noindent
is replaced by a sum of expressions like this:

\vskip 2em
\centerline{\epsfysize=2in\epsfbox{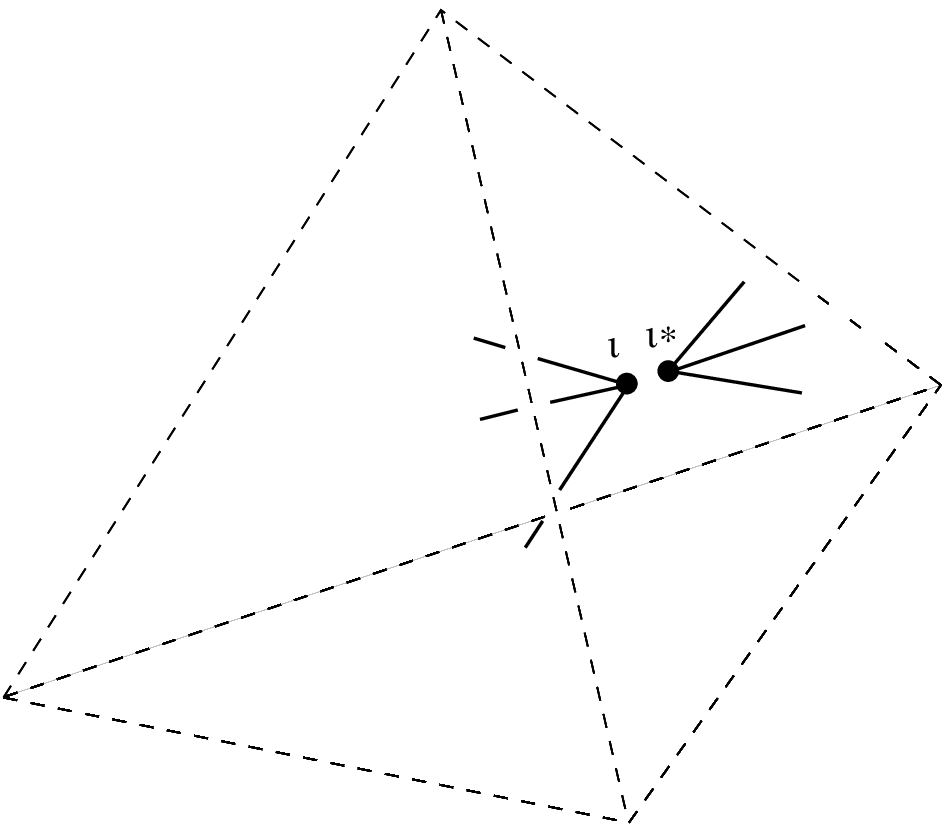}} \medskip

\noindent 
(We have not bothered to show the orientation of the edges in these
pictures, since they depend on how we orient the edges of the dual 
2-skeleton.)  When we do this for all the triangular faces of a given 
tetrahedron, we obtain a little tetrahedral spin network like this:

\vskip 2em
\centerline{\epsfysize=2in\epsfbox{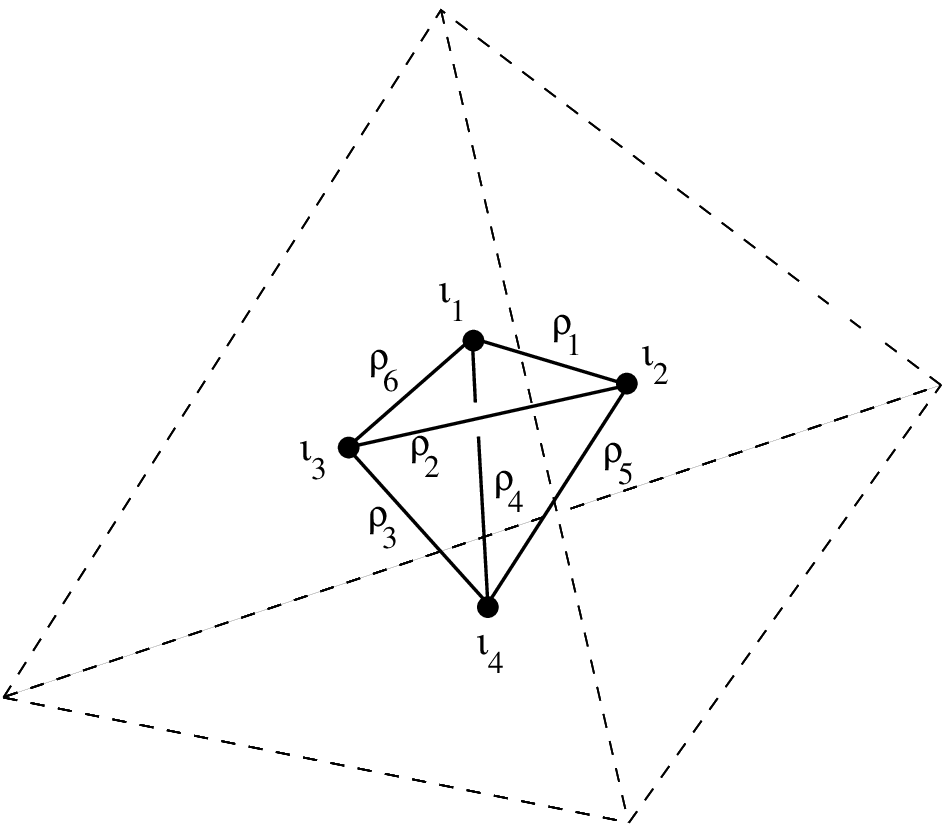}} \medskip

\noindent
which we can evaluate in the usual way.   This tetrahedral spin
network is `dual' to the original tetrahedron in the triangulation 
of $M$: its vertices (resp.\ edges, faces) correspond to faces (resp.\
edges, vertices) of the original tetrahedron.

We thus obtain the following formula for the partition function 
in 3-dimensional $BF$ theory:

\vskip 2em
\begin{center}
\setlength{\unitlength}{0.00083300in}%
\begingroup\makeatletter\ifx\SetFigFont\undefined
\def\x#1#2#3#4#5#6#7\relax{\def\x{#1#2#3#4#5#6}}%
\expandafter\x\fmtname xxxxxx\relax \def\y{splain}%
\ifx\x\y   
\gdef\SetFigFont#1#2#3{%
  \ifnum #1<17\tiny\else \ifnum #1<20\small\else
  \ifnum #1<24\normalsize\else \ifnum #1<29\large\else
  \ifnum #1<34\Large\else \ifnum #1<41\LARGE\else
     \huge\fi\fi\fi\fi\fi\fi
  \csname #3\endcsname}%
\else
\gdef\SetFigFont#1#2#3{\begingroup
  \count@#1\relax \ifnum 25<\count@\count@25\fi
  \def\x{\endgroup\@setsize\SetFigFont{#2pt}}%
  \expandafter\x
    \csname \romannumeral\the\count@ pt\expandafter\endcsname
    \csname @\romannumeral\the\count@ pt\endcsname
  \csname #3\endcsname}%
\fi
\fi\endgroup
\begin{picture}(4650,1725)(3301,-5491)
\thicklines
\put(7201,-3961){\circle*{68}}
\put(6601,-4861){\circle*{68}}
\put(7501,-5161){\circle*{68}}
\put(7801,-4561){\circle*{68}}
\put(7201,-3961){\line( 1,-4){300}}
\put(7501,-5161){\line(-3, 1){900}}
\put(6601,-4861){\line( 2, 3){600}}
\put(7201,-3961){\line( 1,-1){600}}
\put(7801,-4561){\line(-1,-2){300}}
\put(7801,-4561){\line(-5,-1){346.154}}
\put(6601,-4861){\line( 4, 1){674.118}}
\put(3000,-4636){\makebox(0,0)[lb]{\smash{$\displaystyle
Z(M) =  \sum_{\rho \maps \F \to \Irrep(G)} \sum_\iota\; \prod_{f \in \F} 
\dim(\rho_f)\; \prod_{v \in \V}$}}}
\put(7051,-3880){\makebox(0,0)[lb]{\smash{$\iota_1$}}}
\put(7900,-4531){\makebox(0,0)[lb]{\smash{$\iota_2$}}}
\put(7520,-5391){\makebox(0,0)[lb]{\smash{$\iota_3$}}}
\put(6398,-4989){\makebox(0,0)[lb]{\smash{$\iota_4$}}}
\put(7576,-4261){\makebox(0,0)[lb]{\smash{$\rho_1$}}}
\put(7726,-5011){\makebox(0,0)[lb]{\smash{$\rho_2$}}}
\put(6976,-5236){\makebox(0,0)[lb]{\smash{$\rho_3$}}}
\put(6676,-4411){\makebox(0,0)[lb]{\smash{$\rho_4$}}}
\put(7126,-4486){\makebox(0,0)[lb]{\smash{$\rho_6$}}}
\put(7126,-4836){\makebox(0,0)[lb]{\smash{$\rho_5$}}}
\end{picture}
\end{center}
\medskip

\noindent  Here for each labelling $\rho \maps \F \to \Irrep(G)$,  we
take a sum over labellings $\iota$ of dual edges by intertwiners  taken
from the appropriate bases.   For each dual vertex  $v$, the tetrahedral
spin network shown above is built using the representations $\rho_i$
labelling the 6 dual faces incident to $v$ and the intertwiners
$\iota_i$ labelling the 4 dual  edges incident to $v$.  When $G =
\SU(2)$ or $\SO(3)$, the labelling by intertwiners is trivial, so the
tetrahedral spin network depends only on 6 spins.  Using our graphical
notation, it is not hard to express the value of this spin network in
terms of the $6j$ symbols described in the previous section.  We leave
this as an exercise for the reader.     

The calculation in 4 dimensions is similar, but now 4 dual faces 
share each dual edge, so we need to use the formula 
\[     \int dg \; \rho_1(g) \tensor \rho_2(g) \tensor \rho_3(g)
\tensor \rho_4(g) = \sum_\iota \iota \iota^\ast  \]
where now the sum ranges over a basis of intertwiners
$\iota \maps \rho_1 \tensor \rho_2 \tensor \rho_3 \tensor \rho_4 \to \C$,
normalized so that $\tr(\iota_1 \iota_2^*) = \delta_{\iota_1 \iota_2}$ 
for any intertwiners $\iota_1, \iota_2$ in the basis.  
Again both sides describe the projection on the subspace of vectors
that transform in the trivial representation, and again we can write
the formula as a generalized skein relation:

\vskip 2em
\begin{center}
\setlength{\unitlength}{0.00083300in}%
\begingroup\makeatletter\ifx\SetFigFont\undefined
\def\x#1#2#3#4#5#6#7\relax{\def\x{#1#2#3#4#5#6}}%
\expandafter\x\fmtname xxxxxx\relax \def\y{splain}%
\ifx\x\y   
\gdef\SetFigFont#1#2#3{%
  \ifnum #1<17\tiny\else \ifnum #1<20\small\else
  \ifnum #1<24\normalsize\else \ifnum #1<29\large\else
  \ifnum #1<34\Large\else \ifnum #1<41\LARGE\else
     \huge\fi\fi\fi\fi\fi\fi
  \csname #3\endcsname}%
\else
\gdef\SetFigFont#1#2#3{\begingroup
  \count@#1\relax \ifnum 25<\count@\count@25\fi
  \def\x{\endgroup\@setsize\SetFigFont{#2pt}}%
  \expandafter\x
    \csname \romannumeral\the\count@ pt\expandafter\endcsname
    \csname @\romannumeral\the\count@ pt\endcsname
  \csname #3\endcsname}%
\fi
\fi\endgroup
\begin{picture}(4005,3765)(4576,-5119)
\thicklines
\put(5701,-3211){\circle{848}}
\put(8101,-3061){\circle*{88}}
\put(8101,-3361){\circle*{88}}
\put(6001,-1561){\line( 0,-1){1500}}
\put(6001,-3361){\line( 0,-1){1500}}
\put(5401,-3361){\line( 0,-1){1500}}
\put(5401,-1561){\line( 0,-1){1500}}
\put(7801,-1561){\line( 1,-5){300}}
\put(8101,-3361){\line( 2,-5){600}}
\put(5589,-1569){\line( 0,-1){1350}}
\put(5776,-1561){\line( 0,-1){1335}}
\put(5596,-3541){\line( 0,-1){1335}}
\put(5806,-3511){\line( 0,-1){1358}}
\put(8101,-3361){\line(-1,-5){302.115}}
\put(8101,-3361){\line( 1,-5){300.769}}
\put(8101,-3061){\line( 2, 5){605.172}}
\multiput(5326,-4636)(3.75000,-7.50000){21}{\makebox(6.6667,10.0000){\SetFigFont{7}{8.4}{rm}.}}
\multiput(5401,-4786)(3.75000,7.50000){21}{\makebox(6.6667,10.0000){\SetFigFont{7}{8.4}{rm}.}}
\multiput(5926,-4636)(3.75000,-7.50000){21}{\makebox(6.6667,10.0000){\SetFigFont{7}{8.4}{rm}.}}
\multiput(6001,-4786)(3.75000,7.50000){21}{\makebox(6.6667,10.0000){\SetFigFont{7}{8.4}{rm}.}}
\multiput(5724,-4681)(3.75000,-7.50000){21}{\makebox(6.6667,10.0000){\SetFigFont{7}{8.4}{rm}.}}
\multiput(5799,-4831)(3.75000,7.50000){21}{\makebox(6.6667,10.0000){\SetFigFont{7}{8.4}{rm}.}}
\multiput(5521,-4681)(3.75000,-7.50000){21}{\makebox(6.6667,10.0000){\SetFigFont{7}{8.4}{rm}.}}
\multiput(5596,-4831)(3.75000,7.50000){21}{\makebox(6.6667,10.0000){\SetFigFont{7}{8.4}{rm}.}}
\put(8101,-3361){\line(-2,-5){600}}
\multiput(7572,-4482)(1.41394,-8.48365){20}{\makebox(6.6667,10.0000){\SetFigFont{7}{8.4}{rm}.}}
\multiput(7594,-4644)(5.50043,6.60052){20}{\makebox(6.6667,10.0000){\SetFigFont{7}{8.4}{rm}.}}
\multiput(8626,-4460)(-1.41394,-8.48365){20}{\makebox(6.6667,10.0000){\SetFigFont{7}{8.4}{rm}.}}
\multiput(8604,-4622)(-5.50043,6.60052){20}{\makebox(6.6667,10.0000){\SetFigFont{7}{8.4}{rm}.}}
\multiput(7765,-4634)(3.12159,-7.80399){20}{\makebox(6.6667,10.0000){\SetFigFont{7}{8.4}{rm}.}}
\multiput(7820,-4784)(5.50043,6.60052){20}{\makebox(6.6667,10.0000){\SetFigFont{7}{8.4}{rm}.}}
\multiput(8439,-4621)(-3.12159,-7.80399){20}{\makebox(6.6667,10.0000){\SetFigFont{7}{8.4}{rm}.}}
\multiput(8384,-4771)(-5.50043,6.60052){20}{\makebox(6.6667,10.0000){\SetFigFont{7}{8.4}{rm}.}}
\put(8101,-3061){\line(-2, 5){605.172}}
\put(8401,-1561){\line(-1,-5){300}}
\multiput(8537,-1768)(1.41394,-8.48365){20}{\makebox(6.6667,10.0000){\SetFigFont{7}{8.4}{rm}.}}
\multiput(8559,-1930)(5.50043,6.60052){20}{\makebox(6.6667,10.0000){\SetFigFont{7}{8.4}{rm}.}}
\multiput(8296,-1651)(3.12159,-7.80399){20}{\makebox(6.6667,10.0000){\SetFigFont{7}{8.4}{rm}.}}
\multiput(8351,-1801)(5.50043,6.60052){20}{\makebox(6.6667,10.0000){\SetFigFont{7}{8.4}{rm}.}}
\multiput(7902,-1644)(-3.12159,-7.80399){20}{\makebox(6.6667,10.0000){\SetFigFont{7}{8.4}{rm}.}}
\multiput(7847,-1794)(-5.50043,6.60052){20}{\makebox(6.6667,10.0000){\SetFigFont{7}{8.4}{rm}.}}
\multiput(7689,-1809)(-1.41394,-8.48365){20}{\makebox(6.6667,10.0000){\SetFigFont{7}{8.4}{rm}.}}
\multiput(7667,-1971)(-5.50043,6.60052){20}{\makebox(6.6667,10.0000){\SetFigFont{7}{8.4}{rm}.}}
\put(5648,-3248){\makebox(0,0)[lb]{\smash{\SetFigFont{12}{14.4}{it}
g
}}}
\put(6559,-3286){\makebox(0,0)[lb]{\smash{\SetFigFont{14}{16.8}{rm}
=
}}}
\put(7050,-3286){\makebox(0,0)[lb]{\smash{\large $\displaystyle \sum_{\iota} $}}}
\put(4576,-3286){\makebox(0,0)[lb]{\smash{\large $\int dg$}}}
\put(7876,-3136){\makebox(0,0)[lb]{\smash{$\iota$}}}
\put(7876,-3436){\makebox(0,0)[lb]{\smash{$\iota^*$}}}
\put(5176,-4733){\makebox(0,0)[lb]{\smash{$\rho_1$}}}
\put(5527,-5015){\makebox(0,0)[lb]{\smash{$\rho_2$}}}
\put(5759,-5011){\makebox(0,0)[lb]{\smash{$\rho_3$}}}
\put(6159,-4715){\makebox(0,0)[lb]{\smash{$\rho_4$}}}
\put(7360,-4565){\makebox(0,0)[lb]{\smash{$\rho_1$}}}
\put(7790,-4982){\makebox(0,0)[lb]{\smash{$\rho_2$}}}
\put(8394,-4968){\makebox(0,0)[lb]{\smash{$\rho_3$}}}
\put(8746,-4566){\makebox(0,0)[lb]{\smash{$\rho_4$}}}
\put(7406,-1921){\makebox(0,0)[lb]{\smash{$\rho_1$}}}
\put(7761,-1482){\makebox(0,0)[lb]{\smash{$\rho_2$}}}
\put(8401,-1467){\makebox(0,0)[lb]{\smash{$\rho_3$}}}
\put(8731,-1895){\makebox(0,0)[lb]{\smash{$\rho_4$}}}
\end{picture}
\end{center}
\medskip

\noindent 
We use this formula once for each dual edge --- or equivalently, once
for each tetrahedron in the triangulation --- to do the integral
over all group elements in the partition function.   Each time we do so,
we introduce an intertwiner labelling the dual edge in question.  We
obtain

\begin{center}
\setlength{\unitlength}{0.00083300in}%
\begingroup\makeatletter\ifx\SetFigFont\undefined
\def\x#1#2#3#4#5#6#7\relax{\def\x{#1#2#3#4#5#6}}%
\expandafter\x\fmtname xxxxxx\relax \def\y{splain}%
\ifx\x\y   
\gdef\SetFigFont#1#2#3{%
  \ifnum #1<17\tiny\else \ifnum #1<20\small\else
  \ifnum #1<24\normalsize\else \ifnum #1<29\large\else
  \ifnum #1<34\Large\else \ifnum #1<41\LARGE\else
     \huge\fi\fi\fi\fi\fi\fi
  \csname #3\endcsname}%
\else
\gdef\SetFigFont#1#2#3{\begingroup
  \count@#1\relax \ifnum 25<\count@\count@25\fi
  \def\x{\endgroup\@setsize\SetFigFont{#2pt}}%
  \expandafter\x
    \csname \romannumeral\the\count@ pt\expandafter\endcsname
    \csname @\romannumeral\the\count@ pt\endcsname
  \csname #3\endcsname}%
\fi
\fi\endgroup
\begin{picture}(5198,2182)(3301,-5374)
\thicklines
\put(8101,-5161){\circle*{68}}
\put(8401,-3961){\circle*{68}}
\put(6601,-3961){\circle*{68}}
\put(7501,-3361){\circle*{68}}
\put(6901,-5161){\circle*{68}}
\put(8101,-5161){\line( 1, 4){300}}
\put(7501,-3361){\line( 3,-2){900}}
\put(6601,-3961){\line( 3, 2){900}}
\put(8101,-5161){\line(-1, 0){1200}}
\put(6901,-5161){\line(-1, 4){300}}
\put(7876,-4479){\line( 1,-3){225.800}}
\put(7089,-4351){\line(-6, 5){491.312}}
\put(7801,-3961){\line( 1, 0){600}}
\put(6901,-5161){\line( 1, 3){383.100}}
\put(7321,-3871){\line( 1, 3){173.100}}
\put(7501,-3354){\line( 1,-3){325.500}}
\put(7456,-4726){\line(-4,-3){581.920}}
\put(8394,-3961){\line(-5,-4){828.902}}
\put(7569,-4629){\line( 0,-1){  7}}
\put(6601,-3969){\line( 1, 0){1020}}
\put(8109,-5161){\line(-5, 4){867.073}}
\put(3200,-4406){\makebox(0,0)[lb]{\smash{$\displaystyle Z(M) =  \sum_{\rho \maps \F \to \Irrep(G)} \sum_\iota\; \prod_{f \in \F}
\dim(\rho_f)\; \prod_{v \in \V}$}}}
\put(7456,-3264){\makebox(0,0)[lb]{\smash{$\iota_1$}}}
\put(8499,-3961){\makebox(0,0)[lb]{\smash{$\iota_2$}}}
\put(8229,-5326){\makebox(0,0)[lb]{\smash{$\iota_3$}}}
\put(6399,-3969){\makebox(0,0)[lb]{\smash{$\iota_5$}}}
\put(6714,-5349){\makebox(0,0)[lb]{\smash{$\iota_4$}}}
\put(7951,-3571){\makebox(0,0)[lb]{\smash{$\rho_1$}}}
\put(8379,-4644){\makebox(0,0)[lb]{\smash{$\rho_2$}}}
\put(7464,-5356){\makebox(0,0)[lb]{\smash{$\rho_3$}}}
\put(6504,-4696){\makebox(0,0)[lb]{\smash{$\rho_4$}}}
\put(6929,-3571){\makebox(0,0)[lb]{\smash{$\rho_5$}}}
\put(7599,-4261){\makebox(0,0)[lb]{\smash{$\rho_6$}}}
\put(7339,-4492){\makebox(0,0)[lb]{\smash{$\rho_8$}}}
\put(7266,-4284){\makebox(0,0)[lb]{\smash{$\rho_9$}}}
\put(7576,-4463){\makebox(0,0)[lb]{\smash{$\rho_7$}}}
\put(7397,-4110){\makebox(0,0)[lb]{\smash{$\rho_{10}$}}}
\end{picture}
\end{center}

\noindent
The 4-simplex in this formula is dual to the 4-simplex in the original
triangulation that contains $v \in \V$.   Its edges are labelled by the
representations labelling the 10 dual faces incident to $v$, and its
vertices are labelled by the intertwiners labelling the 5 dual edges
incident to $v$.    

People often rewrite this formula for the partition function by
splitting each 4-valent vertex into two trivalent vertices using the
skein relations described in Section \ref{canonical.quantization}.  The
resulting equation involves a trivalent spin network with 15 edges.  In
the $\SU(2)$ case this trivalent spin network is called a `$15j$
symbol', since it depends on 15 spins. 

Having computed the $BF$ theory partition function in 2, 3, and 4 
dimensions, it should be clear that the same basic idea works in all
higher dimensions, too.  We always get a formula for the partition
function as  a sum over ways of labelling dual faces by representations
and dual edges by intertwiners.  There is, however, a problem.  The sum
usually diverges!   The only cases I know where it converges are when
$G$ is a finite group (see Remark 2 below), when $M$ is 0- or
1-dimensional, or when $M$ is 2-dimensional with $\chi(M) < 0$.  Not
surprisingly, these are a subset of the cases when the moduli space of
flat connections on $M$ has a natural measure.  In other cases, it seems
there are too many delta functions in the expression
\[        Z(M) = \int \prod_{e \in \E} dg_e\;
\prod_{f \in \F}\; \delta(g_{e_1 f} \cdots g_{e_N f}) \]
to extract a meaningful answer.  We discuss this problem further in 
Section \ref{q-deformation}.

Of course, there is more to dynamics than the partition function.  For
example, we also want to compute vacuum expectation values of
observables, and transition amplitudes between states.   It is not hard
to generalize the formulas above to handle these more complicated
calculations.  However, at this point it helps to explicitly introduce
the concept of a `spin foam'.

\subsubsection*{Remarks}

1.  Ponzano and Regge gave a formula for a discretized version of 
the action in 3-dimensional Riemannian general relativity.  In their
approach the spacetime manifold $M$ is triangulated and each edge is
assigned a length.   The Ponzano-Regge action is the sum over all 
tetrahedra of the quantity:
\[   S = \sum_e \ell_e \theta_e  \]
where the sum is taken over all 6 edges, $l_e$ is the length of the edge
$e$, and $\theta_e$ is the dihedral angle of the edge $e$, that is, the
angle between the outward normals of the two faces incident to this
edge.  One can show that in a certain precise their action is an
approximation to the integral of the Ricci scalar curvature.   In the
limit of large spins, the value of the tetrahedral spin network 
described above is asymptotic to 
\[  \sqrt{2\over 3\pi V} \cos(S + {\pi\over 4}) .\]
where the lengths $\ell_e$ are related to the spins $j_e$ labelling the
tetrahedron's edges by $\ell = j + 1/2$, and $V$ is the volume of the
tetrahedron.  Naively one might have hoped to get $\exp(iS)$.  That one
gets a cosine instead can be traced back to the fact that the lengths of
the edges of a tetrahedron only determine its geometry modulo rotation
{\it and reflection}.    The phase ${\pi\over 4}$ shows up because
calculating the asymptotics of the tetrahedral spin network involves 
a stationary phase approximation.  

\noindent   
2.  Ever since Section 4 we have been assuming that $G$ is connected. 
The main reason for this is that it ensures the map from $\A$ to
$\A_\gamma$ is onto for any graph $\gamma$ in $S$, so that we have
inclusions $L^2(\A_\gamma) \hookrightarrow L^2(\A)$ and 
$L^2(\A_\gamma/\G_\gamma) \hookrightarrow L^2(\A/\G)$.   When $G$ is 
not connected, these maps are usually not one-to-one.  

Requiring that $G$ be connected rules out all nontrivial finite 
groups.   However, our formula for the $BF$ theory partition function 
makes equally good sense for groups that are not connected.  In
fact, when $G$ is finite, the partition function is convergent
regardless of the dimension of $M$, and when a suitable normalization
factor is included it becomes triangulation-independent.  This is a
special case of the `Dijkgraaf-Witten model.'    

In this model, the path integral is not an integral over flat
connections on a fixed $G$-bundle over $M$, but rather a sum over
isomorphism classes of $G$-bundles.  In fact, our discretized formula
for the path integral in $BF$ theory always implicitly includes a sum
over isomorphism classes of $G$-bundles, because it corresponds to an
integral over the whole moduli space of flat $G$-bundles over $M$,
rather than the moduli space of flat connections on a fixed $G$-bundle.
(For the relation between these spaces, see Remark 2 in Section 3.) 
When $G$ is a finite group, the moduli space of flat $G$-bundles is
discrete, with one point for each isomorphism class of $G$-bundle.

\section{Spin Foams} \label{spin.foams}

We have seen that in $BF$ theory the partition function can be computed
by triangulating spacetime and considering all ways of labelling dual
faces by irreducible representations and dual edges by intertwiners. 
For each such labelling, we compute an `amplitude' as a product of
amplitudes for dual faces, dual edges, and dual vertices.    (By
cleverly normalizing our intertwiners we were able to make the edge
amplitudes equal 1, rendering them invisible, but this was really just a
cheap trick.)  We then take a sum over all labellings to obtain the
partition function.   

To formalize this idea we introduce the concept of a `spin foam'.   A
spin foam is the 2-dimensional analog of a spin network.  Just as a spin
network is a graph with edges labelled by irreducible representations
and vertices labelled by intertwiners, a spin foam is a 2-dimensional
complex with faces labelled by irreducible representations and edges
labelled by intertwiners.   Of course, to make this precise we need a
formal definition of `2-dimensional complex'.  Loosely, such a thing
should consist of vertices, edges, and polygonal faces.   There is some
flexibility about the details.   However, we certainly want the dual
2-skeleton of a triangulated manifold to qualify.  Since topologists
have already studied such things, this suggests that we take a
2-dimensional complex to be what they call a `2-dimensional piecewise
linear cell complex'.   

The precise definition of this concept is somewhat technical, so we banish
it to the Appendix and only state what we need here.   A 2-dimensional
complex has a finite set $\V$ of vertices, a finite set $\E$ of edges, and
a finite set $\F_N$ of $N$-sided faces for each $N \ge 3$, with only
finitely many $\F_N$ being nonempty.  In fact, we shall work with
`oriented' 2-dimensional complexes, where each edge and each face has 
an orientation.   The orientations of the edges give maps
\[         s,t \maps \E \to \V \]
assigning to each edge its source and target.   The orientation of each
face gives a cyclic ordering to its edges and vertices.  Suppose we
arbitrarily choose a distinguished vertex for each face $f \in \F_N$.  
Then we may number all its vertices and edges from $1$ to $N$.   If
we think of these numbers as lying in $\Z_N$, we obtain maps 
\[     e_i \maps \F_N \to \E, \quad v_i \maps \F_N \to \V  
\qquad \qquad i \in \Z_N. \] 
We say $f$ is `incoming' to $e$ when the orientation of $e$ agrees with
the orientation it inherits from $f$, and `outgoing' when these
orientations do not agree:

\vskip 2em
\centerline{\epsfysize=1.0in\epsfbox{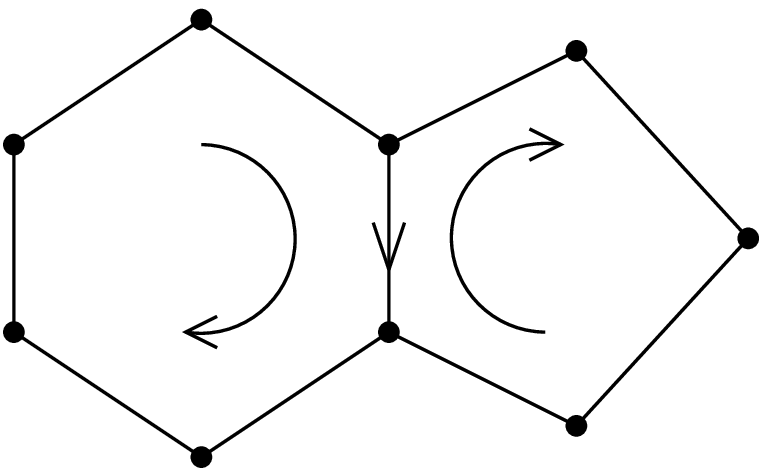}} \medskip

With this business taken care of, we can define spin foams.  The
simplest kind is a `closed' spin foam.  This is the sort we sum
over when computing partition functions in $BF$ theory.  

\begin{defn}\et 
A {\rm closed spin foam} $F$ is a triple $(\kappa,\rho,\iota)$
consisting of:
\begin{enumerate}
\item a 2-dimensional oriented complex $\kappa$, 
\item a labelling $\rho$ of each face $f$ of $\kappa$ by an 
irreducible representation $\rho_f$ of $G$,
\item a labelling $\iota$ of each edge $e$ of $\kappa$ by an intertwiner
\[       \iota_e \maps \rho_{f_1} \tensor \cdots \tensor \rho_{f_n} 
\to \rho_{f'_1} \tensor \cdots \tensor \rho_{f'_m}  \]
where $f_1,\dots,f_n$ are the faces incoming to $e$ and
$f'_1,\dots,f'_m$ are the faces outgoing from $e$.  
\end{enumerate}
\end{defn}

Note that this definition is exactly like that of a spin network, but
with everything one dimension higher!   This is why a generic slice of a
spin foam is a spin network.   We can formalize this using the 
notion of a spin foam $F \maps \Psi \to \Psi'$ going from a
spin network $\Psi$ to a spin network $\Psi'$:

\vskip 2em
\centerline{\epsfysize=2.5in\epsfbox{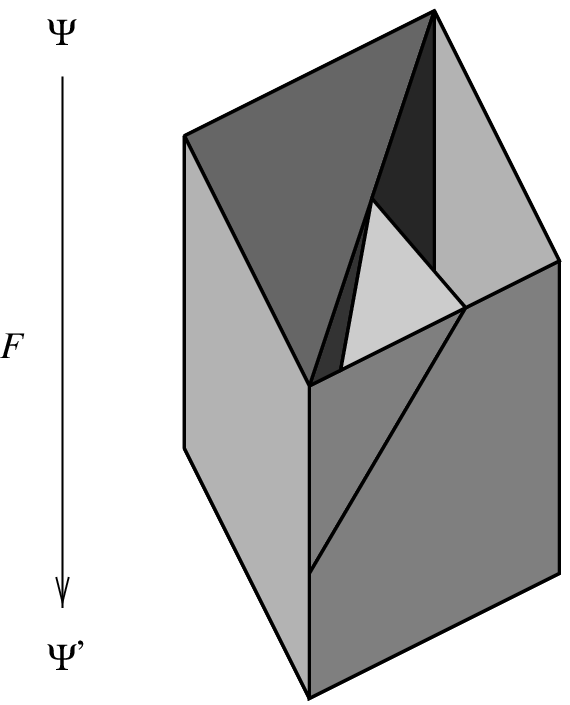}} \medskip

\noindent 
This is the sort we sum over when computing transition amplitudes in
$BF$ theory.  (To reduce clutter, we have not drawn the labellings 
of edges and faces in this spin foam.)  
In this sort of spin foam, the edges that lie in $\Psi$
and $\Psi'$ are not labelled by intertwiners.  Also, the edges ending at
spin network vertices must be labelled by intertwiners that match those
labelling the spin network vertices.  These extra requirements are
lacking for closed spin foams, because a closed spin foam is just one of
the form $F \maps \emptyset \to \emptyset$, where $\emptyset$ is the
`empty spin network': the spin network with no vertices and no edges.  

To make this more precise, we need to define what it means for a graph
$\gamma$ to `border' a 2-dimensional oriented complex $\kappa$.  The
reader can find this definition in Appendix A.  What matters here is
that if $\gamma$ borders $\kappa$, then each vertex $v$ of $\gamma$ is
the source or target of a unique  edge $\tilde v$ of $\kappa$, and each
edge $e$ of $\gamma$ is the edge of a unique face $\tilde e$ of
$\kappa$.   Using these ideas, we first define spin foams of the form 
$F \maps \emptyset \to \Psi$:

\begin{defn}\et
Suppose that $\Psi = (\gamma,\rho,\iota)$ is a spin network.  
A {\rm spin foam} $F \maps \emptyset \to \Psi$ 
is a triple $(\kappa,\tilde \rho,\tilde \iota)$ consisting of:
\begin{enumerate}
\item a 2-dimensional oriented complex $\kappa$ such that
$\gamma$ borders $\kappa$, 
\item a labeling $\tilde \rho$ of each face $f$ of $\kappa$ by 
an irreducible representation $\tilde \rho_f$ of $G$,
\item a labeling $\tilde \iota$ of each edge $e$ of $\kappa$ not 
lying in $\gamma$ by an intertwiner
\[      \tilde\iota_e \maps \rho_{f_1} \tensor \cdots \tensor \rho_{f_n}
\to \rho_{f'_1} \tensor \cdots \tensor \rho_{f'_m}  \]
where $f_1,\dots,f_n$ are the faces incoming to $e$ and
$f'_1,\dots,f'_m$ are the faces outgoing from $e$,
\end{enumerate}
such that the following hold:
\begin{enumerate}
\item For any edge $e$ of $\gamma$, $\tilde \rho_{\tilde e} = \rho_e$
if $\tilde e$ is incoming to $e$, while $\tilde \rho_{\tilde e} = 
(\rho_e)^\ast$ if $\tilde e$ is outgoing from $e$.
\item For any vertex $v$ of $\gamma$, $\tilde \iota_{\tilde e}$ 
equals $\iota_e$ after appropriate dualizations.  
\end{enumerate}
\end{defn}

Finally, to define general spin foams, we need the notions of `dual' and
`tensor product' for spin networks.  The dual of a spin network $\Psi =
(\gamma,\rho,\iota)$ is the spin network $\Psi^\ast$ with the same
underlying graph, but with each edge $e$ labelled by the dual
representation $\rho_e^\ast$, and each vertex $v$ labelled by the
appropriately dualized form of the intertwining operator $\iota_v$. 
Given spin networks $\Psi = (\gamma,\rho,\iota)$ and  $\Psi' = (\gamma',
\rho', \iota')$, their tensor product $\Psi \tensor \Psi'$ is defined to
be the spin network whose underlying graph is the disjoint union of
$\gamma$ and $\gamma'$, with edges and vertices labelled by
representations and intertwiners using $\rho,\rho'$ and $\iota,\iota'$. 
As usual, duality allows us to think of an input as an output:
 
\begin{defn}\et Given spin networks $\Psi$ and $\Psi'$, a {\rm spin
foam} $F \maps \Psi \to \Psi'$ is defined to be a spin foam $F \maps
\emptyset \to \Psi^\ast \tensor \Psi'$. \end{defn}
 
Here is how we compute transition amplitudes in $BF$ theory as a sum
over spin foams.  Suppose spacetime is given by a compact oriented
cobordism $M \maps S \to S'$, where $S$ and $S'$ are compact oriented
manifolds of dimension $n-1$:

\vskip 2em
\centerline{\epsfysize=1.5in\epsfbox{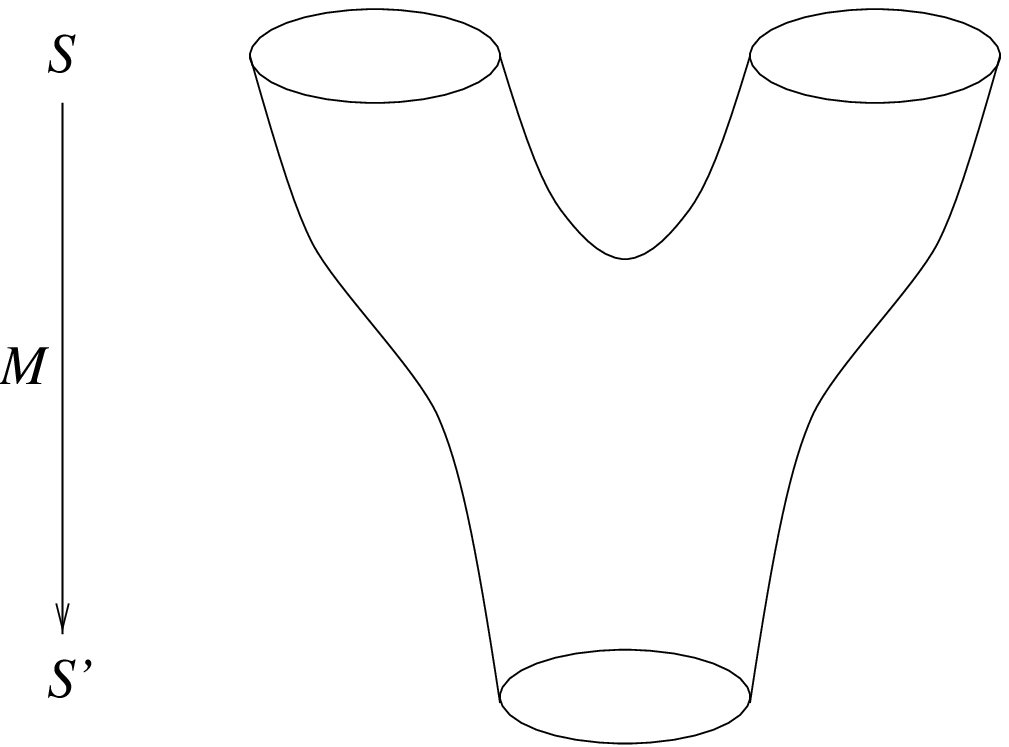}} \medskip

\noindent  
Choose a triangulation of $M$.  This induces triangulations of $S$ and
$S'$ with dual 1-skeletons $\gamma$ and $\gamma'$, respectively.   As
described in Section \ref{triangulations}, in this triangulated
context we can use $L^2(\A_\gamma/\G_\gamma)$ as the gauge-invariant
Hilbert space for $S$.  This Hilbert space has a
basis given by spin networks whose underlying graph is the dual
1-skeleton of $S$.  Similarly, we use $L^2(\A_{\gamma'}/\G_{\gamma'})$
as the space of gauge-invariant states on $S'$.  We describe time evolution
as an operator 
\[ Z(M) \maps L^2(\A_\gamma/\G_\gamma) \to L^2(\A_{\gamma'}/\G_{\gamma'}). \]
To specify this operator, it suffices to describe the transition
amplitudes $\langle \Psi', Z(M) \Psi\rangle$ when $\Psi,\Psi'$ are 
spin network states.  We write this transition amplitude as a sum over spin 
foams going from $\Psi$ to $\Psi'$:
\[  \langle \Psi', Z(M) \Psi\rangle  = \sum_{F \maps \Psi \to \Psi'} Z(F) \]
Since we are working with a fixed triangulation of $M$, we restrict the
sum  to spin foams whose underlying complex is the dual 2-skeleton of
$M$.   The crucial thing is the formula for the amplitude $Z(F)$ of a
given spin foam $F$.  

We have already given a formula for the amplitude of a closed spin foam
in the previous section: it is computed as a product of amplitudes for
spin foam faces, edges and vertices.   A similar formula works for any
spin foam $F \maps \Psi \to \Psi'$, but we need to make a few adjustments.
First, when we take the product over faces, edges and vertices, we 
exclude edges and vertices that lie in $\Psi$ and $\Psi'$.  Second,  
we use the square root of the usual edge amplitude for edges of the form
$\tilde v$, where $v$ is a vertex of $\Psi$ or $\Psi'$.  Third, we use
the square root of the usual face amplitudes for faces of the form 
$\tilde e$, where $e$ is an edge of $\Psi$ or $\Psi'$.  The reason for
these adjustments is that we want to have
\[        Z(M')Z(M) = Z(M'M)   \]
when $M \maps S \to S'$ and $M' \maps S' \to S''$ are composable cobordisms
and $M'M \maps S \to S''$ is their composite:

\vskip 2em
\centerline{\epsfysize=2.5in\epsfbox{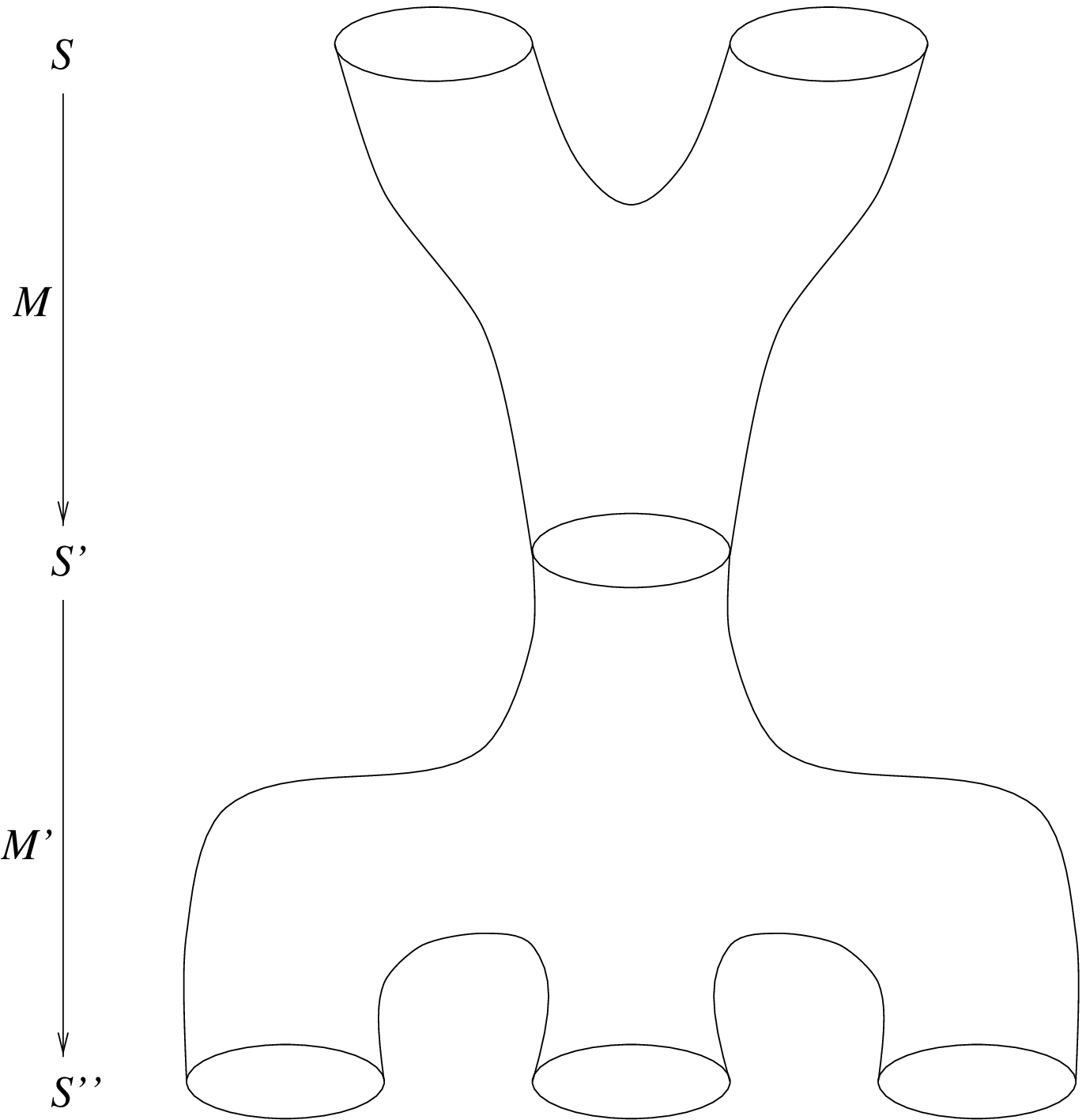}} \medskip
 
\noindent 
For this to hold, we want
\[        Z(F')Z(F) = Z(F'F)   \]
whenever $F'F \maps \Psi \to \Psi''$ is the spin foam formed by gluing
together $F\maps \Psi \to \Psi'$ and $F' \maps \Psi' \to \Psi''$ 
along their common border $\Psi'$ and erasing the vertices and edges
that lie in $\Psi'$.  The adjustments described above make this equation
true.  Of course, the argument that $Z(F')Z(F) = Z(F'F)$ implies 
$Z(M')Z(M) = Z(M'M)$ is merely formal unless the sums over spin foams 
used to define these time evolution operators converge in a sufficiently
nice way.  

Let us conclude with some general remarks on the meaning of the spin
foam formalism.  Just as spin networks are designed to merge the
concepts of {\it quantum state} and the {\it geometry of space}, spin
foams are designed to merge the concepts of {\it quantum history} and
the {\it geometry of spacetime}.   However, the concept of `quantum
history' is a bit less familiar than the concept of `quantum state', so
it deserves some comment.  Perhaps the most familiar example of a
quantum history is a Feynman diagram.  A Feynman diagram determines an
operator on Fock space, but there is more information in the diagram
than this operator, since besides telling us transition  amplitudes
between states, the diagram also tells a story of  `how the transition
happened'.   In other words, the internal edges and vertices of the
diagram describe a `quantum history' in which various virtual particles
are created and annihilated.  

Similarly, spin foams can be used to describe operators, but they 
contain extra information.  If $\Psi$ and $\Psi'$ are spin networks with
underlying graphs $\gamma$ and $\gamma'$, respectively, then any spin
foam $F \maps \Psi \to \Psi'$ determines an operator from
$L^2(\A_\gamma/\G_\gamma)$ to $L^2(\A_{\gamma'}/\G_{\gamma'})$, which
we also denote by $F$, such that
\[ \langle \Phi', F\Phi \rangle = 
\langle \Phi',\Psi' \rangle \langle \Psi,\Phi\rangle \]
for any states $\Phi,\Phi'$.   The time evolution operator $Z(M)$ is a
linear combination of these operators weighted by the amplitudes $Z(F)$.
But a spin foam contains more information than the operator it
determines, since the operator depends only on the initial state $\Psi$
and the final state $\Psi'$, not on the details of the spin foam 
at intermediate times.   This extra information is what we call a 
`quantum history'.

How exactly does a spin foam describe the geometry of spacetime?  In
part, this follows from how spin networks describe the geometry of
space.  Consider, for example, 4d $BF$ theory with gauge group $\SU(2)$.
Spin network edges give area to surfaces they puncture, while spin
network vertices give volume to regions of space in which they lie.  But
a spin network edge is really just a slice of a spin foam face, and a
spin network vertex is a slice of a spin foam edge.  Thus in the
spacetime context, spin foam faces give area to surfaces they intersect,
while spin foam edges give 3-volume to 3-dimensional submanifolds they
intersect.   Continuing the pattern, one expects that spin foam vertices
give 4-volume to regions of spacetime in which they lie.  However,
calculations have not yet been done to confirm this, in part because a
thorough picture of the metric geometry of spacetime in 4 dimensions
requires that one impose constraints on the $E$ field.   We discuss this
a bit more in Section \ref{gr}.

A similar story holds for 3d $BF$ theory with gauge group $\SU(2)$, or
in other words, Riemannian quantum gravity in 3 dimensions.  In this
case, spin foam faces give length to curves they intersect and spin foam
edges give area to surfaces they intersect.   We expect that spin foam 
vertices give volume to regions of spacetime in which they lie, but so
far the calculations remain a bit problematic.

\subsubsection*{Remarks}

1.  The notation $F \maps \Psi \to \Psi'$ is meant to suggest that there
is a category with spin networks as objects and spin foams as morphisms.
For this, we should be able to compose spin foams $F \maps \Psi
\to \Psi'$ and $F' \maps \Psi' \to \Psi''$ and obtain a spin foam $F'F
\maps \Psi \to \Psi''$.   This composition should be associative, and
for each spin network $\Psi$ we want a spin foam $1_\Psi \maps \Psi \to
\Psi$ serving as a left and right unit for composition.   

To get this to work, we actually need to take certain equivalence
classes of spin foams as morphisms.  In my previous paper on this
subject, the equivalence relation described was actually not coarse
enough to prove associativity and the left and right unit laws.  The
quickest way to fix this problem is to simply impose extra equivalence
relations of the form $F(GH) \sim (FG)H$ and $1_\Psi F \sim F \sim
1_{\Psi'}$, to ensure that these laws hold.   

2.  The physical meaning of the time evolution operators
\[       Z(M) \maps L^2(\A_\gamma/\G_\gamma) \to 
      L^2(\A_{\gamma'}/\G_{\gamma'})   \]
is somewhat subtle in a background-independent theory.  For example,
when $M = S \times [0,1]$ is a cylinder cobordism from $S$ to itself,
we should have $Z(M)^2 = Z(M)$.   In this case $Z(M)$ should represent
the projection from the gauge-invariant Hilbert space to the space
of physical states.

\section{$q$-Deformation and the Cosmological Constant} 
\label{q-deformation}

As we have seen, $BF$ theory leads to a beautiful interplay between
representation theory and geometry, in which the distinction between the
two subjects gradually fades away.  In the end, spin networks  serve
simultaneously as a tool for calculations in representation theory and
as a description of the quantum geometry of space.  Spin foams extend
this idea to the geometry of spacetime.  This is exactly the sort of
thing one would hope for in a theory of quantum gravity, since quantum
mechanics is largely based on representation theory, while general
relativity is founded on differential geometry.

But so far, our treatment has been plagued by a serious technical
problem.   Mathematically, the problem is that the moduli space of flat
connections only has a natural measure in dimensions 2 or less.   We
need this measure to define the physical Hilbert space, so canonical
quantization only works when the dimension of {\it space} is at most 2. 
But we also need this measure to do path integrals in $BF$ theory, so
transition amplitudes between states are only well-defined when the
dimension of {\it spacetime} is at most 2.  Physically, the problem is
the presence of infrared divergences.   For example, in 3-dimensional
Riemannian quantum gravity, spin networks describe the geometry of
space, while spin foams describe the geometry of spacetime.   When 
we compute a transition amplitude from one spin network to another, we
sum over spin foams going between them.  The transition amplitude
diverges because we are summing over spin foams with faces labelled 
by arbitrarily high spins.   These correspond to arbitrarily large
spacetime geometries.  

In quantum field theory, one can often learn to live with infrared 
divergences by restricting the set of questions one expects the theory
to answer.  Crudely speaking, the idea is that we can ignore the
behavior of a theory on length scales greatly exceeding the
characteristic length scale of the experiment whose outcome we are
seeking to predict.  For example, certain infrared divergences in 
quantum electrodynamics can be ignored if we assume our apparatus is
unable to detect `soft photons', i.e., those with very long 
wavelengths.    Similarly, one can argue that the possibility of 
arbitrarily large spacetime geometries should not affect the outcome of
an experiment that occurs within a bounded patch of spacetime.   Thus it
is quite possible that with a little cleverness we can learn to extract
extra sensible physics from spin foam models with infrared divergences. 

Luckily, when it comes to $BF$ theory, we have another option: we
can completely {\it eliminate} the infrared divergences by adding an
extra term to the Lagrangian of our theory, built using only the  $E$
field.  This trick only works when spacetime has dimension 3 or 4.   In
dimension 3, the modified Lagrangian is
\[        \L = \tr(E \we F + {\Lambda \over 6} E \we E \we E), \]
while in dimension 4 it is 
\[        \L = \tr(E \we F + {\Lambda \over 12} E \we E) .\]
For reasons that will become clear, the coupling constant $\Lambda$ 
is called the `cosmological constant'.  We only consider the case
$\Lambda > 0$.  

Adding this `cosmological term' has a profound effect on $BF$ theory: it
changes all our calculations involving the representation theory of the
gauge group into analogous calculations involving the representation
theory of the corresponding quantum group.  This gives us a well-defined
and {\it finite-dimensional} physical Hilbert space, and turns the
divergent sum over spin foams into a {\it finite} sum for the transition
amplitudes between states.  This process is known as `$q$-deformation', 
because the quantum group depends on a parameter $q$, and reduces to the
original group at $q = 1$.  Often people think of $q$ as a function of 
$\hbar$, but for us it is a function of $\Lambda$, and we have $q = 1$
when $\Lambda = 0$.  Thus, at least in the present context, quantum
groups should really be called `cosmological groups'!  

To understand how quantum groups are related to $BF$ theory with a 
cosmological term, we need to exploit its ties to Chern-Simons
theory.   This is a background-free gauge theory in 3 dimensions whose
action depends only the connection $A$:
\[ S_{CS}(A) = 
{k \over 4 \pi} \int_M \tr(A \we dA + {2\over 3} A \we A \we A) \]
This formula only makes sense after we have chosen a trivialization of
$P$.  Luckily, if we assume $G$ is simply connected, every $G$-bundle
over a 3-manifold admits a trivialization.  The Chern-Simons action is
not invariant under large gauge transformations.    However, if we also
assume that $G$ is semisimple and `$\tr$' is defined using the Killing 
form, then the Chern-Simons action changes by an integer multiple 
of $2\pi k$ when we do a large gauge transformation.  This implies that 
$\exp(iS_{CS}(A))$ is gauge-invariant when the quantity $k$, called the
`level', is an integer.   Since this exponential of the action is what
actually appears in the path integral, one might hope that Chern-Simons
theory admits a reasonable quantization  in this case.  And indeed this
is so --- at least when $G$ is compact.  Unfortunately, Chern-Simons theory 
with noncompact gauge group is still poorly understood.

The vacuum expectation values of spin network observables are very
interesting in Chern-Simons theory.  Suppose $\Psi$ is a spin network
in $M$.   We can try to compute the vacuum expectation value
\[     \langle \Psi \rangle = 
{ \int \Psi(A) \, e^{iS_{CS}(A)}\, {\cal D}A  
\over \int e^{iS_{CS}(A)}\, {\cal D}A  }. \]
Naively, we would expect from the diffeomorphism-invariance of the 
Chern-Simons action that $\langle \Psi \rangle$ remains unchanged when
we apply a diffeomorphism to $\Psi$.  In fact, this expectation value is
ill-defined until we smear $\Psi$ by equipping it with a `framing'. 
Roughly, this means that we thicken each edge of $\Psi$ into a ribbon,
put a small disc at each vertex, and demand that the ribbons merge with
the discs smoothly at each vertex to form an orientable surface with
boundary.  The expectation values of these framed spin networks are
diffeomorphism-invariant, and they satisfy skein relations which allow
one to calculate them in a completely combinatorial way.  

The reader will recall that in $BF$ theory without cosmological term,
spin network observables also satisfied skein relations.  In that case,
the skein relations encoded the representation theory of $G$.  That is
what allowed us to give a purely combinatorial, or algebraic, 
description of the theory.  Marvelously, a similar thing is true in
Chern-Simons theory!  In Chern-Simons theory, however, the skein
relations encode the representation theory of the quantum group $U_q
\g$.  This is an algebraic gadget depending on a parameter $q$ which is
related to $k$ by the formula
\[              q = \exp(2 \pi i / (k + h))  \]
where $h$ is the value of the Casimir in the adjoint representation of
$\g$.   Alas, it would vastly expand the size of this paper to really 
explain what quantum groups are, and how they arise from Chern-Simons 
theory.   To learn these things, the reader must turn to the references
in the Notes. For our purposes, the most important thing is that the
representation theory of $U_q \g$ closely resembles that of $G$.   In
particular, each representation of $G$ gives a representation of $U_q
\g$.  This lets us think of spin network edges as labelled by
representations of the quantum group rather than the group.   However,
only finitely many irreducible representations of the group give
irreducible representations of the quantum group with nice algebraic
properties.  We shall call these `good' representations.  For example,
when $G = \SU(2)$, only the representations of spin  $j = 0, {1\over 2},
\dots, {k \over 2}$ give good representations of $U_q \g$.  It turns
out that Chern-Simons theory admits an algebraic formulation involving
only the good representations of $U_q \g$.  
 
With this information in hand, let us turn to 3-dimensional $BF$ theory
with cosmological term.  Starting from the action one can derive the
classical field equations: 
\[     F + {\Lambda \over 2} E \we E = 0 , \qquad d_A E = 0.\] 
For $G = \SO(2,1)$, these are equivalent to the vacuum Einstein 
equations {\it with a cosmological constant} when $E$ is one-to-one.   
One can show this using the same sort of argument we gave in Section
\ref{equations} for the case $\Lambda = 0$.  This reason this works is
that $\tr(E \we E \we E)$ is proportional to the volume form coming from
the metric defined by $E$.  Up to a constant factor, it is therefore
just a rewriting of the usual cosmological term in the action for
general relativity.     Similar remarks apply to $G = \SO(3)$, which
gives us Riemannian general relativity with cosmological constant.  We
can also use the double covers of these gauge groups without affecting
the classical theory.  

The relation between 3d $BF$ theory with cosmological term and Chern-Simons
theory is as follows.  Starting from the $A$ and $E$ fields in $BF$
theory, we can define two new connections $A_\pm$ as follows:
\[              A_\pm = A \pm \sqrt{\Lambda} E . \]
Ignoring boundary terms, we then have
\[    \int_M \tr(E \we F + {\Lambda \over 6} E \we E \we E) = 
S_{CS}(A_+) - S_{CS}(A_-)  \]
where 
\[         k = {4 \pi \over \sqrt{\Lambda}}  .\]
In short, the action for 3d $BF$ theory with cosmological term is a
difference of two Chern-Simons actions.   Thus we can quantize this $BF$
theory whenever we can quantize Chern-Simons theory at levels $k$ and
$-k$, and we obtain a theory equivalent to two independent copies of
Chern-Simons theory with these two opposite values of $k$.   The
physical  Hilbert space is thus the tensor product of Hilbert spaces for
two copies of Chern-Simons theory with opposite values of $k$, and a
similar factorization holds for the time evolution operators associated
to cobordisms.   Actually, we can simplify this description using the
fact that the Hilbert space for Chern-Simons theory at level $-k$ is
naturally the dual of the Hilbert space at level $k$.  This let us
describe 3d $BF$ theory with cosmological constant $\Lambda$
completely in terms of Chern-Simons theory at level $k$.  

Using this description together with the formulation of  Chern-Simons
theory in terms of quantum groups, one can derive a formula for the
partition function 3d $BF$ theory with cosmological term.   This formula
is almost identical to the one given in Section \ref{dynamics} for 3d
$BF$ theory with $\Lambda = 0$.   The main difference is that now the
quantum group $U_q \g$ takes over the role of the group $G$.  In other
words, we now label dual faces by good representations of $U_q \g$ and
label dual edges by intertwiners between tensor products of these
representations.  A similar formula holds for transition amplitudes.   
In short, we have a spin foam model of a generalized sort, based on the
representation theory of a quantum group instead of a group.   The
wonderful thing about this spin foam model is that the sums involved are
finite, since there are only finitely many good representations of $U_q
\g$.   With the infrared divergences eliminated, the partition function 
and transition amplitudes are truly well-defined.  Even better, one can
check that they are triangulation-independent!

The first example of this sort of spin foam model is due to Turaev and
Viro, who considered the case $G = \SU(2)$.  As we have seen, this model
corresponds to 3-dimensional Riemannian gravity with cosmological
constant $\Lambda$.  In this case only spins $j \le {k \over 2}$
correspond to good representations of $U_q \g$.  This constraint on the
spins labelling dual faces corresponds to an upper bound on the 
lengths of the edges of the original triangulation.  We thus have, not
only a minimum length due to nonzero Planck's constant, but also a
maximum length due to nonzero cosmological constant!  As $\Lambda \to
0$, this maximum length goes to infinity.  

Now let us turn to 4-dimensional $BF$ theory with cosmological term.
Here the classical field equations are
\[     F + {\Lambda \over 6} E = 0 , \qquad d_A E = 0.\] 
If we canonically quantize the theory, we discover something
interesting: for any compact oriented 3-manifold $S$ representing space,
the space of physical states is 1-dimensional.  To see this, note first
that `kinematical' states should be functions on $\A$, just as we saw in
Section \ref{canonical.quantization} for the case $\Lambda = 0$.  Physical
states are solutions of the constraints
\[     B + {\Lambda \over 6} E = 0 , \qquad d_A E = 0, \] 
where $B$ is the curvature of $A \in \A$.  As before, the constraint
$d_A E = 0$ generates gauge transformations, so imposing this constraint
should restrict us to gauge-invariant functions on $\A$.  But the other
constraint has a very different character when $\Lambda \ne 0$ than it
did for $\Lambda = 0$.  If we naively replace $A$ and $E$ by operators
following the usual rules of canonical quantization, we see that states
satisfying this constraint should be functions $\Psi \maps \A \to \C$
with  
\[   (B_{ij}^a + {\Lambda \over 6i} \epsilon_{ijk}{\delta \over
\delta A_{ka}})\psi = 0 .\] 
For $\Lambda \ne 0$ this equation has just
one solution, the so-called  `Chern-Simons state': 
\[     \psi(A) = e^{-{3i \over \Lambda}\int_S \tr(A \we dA + {2\over 3} A
\we A \we A) }  .\] 
By our previous remarks, if $G$ is simple, connected
and simply-connected and `\tr' is defined using the Killing form, the
Chern-Simons state is gauge-invariant exactly when the quantity 
\[        k = {12 \pi\over \Lambda}  \] 
is an integer.  If in addition $G$ is compact, we can go further: we can
compute expectation values of framed spin networks in the Chern-Simons
state using skein relations.   

How do we describe dynamics in 4-dimensional $BF$ theory with
cosmological term?  Unlike the other cases we have discussed, there is
not yet a plausible `derivation' of a spin foam model for this theory. 
At present, about the best one can do is note the following facts.  
There is a quantum group analog of the spin foam model for 4d $BF$
theory discussed in Section \ref{dynamics}, and this theory has finite
and triangulation-independent partition function and transition
amplitudes.   One can show that this theory has a 1-dimensional 
physical Hilbert space for any compact oriented 3-manifold $S$.  
Moreover, one can compute the expectation values of framed spin networks
in this theory, and one gets the same answers as in the Chern-Simons
state.   Thus it seems plausible that this theory is the correct spin
foam model for 4d $BF$ theory with cosmological term.   However, this
subject deserves further investigation.

\section{4-Dimensional Quantum Gravity} \label{gr}

We finally turn to theory that really motivates the interest in spin foam
models: quantum gravity in 4 dimensions.   Various competing spin foam 
models have been proposed for 4-dimensional quantum gravity --- mainly 
in the Riemannian case so far.   While some of these models are very
elegant, their physical meaning has not really been unravelled, and some
basic problems remain unsolved.  The main reason is that, unlike $BF$
theory, general relativity in 4 dimensions has local degrees of freedom.
In short, the situation is full of that curious mix of promise and
threat so typical of quantum gravity.  In what follows we do not attempt
a full description of the state of the art, since it would soon be
outdated anyway.  Instead, we merely give the reader a taste of the
subject.  For more details, see the Notes!  

We begin by describing the Palatini formulation of general relativity in
4 dimensions.   Let spacetime be given by a 4-dimensional oriented
smooth manifold $M$.  We choose a bundle $\T$ over $M$ that is
isomorphic to the tangent bundle, but not in any canonical way.  This
bundle, or any of its fibers, is called the `internal space'.  We equip
it with an orientation and a metric $\eta$, either Lorentzian or
Riemannian. Let $P$ denote the oriented orthonormal frame bundle of $M$.
This is a principal $G$-bundle, where $G$ is either $\SO(3,1)$ or
$\SO(4)$ depending on the signature of $\eta$.  The basic fields in the
Palatini formalism are:
\begin{itemize}
\item a connection $A$ on $P$,
\item a $\T$-valued $1$-form $e$ on $M$.
\end{itemize}
The curvature of $A$ is an $\ad(P)$-valued 2-form which, as usual, we
call $F$.  Note however that the bundle $\ad(P)$ is isomorphic to the
second exterior power $\Lambda^2 \T$.   Thus we are free to switch
between thinking of $F$ as an $\ad(P)$-valued 2-form and a $\Lambda^2
\T$-valued 2-form.  The same is true for the field $e \we e$.

The Lagrangian of the theory is
\[   \L =  \tr(e \we e \we F) . \]
Here we first take the wedge products of the differential form parts of
$e \we e$ and $F$ while simultaneously taking the wedge products of
their `internal' parts, obtaining the $\Lambda^4 \T$-valued 4-form $e
\we e \we F$.  The metric and orientation on $\T$ give us an `internal
volume form', that is, a nowhere vanishing section of $\Lambda^4 \T$.    
We can write $e \we e \we F$ as this volume form times an ordinary 
4-form, which we call $\tr(e \we e \we F)$.  

To obtain the field equations, we set the variation of the action to
zero:
\ban   0 &=& \delta \int_M \L       \\
&=&  \int_M \tr(\delta e \we e \we F + e \we \delta e \we F + 
e \we e \we \delta F)   \\
&=& \int_M \tr (2 \delta e \we e \we F + e \we e \we d_A \delta A) \\ 
&=& \int_M \tr(2\delta e \we e \we F - d_A(e \we e) \we \delta A).
\ean 
The field equations are thus
\[   e \we F = 0, \qquad d_A(e \we e) = 0.\]
These equations are really just an extension of the vacuum Einstein
equation to the case of degenerate metrics.  To see this, first 
define a metric $g$ on $M$ by 
\[   g(v,w) = \eta(ev,ew) .  \]
When $e \maps TM \to \T$ is one-to-one, $g$ is nondegenerate,  with the
same signature as $\eta$.   The equation $d_A(e \we e) = 0$ is
equivalent to $e \we d_Ae = 0$, and when $e$ is one-to-one this implies
$d_A e = 0$.  If we use $e$ to pull back $A$ to a metric-preserving
connection $\Gamma$ on the tangent bundle, the equation $d_A e = 0$ says
that $\Gamma$ is torsion-free, so $\Gamma$ is the Levi-Civita connection
of $g$.  This lets us rewrite $e \we F$ in terms of the Riemann tensor. 
In fact, $e \we F$ is proportional to the Einstein tensor, so $e \we F
= 0$ is equivalent to the vacuum Einstein equation. 

There are a number of important variants of the Palatini formulation
which give the same classical physics (at least for nondegenerate
metrics) but suggest different approaches to quantization.  Most simply,
we can pick a spin structure on $M$ and use the double cover $\Spin(3,1)
\iso \SL(2,\C)$ or $\Spin(4) \iso \SU(2) \times \SU(2)$ as gauge group.
A subtler trick is to work with the `self-dual' or `left-handed' part of
the spin connection.  In the Riemannian case this amounts to using only
one of the $\SU(2)$ factors of $\Spin(4)$ as gauge group; in the
Lorentzian case we need to complexify $\Spin(3,1)$ first, obtaining
$\SL(2,\C) \times \SL(2,\C)$, and then use one of these $\SL(2,\C)$
factors.  It it not immediately obvious that one can formulate general
relativity using only the left-handed part of the connection, but the
great discovery of Plebanski and Ashtekar is that one can.   A further
refinement of this trick allows one to formulate the canonical
quantization of Lorentzian general relativity in terms of the $e$ field
and an $\SU(2)$ connection.  These so-called `real Ashtekar variables'
play a crucial role in most work on loop quantum gravity.   Indeed, much
of the spin network technology described in this paper was first
developed for use with the real Ashtekar variables.  However, to keep
the discussion focused, we only discuss the Palatini formulation
in what follows.

The Palatini formulation of general relativity brings out its similarity
to $BF$ theory.  In fact, if we set $E = e \we e$, the Palatini 
Lagrangian looks exactly like the $BF$ Lagrangian.  The big difference,
of course, is that not every $\ad(P)$-valued 2-form $E$ is of the form
$e \we e$.  This restricts the allowed variations of the $E$ field when
we compute the variation of the action in general relativity.  As a 
result, the equations of general relativity in 4 dimensions:
\[   e \we F = 0, \qquad d_A E = 0 \] 
are weaker than the $BF$ theory equations:
\[   F = 0, \qquad d_A E = 0. \]
Another, subtler difference is that, even when $E$ is of the form $e \we
e$, we cannot uniquely recover $e$ from $E$.  In the nondegenerate case
there is only a sign ambiguity: both $e$ and $-e$ give the same $E$.  
Luckily, changing the sign of $e$ does not affect the metric.  In the
degenerate case the ambiguity is greater, but we need not be unduly
concerned about it, since we do not really know the `correct'
generalization of Einstein's equation to degenerate metrics.  

The relation between the Palatini formalism and $BF$ theory suggests
that one develop a spin foam model of quantum gravity by taking the spin
foam model for $BF$ theory and imposing extra constraints: quantum
analogues of the constraint that $E$ be of the form $e \we e$.  However,
there are some obstacles to doing this.  First, $BF$ theory is only
well-understood when the gauge group is compact.  If we work with a
compact gauge group, we are limited to Riemannian quantum gravity. 
Of course, this simply means that we should work harder and try to
understand $BF$ theory with noncompact gauge group.  Work on this
is currently underway, but the picture is still rather murky, and a 
fair amount of new mathematics will need to be developed before it 
clears up.   For this reason, we only consider the Riemannian quantum
gravity in what follows.

Second, when computing transition amplitudes in $BF$ theory, we only
summed over spin foams living in the dual 2-skeleton of a fixed
triangulation of spacetime.  This was acceptable because we could later
show triangulation-independence.   But triangulation-independence is
closely related to the fact that $BF$ theory lacks local degrees of
freedom: if we study $BF$ theory on a triangulated manifold, subdividing
the triangulation changes the gauge-invariant Hilbert space, but it does
not increase the number of physical degrees of freedom.   There is no
particular reason to expect something like this to hold in 4d quantum
gravity, since general relativity in 4 dimensions {\it does} have local
degrees of freedom.  So what should we do?   Nobody knows!  This problem
requires careful thought and perhaps some really new ideas.  In what
follows, we simply ignore it and restrict attention to spin foams lying
in the dual 2-skeleton of a fixed triangulation, for no particular good
reason.  

We begin by considering at the classical level the constraints that must
hold for the $E$ field to be of the form $e \we e$.  We pick a spin
structure for spacetime and take the double cover $\Spin(4)$ as our
gauge group.  Locally we may think of the $E$ field as taking
values in the Lie algebra $\so(4)$, but the splitting
\[      \so(4) \iso \so(3) \oplus \so(3) \]
lets us write $E$ as the sum of left-handed and right-handed parts
$E^\pm$ taking values in $\so(3)$.  If $E = e \we e$, 
the following constraint holds for all vector fields $v,w$ on $M$:
\[     | E^+(v,w) | = | E^-(v,w)|   \]
where $| \cdot |$ is the norm on $\so(3)$ coming from the Killing
form.   In fact, this constraint is almost sufficient to guarantee that
$E$ is of the form $e \we e$.  Unfortunately, in addition to solutions
of the desired form, there are also solutions of the form $-e \we e$,
$\ast(e \we e)$, and $-\!\ast\!(e \we e)$, where $\ast$ is the Hodge star
operator on $\Lambda^2 \T$.  

If we momentarily ignore this problem and work with the constraint as
described, we must next decide how to impose this constraint in a spin
foam model.  First recall some facts about 4d $BF$ theory with gauge
group $\SU(2)$.  In this theory, a spin foam in the dual 2-skeleton of a
triangulated 4-manifold is given by labelling each dual face with a spin
and each dual edge with an intertwiner.  This is equivalent to labelling
each triangle with a spin and each tetrahedron with an intertwiner.  We
can describe these intertwiners by chopping each tetrahedra in half
with a parallelogram and labelling all these parallelograms with spins.  
Then all the data is expressed in terms of spins labelling surfaces, and
each spin describes the integral of $|E|$ over the surface it labels. 

Now we are trying to describe 4-dimensional Riemannian quantum gravity 
as a $BF$ theory with extra constraints, but now the gauge group is
$\Spin(4)$.  Since $\Spin(4)$ is isomorphic to $\SU(2) \times \SU(2)$,
irreducible representation of this group are of the form $j^+ \tensor
j^-$ for arbitrary spins $j^+, j^-$.  Thus, before we take the
constraints into account, a spin foam with gauge group $\Spin(4)$ can
be given by labelling each triangle and parallelogram with a {\it pair}
of spins.  These spins describe the integrals of $|E^+|$ and 
$|E^-|$, respectively, over the surface in question.  Thus, to impose the 
constraint 
\[     | E^+(v,w) | = | E^-(v,w) |   \]
at the quantum level, it is natural to restrict ourselves to labellings
for which these spins are equal.    This amounts to labelling each
triangle with a representation of the form $j \tensor j$ and each
tetrahedron with an intertwiner of the form $\iota_j \tensor \iota_j$,
where $\iota_j \maps j_1 \tensor j_2 \to j_3 \tensor j_4$ is given 
in our graphical notation by:

\begin{center}
\setlength{\unitlength}{0.00046600in}%
\begingroup\makeatletter\ifx\SetFigFont\undefined
\def\x#1#2#3#4#5#6#7\relax{\def\x{#1#2#3#4#5#6}}%
\expandafter\x\fmtname xxxxxx\relax \def\y{splain}%
\ifx\x\y   
\gdef\SetFigFont#1#2#3{%
  \ifnum #1<17\tiny\else \ifnum #1<20\small\else
  \ifnum #1<24\normalsize\else \ifnum #1<29\large\else
  \ifnum #1<34\Large\else \ifnum #1<41\LARGE\else
     \huge\fi\fi\fi\fi\fi\fi
  \csname #3\endcsname}%
\else
\gdef\SetFigFont#1#2#3{\begingroup
  \count@#1\relax \ifnum 25<\count@\count@25\fi
  \def\x{\endgroup\@setsize\SetFigFont{#2pt}}%
  \expandafter\x
    \csname \romannumeral\the\count@ pt\expandafter\endcsname
    \csname @\romannumeral\the\count@ pt\endcsname
  \csname #3\endcsname}%
\fi
\fi\endgroup
\begin{picture}(1844,3044)(8094,-4598)
\thicklines
\put(9016,-2476){\circle*{88}}
\put(9016,-3676){\circle*{88}}
\put(9916,-1576){\line(-1,-1){900}}
\put(8116,-1576){\line( 1,-1){900}}
\put(9016,-3676){\line( 0, 1){1200}}
\put(8116,-4576){\line( 1, 1){900}}
\put(9016,-3676){\line( 1,-1){900}}
\multiput(8949,-3076)(3.92222,-7.84444){19}{\makebox(6.6667,10.0000){\SetFigFont{7}{8.4}{rm}.}}
\multiput(9016,-3219)(3.75000,7.50000){21}{\makebox(6.6667,10.0000){\SetFigFont{7}{8.4}{rm}.}}
\multiput(8529,-4059)(-2.75000,-8.25000){19}{\makebox(6.6667,10.0000){\SetFigFont{7}{8.4}{rm}.}}
\multiput(8484,-4209)(8.38754,2.09689){18}{\makebox(6.6667,10.0000){\SetFigFont{7}{8.4}{rm}.}}
\multiput(8581,-1937)(2.75000,-8.25000){19}{\makebox(6.6667,10.0000){\SetFigFont{7}{8.4}{rm}.}}
\multiput(8626,-2087)(-8.38754,2.09689){18}{\makebox(6.6667,10.0000){\SetFigFont{7}{8.4}{rm}.}}
\multiput(9488,-4044)(2.75000,-8.25000){19}{\makebox(6.6667,10.0000){\SetFigFont{7}{8.4}{rm}.}}
\multiput(9533,-4194)(-8.38754,2.09689){18}{\makebox(6.6667,10.0000){\SetFigFont{7}{8.4}{rm}.}}
\multiput(9459,-1937)(-2.75000,-8.25000){19}{\makebox(6.6667,10.0000){\SetFigFont{7}{8.4}{rm}.}}
\multiput(9414,-2087)(8.38754,2.09689){18}{\makebox(6.6667,10.0000){\SetFigFont{7}{8.4}{rm}.}}
\put(9676,-2116){\makebox(0,0)[lb]{\smash{$j_2$}}}
\put(9623,-4089){\makebox(0,0)[lb]{\smash{$j_4$}}}
\put(8200,-4066){\makebox(0,0)[lb]{\smash{$j_3$}}}
\put(8259,-2116){\makebox(0,0)[lb]{\smash{$j_1$}}}
\put(8723,-3145){\makebox(0,0)[lb]{\smash{$j$}}}
\end{picture}
\end{center}

\noindent
and $j_1,\dots,j_4$ are the spins labelling the 4 triangular faces
of the tetrahedron.  More generally, we can label the tetrahedron
by any intertwiner of the form $\sum_j c_j (\iota_j \tensor \iota_j)$.

However, there is a subtlety.  There are three ways to split a tetrahedron in 
half with a parallelogram $P$, and we really want the constraint 
\[    \int_P |E^+| = \int_P |E^-|   \]
to hold for all three.   To achieve this, we must label tetrahedra
with intertwiners of the form $\sum_j c_j (\iota_j \tensor \iota_j)$
that {\it remain} of this form when we switch to a different splitting
using the $6j$ symbols.  Barrett and Crane found an intertwiner with this 
property:
\[   \iota = \sum_{j} (2j + 1) (\iota_j \tensor \iota_j)  .\]
Later, Reisenberger proved that this was the unique solution. Thus, in
this spin foam model for 4-dimensional Riemannian quantum gravity, we
take the partition function to be:

\begin{center}
\setlength{\unitlength}{0.00083300in}%
\begingroup\makeatletter\ifx\SetFigFont\undefined
\def\x#1#2#3#4#5#6#7\relax{\def\x{#1#2#3#4#5#6}}%
\expandafter\x\fmtname xxxxxx\relax \def\y{splain}%
\ifx\x\y   
\gdef\SetFigFont#1#2#3{%
  \ifnum #1<17\tiny\else \ifnum #1<20\small\else
  \ifnum #1<24\normalsize\else \ifnum #1<29\large\else
  \ifnum #1<34\Large\else \ifnum #1<41\LARGE\else
     \huge\fi\fi\fi\fi\fi\fi
  \csname #3\endcsname}%
\else
\gdef\SetFigFont#1#2#3{\begingroup
  \count@#1\relax \ifnum 25<\count@\count@25\fi
  \def\x{\endgroup\@setsize\SetFigFont{#2pt}}%
  \expandafter\x
    \csname \romannumeral\the\count@ pt\expandafter\endcsname
    \csname @\romannumeral\the\count@ pt\endcsname
  \csname #3\endcsname}%
\fi
\fi\endgroup
\begin{picture}(5198,2182)(3301,-5374)
\thicklines
\put(8101,-5161){\circle*{68}}
\put(8401,-3961){\circle*{68}}
\put(6601,-3961){\circle*{68}}
\put(7501,-3361){\circle*{68}}
\put(6901,-5161){\circle*{68}}
\put(8101,-5161){\line( 1, 4){300}}
\put(7501,-3361){\line( 3,-2){900}}
\put(6601,-3961){\line( 3, 2){900}}
\put(8101,-5161){\line(-1, 0){1200}}
\put(6901,-5161){\line(-1, 4){300}}
\put(7876,-4479){\line( 1,-3){225.800}}
\put(7089,-4351){\line(-6, 5){491.312}}
\put(7801,-3961){\line( 1, 0){600}}
\put(6901,-5161){\line( 1, 3){383.100}}
\put(7321,-3871){\line( 1, 3){173.100}}
\put(7501,-3354){\line( 1,-3){325.500}}
\put(7456,-4726){\line(-4,-3){581.920}}
\put(8394,-3961){\line(-5,-4){828.902}}
\put(7569,-4629){\line( 0,-1){  7}}
\put(6601,-3969){\line( 1, 0){1020}}
\put(8109,-5161){\line(-5, 4){867.073}}
\put(3200,-4406){\makebox(0,0)[lb]{\smash{$
\displaystyle Z(M) =  
\sum_{j \maps \F \to \{0,{1\over 2},1,\dots\} }\; \prod_{f \in \F}
(2j_f + 1)\; \prod_{v \in \V}$}}}
\put(7456,-3264){\makebox(0,0)[lb]{\smash{$\iota$}}}
\put(8499,-3961){\makebox(0,0)[lb]{\smash{$\iota$}}}
\put(8229,-5326){\makebox(0,0)[lb]{\smash{$\iota$}}}
\put(6399,-3969){\makebox(0,0)[lb]{\smash{$\iota$}}}
\put(6714,-5349){\makebox(0,0)[lb]{\smash{$\iota$}}}
\put(7951,-3571){\makebox(0,0)[lb]{\smash{$j_1$}}}
\put(8379,-4644){\makebox(0,0)[lb]{\smash{$j_2$}}}
\put(7464,-5356){\makebox(0,0)[lb]{\smash{$j_3$}}}
\put(6504,-4696){\makebox(0,0)[lb]{\smash{$j_4$}}}
\put(6929,-3571){\makebox(0,0)[lb]{\smash{$j_5$}}}
\put(7599,-4261){\makebox(0,0)[lb]{\smash{$j_6$}}}
\put(7339,-4492){\makebox(0,0)[lb]{\smash{$j_8$}}}
\put(7266,-4284){\makebox(0,0)[lb]{\smash{$j_9$}}}
\put(7576,-4463){\makebox(0,0)[lb]{\smash{$j_7$}}}
\put(7397,-4110){\makebox(0,0)[lb]{\smash{$j_{10}$}}}
\end{picture}
\end{center}

\noindent Here $j_1,\dots,j_{10}$ are the spins labelling the dual
faces meeting at the dual vertex in question, and $\iota$ is the
Barrett-Crane intertwiner.   One can also write down a similar formula
for transition amplitudes.  

The sums in these formulas probably diverge, but there is a $q$-deformed
version where they become finite.  This $q$-deformed version appears 
{\it not} to be triangulation-independent.  We expect that it
is related to general relativity with a nonzero cosmological
constant.   As a piece of evidence for this, note that adding a
cosmological term to general relativity in 4 dimensions changes the
Lagrangian to
\[   \L = \tr(e \we e \we F + {\Lambda \over 12}e \we e \we e \we e) .\]
We can think of this as the $BF$ Lagrangian with cosmological term
together with a constraint saying that $E = e \we e$.  

So, where do we stand?  We have a specific proposal for a spin foam
model of quantum gravity.   In this theory, a quantum state of the
geometry of space is described by a linear combination of spin networks.
Areas and volumes take on a discrete spectrum of quantized values. 
Transition amplitudes between states are computed as sums over spin
foams.   In the $q$-deformed version of the theory these sums are
finite and explicitly computable.  

This sounds very nice, but there are severe problems as well.  The
theory is actually a theory of Riemannian rather than Lorentzian
quantum gravity.  It depends for its formulation on a fixed triangulation
of spacetime.  Even worse, our ability to do computations with the
theory is too poor to really tell if it reduces to classical Riemannian
general relativity in the large-scale limit, i.e.\ the limit of distances
much larger than the Planck length.  We thus face the following tasks:
\begin{itemize}
\item Develop spin foam models of Lorentzian quantum gravity.
\item Determine what role, if any, triangulations or 
related structures should play in spin foam models with local degrees 
of freedom.  
\item Develop computational techniques for studying the large-scale
limit of spin foam models.  
\end{itemize}
Luckily, work on these tasks is already underway.  

\subsubsection*{Remarks}

1.  Regge gave a formula for a discretized version of  the action in
4-dimensional Riemannian general relativity.  In his approach, spacetime
is triangulated and each edge is assigned a length.   The Regge action
is the sum over all 4-simplices of:
\[   S = \sum_t A_t \theta_t  \]
where the sum is taken over the 10 triangular faces $t$, $A_t$ is the
area of the face $t$, and $\theta_t$ is the dihedral angle of $t$, that
is, the angle between the outward normals of the two tetrahedra incident
to this edge.  Calculations suggest that the spin foam vertex amplitudes
in the Barrett-Crane theory are related to the Regge action by a formula
very much like the one relating vertex amplitudes in 3d Riemannian
quantum gravity to the Ponzano-Regge action (see Remark 1 of Section
\ref{dynamics}).  

2.  Our heuristic explanation of the Barrett-Crane model may make
it seem more ad hoc than it actually is.  For a more thorough treatment
one should see the references in the Notes.  At present our best
understanding of this model comes from a 4-dimensional analogue of the
theory of the quantum tetrahedron discussed in Remark 1 of Section
\ref{triangulations}.  In particular, this approach allows a careful
study of the `spurious solutions' to the constraint $|E^+(v,w)| = 
|E^-(v,w)|$.  It appears that at the quantum level, use of the 
Barrett-Crane intertwiner automatically excludes solutions of the form 
$E = \pm \! \ast \! (e \we e)$, but does not exclude solutions of the 
form $E = -e \we e$.  The physical significance of this is still
not clear.

\section*{Appendix: Piecewise linear cell complexes}

Here we give the precise definition of `piecewise linear cell complex'.
A subset $X \subseteq \R^n$ is
said  to be a `polyhedron' if every point $x \in X$ has a
neighborhood in  $X$ of the form
\[       \{ \alpha x + \beta y \,\colon\; \alpha,\beta \ge 0, \;
\alpha + \beta = 1, \; y \in Y\}   \]
where $Y \subseteq X$ is compact.  A compact convex polyhedron $X$ for
which the smallest affine space containing $X$ is of dimension $k$ is
called a `$k$-cell'.  The term `polyhedron' may be somewhat
misleading to the uninitiated; for example, $\R^n$ is a polyhedron, and
any open subset of a polyhedron is a polyhedron.  Cells, on the
other hand, are more special.  For example, every 0-cell is a point, every
1-cell is a compact interval affinely embedded in $\R^n$, and every
2-cell is a convex compact polygon affinely embedded in $\R^n$.

The `vertices' and `faces' of a cell $X$ are defined as follows.  Given
a point $x \in X$, let $\langle x,X\rangle$ be the union of lines $L$
through $x$ such that $L \cap X$ is an interval with $x$ in its
interior.  If there are no such lines, we define $\langle x,X\rangle$ to
be $\{x\}$ and call $x$ a `vertex' of $X$.  One can show that
$\langle x,X\rangle \cap X$ is a cell, and such a cell is called a 
`face' of $X$.   (In the body of this paper we use the words `vertex',
`edge' and `face' to stand for 0-cells, 1-cells and 2-cells, respectively.
This should not be confused with the present use of these terms.)

One can show that any cell $X$ has finitely many vertices $v_i$ and that
$X$ is the convex hull of these vertices, meaning that:
\[      X = \{ \sum \alpha_i v_i \, \colon\; 
\alpha_i \ge 0, \sum \alpha_i = 1\}  .\]
Similarly, any face of $X$ is the convex hull of some subset of the
vertices of $X$.  However, not every subset of the vertices of $X$ has a
face of $X$ as its convex hull.  If the cell $Y$ is a face of $X$ we
write $Y \le X$.  This relation is transitive, and if $Y,Y' \le X$ we
have $Y \cap Y' \le X$.

Finally, one defines a `piecewise linear cell complex', or
`complex' for short, to be a collection $\kappa$ of cells in some
$\R^n$ such that: 
\begin{enumerate}
\item If $X \in \kappa$ and $Y \le X$ then $Y \in \kappa$.
\item If $X,Y \in \kappa$ then $X\cap Y \le X,Y$.  
\end{enumerate}
In this paper we restrict our attention to complexes with finitely many 
cells.  

A complex is `$k$-dimensional' if it has cells of dimension $k$ but
no higher.    A complex is `oriented' if every cell is equipped with
an orientation, with all 0-cells being equipped with the positive
orientation.  The union of the cells of a complex $\kappa$ is a
polyhedron which we denote by $|\kappa|$.  

When discussing spin foams we should really work with spin networks
whose underlying graph is a 1-dimensional oriented complex.    Suppose
$\gamma$ is a 1-dimensional oriented complex and $\kappa$ is a 
2-dimensional oriented complex.  Note that the product $\gamma \times 
[0,1]$ becomes a 2-dimensional oriented complex in a natural way.   We
say $\gamma$ `borders' $\kappa$ if there is a one-to-one affine map $c
\maps |\gamma| \times [0,1] \to |\kappa|$ mapping each cell of $\gamma
\times [0,1]$ onto a unique cell of $\kappa$ in an
orientation-preserving way, such that $c$ maps $\gamma \times [0,1)$
onto an open subset of $|\kappa|$.   Note that in this case, $c$ lets us
regard each $k$-cell of $\gamma$ as the face of a unique $(k+1)$-cell of
$\kappa$.  

\section{Notes}

While long-winded, this bibliography has no pretensions to completeness.  
In particular, as a mathematician by training, my selection of
references inevitably has an emphasis on mathematically rigorous work. 
This gives a somewhat slanted view of the the subject, which is bound to
make some people unhappy.  I apologize for this in advance, and urge the
reader to look at some of the references written by physicists to get a
more balanced picture.  

\subsection*{1  $BF$ Theory: Classical Field Equations}

For all aspects of $BF$ theory, the following papers are invaluable:

\vskip 1em
\noindent
A.\ S.\ Schwartz, The partition function of
degenerate quadratic functionals and Ray-Singer invariants, {\sl Lett.\
Math.\ Phys.\ } {\bf 2} (1978), 247-252.

\noindent
G.\ Horowitz, Exactly soluble diffeomorphism-invariant
theories, {\sl Comm.\ Math.\ Phys.\ }{\bf 125} (1989) 417-437.

\noindent
D.\ Birmingham, M.\ Blau, M.\ Rakowski and G.\ Thompson, Topological
field theories, {\sl Phys.\ Rep.\ }{\bf 209} (1991), 129-340.

\noindent
M.\ Blau and G.\ Thompson, Topological gauge
theories of antisymmetric tensor fields, {\sl Ann.\ Phys.} {\bf
205} (1991), 130-172.
 
\vskip 1em
\noindent
For $BF$ theory on manifolds with boundary, see:

\vskip 1em
\noindent
V.\ Husain and S.\ Major, Gravity and $BF$ theory defined in bounded
regions, {\sl Nucl.\ Phys.\ }{\bf B500} (1997), 381-401. 

\vskip 1em
\noindent
A.\ Momen, Edge dynamics for $BF$ theories and gravity, {\sl Phys.\ 
Lett.\ }{\bf B394} (1997), 269-274.

\subsection*{2  Classical Phase Space}

The space $\A/\G$ and its cotangent bundle have mainly been studied in
the context of Yang-Mills theory:

\vskip 1em
\noindent 
V.\ Moncrief, Reduction of the Yang-Mills equations, in {\sl
Differential Geometrical Methods in Mathematical Physics}, eds.\ P.\
Garcia, A.\ P\'erez-Rend\'on, and J. Souriau, Lecture Notes in
Mathematics 836, Springer-Verlag, New York, 1980, pp.\ 276-291. 

\noindent
P.\ K.\ Mitter, Geometry of the space of gauge orbits and Yang-Mills
dynamical system, in {\sl Recent developments in Gauge Theories}, eds.\
G.\ 't Hooft et al.,  Plenum Press, New York, 1980, pp.\ 265-292.

\vskip 1em
\noindent
The moduli space of flat $G$-bundles and the moduli space of flat
connections on any particular $G$-bundle have been extensively studied
when the base manifold is a Riemann surface.  See for example:

\vskip 1em
\noindent
M.\ Narasimhan and C.\ Seshadri, Stable and unitary vector
bundles on a compact Riemann surface, {\sl Ann.\ Math.\ }{\bf 82} (1965)
540-567.  

\vskip 1em
\noindent 
Later, Goldman and others studied these spaces when the base space is a 
compact 2-dimensional smooth manifold, without any complex structure:

\vskip 1em
\noindent 
W.\ Goldman, The symplectic nature of fundamental groups of
surfaces, {\sl Adv.\ Math.\ }{\bf 54} (1984) 200-225.

\noindent 
W.\ Goldman, Invariant functions on Lie groups and Hamiltonian flows of
surface group representations, {\sl Invent.\ Math.\ }{\bf 83} (1986)
263-302.

\noindent 
W.\ Goldman, Topological components of spaces of representations, {\sl
Invent.\ Math.\ }{\bf 93} (1988) 557-607.

\noindent
A.\ Alekseev, A.\ Malkin,
Symplectic structure of the moduli space of flat connections on a Riemann 
surface, {\sl Commun.\ Math.\ Phys.\ }{\bf 169} (1995), 99-120.

\subsection*{3  Canonical Quantization}

The idea of taking functions of holonomies as the basic observables  or
states in a quantized gauge theory has a long history.  The earliest 
work dealt with Yang-Mills theory and used Wilson loops; later the 
idea was applied to gravity, and the importance of spin networks became
clear still later.   Some good books and review articles include:

\vskip 1em
\noindent
R.\ Gambini and J.\ Pullin, {\sl Loops, Knots, Gauge Theories,
and Quantum Gravity,} Cambridge U.\ Press, Cambridge, 1996.

\noindent
R.\ Loll, Chromodynamics and gravity as theories on loop
space, preprint available as hep-th/9309056.   

\noindent
C.\ Rovelli, Loop quantum gravity, {\sl Living Reviews in Relativity}
(1998), available online at \break 
$\langle$http://www.livingreviews.org$\rangle$. 

\vskip 1em
\noindent
The first really systematic attempt to formulate quantum gravity in terms
of Wilson loops is due to Rovelli and Smolin:

\vskip 1em
\noindent
C.\ Rovelli and L.\ Smolin, Loop representation for
quantum general relativity, {\sl Nucl.\ Phys.\ }{\bf B331} (1990),
80-152.

\vskip 1em
\noindent
An important step towards a rigorous description of the space of
states in loop quantum gravity was made by Ashtekar and Isham:

\vskip 1em
\noindent
A.\ Ashtekar and C.\ J.\ Isham, Representations of the holonomy algebra
of gravity and non-abelian gauge theories,  {\sl Class.\ Quan.\ Grav.\
}{\bf 9} (1992), 1069-1100.

\vskip 1em
\noindent
This work used piecewise smooth loops, which turn out to be technically
difficult to handle, so these authors were unable to construct
$L^2(\A/\G)$ except when $G$ is abelian.  Later, Ashtekar and
Lewandowski used piecewise real-analytic loops to give a rigorous
construction of $L^2(\A/\G)$ for $G = \SU(2)$:

\vskip 1em
\noindent
A.\ Ashtekar and J.\ Lewandowski, Representation theory of analytic
holonomy C*-algebras, in {\sl Knots and Quantum Gravity,} ed.\ J.\ Baez,
Oxford, Oxford U.\ Press, 1994.

\vskip 1em
\noindent
Then graphs with real-analytic edges were introduced, and used to
construct $L^2(\A/\G)$ for more general groups:

\vskip 1em
\noindent
J.\ Baez, Diffeomorphism-invariant generalized measures on the space of
connections modulo gauge transformations, in {\it Proceedings of the
Conference on Quantum Topology,} ed.\ D.\ Yetter, World Scientific,
Singapore, 1994.  

\vskip 1em
\noindent
Later graphs were used to construct the space $L^2(\A)$:

\vskip 1em
\noindent
J.\ Baez, Generalized measures in gauge theory, {\sl Lett.\ Math.\
Phys.\ }{\bf 31} (1994), 213-223.

\vskip 1em
\noindent
The use of graphs for integral and differential calculus on $\A$ and
$\A/\G$ is systematically developed in the following papers:

\vskip 1em
\noindent
A.\ Ashtekar and J.\ Lewandowski, Projective techniques and functional
integration, {\sl Jour.\ Math.\ Phys.\ }{\bf 36} (1995), 2170-2191.

\noindent
A.\ Ashtekar and J.\ Lewandowski,  Differential geometry for spaces of
connections via graphs and projective limits, {\sl Jour.\ Geom.\ Phys.\
}{\bf 17} (1995), 191-230. 

\vskip 1em
\noindent
The history of spin networks is rather complicated and I cannot
do justice to it here.  For a good introduction see:

\vskip 1em
\noindent
L.\ Smolin, The future of spin networks, preprint available
as gr-qc/9702030.

\vskip 1em
\noindent
Briefly, spin networks were first invented by Penrose:

\vskip 1em
\noindent
R.\ Penrose, Angular momentum: an approach to
combinatorial space-time, in {\sl Quantum Theory and Beyond,}
ed.\ T.\ Bastin, Cambridge U.\ Press, Cambridge, 1971, pp.\ 151-180.

\noindent
R.\ Penrose, Applications of negative dimensional tensors, in {\sl
Combinatorial Mathematics and its Applications,} ed.\ D.\ Welsh,
Academic Press, New York, 1971, pp.\ 221-244.

\noindent 
R.\ Penrose, On the nature of quantum geometry, in {\sl Magic Without
Magic,} ed.\ J.\ Klauder, Freeman, San Francisco, 1972, pp.\ 333-354.

\noindent
R.\ Penrose, Combinatorial quantum theory and quantized directions, in
{\sl Advances in Twistor Theory,} eds.\ L.\ Hughston and R.\ Ward,
Pitman Advanced Publishing Program, San Francisco, 1979, pp.\ 301-317.

\vskip 1em
\noindent
Penrose considered trivalent graphs labelled by spins.   He wanted to use
these as the basis for a purely combinatorial approach to spacetime. 
The following thesis is still invaluable for anyone interested in these
ideas:

\vskip 1em
\noindent
J.\ Moussouris, Quantum models of space-time based on recoupling theory,
Ph.D.\ thesis, Department of Mathematics, Oxford University, 1983.

\vskip 1em
\noindent
Later, as part of an attempt to understand the Jones polynomial and
related knot invariants, the notion of spin network was generalized to
include arbitrary graphs labelled by representations of any quantum
group:

\vskip 1em
\noindent
N.\ Reshetikhin and V.\ Turaev, Ribbon graphs and their invariants
derived from quantum groups, {\sl Comm.\ Math.\ Phys.\ }{\bf 127}
(1990), 1-26.

\vskip 1em
\noindent
In this more general context a framing of the graph is required, hence
the term `ribbon graph'.  Spin networks were introduced into loop
quantum gravity by Rovelli and Smolin:

\vskip 1em
\noindent
C.\ Rovelli and L.\ Smolin,  Spin networks in quantum gravity, {\sl
Phys.\ Rev.\ }{\bf D52} (1995), 5743-5759.

\vskip 1em
\noindent
The fact that spin network states span $L^2(\A/\G)$ was shown in:

\vskip 1em
\noindent
J.\ Baez, Spin networks in gauge theory,  {\sl Adv.\ Math.\ }{\bf 117}
(1996), 253-272.

\vskip 1em
\noindent
For an expository account of this proof and a general introduction
to quantum gravity, try:

\vskip 1em
\noindent
J.\ Baez, Spin networks in nonperturbative quantum gravity,
in {\sl The Interface of Knots and Physics}, ed.\ L.\ Kauffman, American
Mathematical Society, Providence, Rhode Island, 1996.

\vskip 1em
\noindent
For a rigorous approach to the canonical quantization of
diffeomorphism-invariant gauge theories using spin networks, see:

\vskip 1em
\noindent
A.\ Ashtekar, J.\ Lewandowski, D.\ Marolf, J.\ Mour\~ao, 
and T.\ Thiemann, Quantization of diffeomorphism invariant theories
of connections with local degrees of freedom, {\sl Jour.\ Math.\ Phys.\
}{\bf 36} (1995), 6456-6493.

\vskip 1em
\noindent
For the theory of $L^2(\A)$ and $L^2(\A/\G)$ in the smooth 
context, which involves the notion of `webs', see:

\vskip 1em
\noindent
J.\ Baez and S.\ Sawin, Functional integration on
spaces of connections, {\sl Jour.\ Funct.\ Anal.\ }{\bf 150} (1997), 1-27.

\noindent
J.\ Baez and S.\ Sawin, Diffeomorphism-invariant
spin network states, {\sl Jour.\ Funct.\ Anal.\ }{\bf 158} (1998), 253-266.

\noindent
J.\ Lewandowski and T.\ Thiemann, Diffeomorphism invariant
quantum field theories of connections in terms of webs, preprint
available as gr-qc/9901015.

\vskip 1em
\noindent
For the canonical quantization of 3-dimensional general relativity, see:

\vskip 1em
\noindent
E.\ Witten, 2+1 dimensional gravity as an exactly
soluble system, {\sl Nucl.\ Phys.\ }{\bf B311} (1988), 46-78.

\noindent
A.\ Ashtekar, V.\ Husain, C.\ Rovelli, J.\ Samuel and
L.\ Smolin, 2+1 gravity as a toy model for the 3+1 theory, {\sl Class.\
Quant.\ Grav.\ }{\bf 6} (1989), L185-L193. 

\noindent
A.\ Ashtekar, Lessons from (2+1)-dimensional quantum gravity,
{\sl Strings 90}, World Scientific, Singapore, 1990, pp.\ 71-88.

\noindent
A.\ Ashtekar, R.\ Loll, New loop representations for 2+1 gravity,
{\sl Class.\ Quant.\ Grav.\ }{\bf 11} (1994), 2417-2434.

\noindent
S.\ Carlip, {\sl Quantum Gravity in 2+1 Dimensions,} Cambridge U.\ 
Press, Cambridge, 1998.

\vskip 1em 
\noindent 
For a discussion of torsion and $BF$ theory, see:

\vskip 1em 
\noindent 
M.\ Blau and G.\ Thompson, A new class of topological field theories and
the Ray-Singer torsion, {\sl Phys.\ Lett.} {\bf B228} (1989), 64-68.

\subsection*{4  Observables}

The first calculation of area and volume operators in loop quantum
gravity was by Rovelli and Smolin:

\vskip 1em
\noindent
C.\ Rovelli and L.\ Smolin, Discreteness of area and
volume in quantum gravity, {\sl Nucl.\ Phys.\ }{\bf B442} (1995),
593-622.  Erratum, {\sl ibid.}  {\bf B456} (1995), 753.

\vskip 1em
\noindent
A rigorous construction and analysis of area and volume operators on
$L^2(\A/\G)$, using a somewhat different quantization scheme, was given
in the following series of papers:

\vskip 1em
\noindent
A.\ Ashtekar and J.\ Lewandowski, Quantum theory of geometry I: area
operators, {\sl Class.\ Quantum Grav.\ }{\bf 14} (1997), A55-A81.

\noindent
A.\ Ashtekar and J.\ Lewandowski, Quantum theory of
geometry II: volume operators, {\sl Adv.\ Theor.\ Math.\ Phys.\ }{\bf 1}
(1998), 388-429.

\noindent
A.\ Ashtekar, A.\ Corichi and J.\ Zapata, Quantum theory
of geometry III: non-commutativity of Riemannian structures, 
{\sl Class.\ Quantum Grav.\ }{\bf 15} (1998), 2955-2972.

\vskip 1em
\noindent 
The area operator considered in these papers is the same as the 
operator ${\cal E}(\Sigma)$ in the special case when space  is
3-dimensional and the gauge group is $\SU(2)$; however, the 
generalization to other dimensions and gauge groups is straightforward. 
For a simplified derivation of the area operator, see:

\vskip 1em
\noindent
C.\ Rovelli and P.\ Upadhya, Loop quantum gravity and
quanta of space: a primer, preprint available as gr-qc/9806079.

\vskip 1em
\noindent
For attempts to compute the entropy of black holes in loop quantum
gravity, see:

\vskip 1em
\noindent
L.\ Smolin, Linking topological quantum field theory and nonperturbative
quantum gravity, {\sl Jour.\ Math.\ Phys.\ }{\bf 36} (1995) 6417-6455.

\noindent
C.\ Rovelli, Loop quantum gravity and black hole physics,
{\sl Helv.\ Phys.\ Acta} {\bf 69} (1996), 582-611.

\noindent
K.\ Krasnov, Counting surface states in loop quantum gravity,
{\sl Phys.\ Rev.\ }{\bf D55} (1997), 3505-3513.

\noindent
K.\ Krasnov, On quantum statistical mechanics of a
Schwarzschild black hole, {\sl Gen.\ Rel.\ Grav.\ }{\bf 30} (1998), 53-68.

\noindent
A.\ Ashtekar, J.\ Baez, A.\ Corichi and K.\ Krasnov,
Quantum geometry and black hole entropy, {\sl Phys.\ Rev.\ Lett.\ }{\bf 80} 
(1998), 904-907.

\noindent
A.\ Ashtekar, A.\ Corichi and K.\ Krasnov,
Isolated black holes: the classical phase space, to appear.

\noindent
A.\ Ashtekar, J.\ Baez, and K.\ Krasnov, Quantum 
geometry of black hole horizons, to appear.

\subsection*{5  Canonical Quantization via Triangulations}

The relation between canonical quantum gravity on a triangulated manifold
and other simplicial approaches to quantum gravity was noted by Rovelli:

\vskip 1em
\noindent  
C.\ Rovelli, The basis of the Ponzano-Regge-Turaev-Viro-Ooguri 
model is the loop representation basis, {\sl Phys.\ Rev.\ }{\bf D48} 
(1993), 2702-2707.

\vskip 1em
\noindent
In a series of papers, Loll developed a version of loop quantum gravity
on a cubical lattice:

\vskip 1em
\noindent
R.\ Loll, Non-perturbative solutions for lattice quantum
gravity, {\sl Nucl.\ Phys.\ }{\bf B444} (1995), 619-640. 

\noindent  
R.\ Loll, The volume operator in discretized quantum
gravity, {\sl Phys.\ Rev.\ Lett.\ }{\bf 75} (1995) 3048-3051.

\noindent  
R.\ Loll, Spectrum of the volume operator in quantum
gravity, {\sl Nucl.\ Phys.\ }{\bf B460} (1996) 143-154. 

\noindent  
R.\ Loll, Further results on geometric operators
in quantum gravity, {\sl Class.\ Quantum\ Grav.\ }{\bf 14} (1997), 
1725-1741.

\noindent  
R.\ Loll, Imposing ${\rm det}E > 0$ in discrete quantum gravity,
{\sl Phys.\ Lett.\ }{\bf B399} (1997), 227-232.

\vskip 1em
\noindent
For a definition of $L^2(\A)$ and $L^2(\A/\G)$ in the piecewise-linear
context, see:

\vskip 1em
\noindent  
J.\ A.\ Zapata, A combinatorial approach to diffeomorphism
invariant quantum gauge theories, {\sl Jour.\ Math.\ Phys.\ }{\bf 38} (1997),
5663-5681.  

\noindent 
J.\ A.\ Zapata, Combinatorial space from loop quantum gravity,
{\sl Gen.\ Rel.\ Grav.\ }{\bf 30} (1998), 1229-1245.   

\vskip 1em
\noindent
The study of the quantum tetrahedron was initiated by Barbieri:

\vskip 1em
\noindent
A.\ Barbieri, Quantum tetrahedra and simplicial spin
networks, {\sl Nucl.\ Phys.\ }{\bf B518} (1998) 714-728.

\vskip 1em
\noindent
For a treatment of the quantum tetrahedron using geometric quantization,
see:

\vskip 1em
\noindent 
J.\ Baez and J.\ Barrett, The quantum tetrahedron in 3 and 4 dimensions,
preprint available as gr-qc/9903060.

\subsection*{6  Dynamics}

The formulation of 3d Riemannian quantum gravity as a sum over
labellings of the edges of a triangulated 3-manifold by spins was first
given by Ponzano and Regge:

\vskip 1em
\noindent
G.\ Ponzano and T.\ Regge, Semiclassical limit of
Racah coefficients, in {\sl Spectroscopic and Group Theoretical
Methods in Physics,} ed.\ F.\ Bloch, North-Holland, New York, 1968.

\vskip 1em
\noindent
The relation to Penrose's spin networks was noted by Hasslacher and Perry:

\vskip 1em
\noindent
B.\ Hasslacher and M.\ Perry, Spin networks are
simplicial quantum gravity, {\sl Phys.\ Lett.\ }{\bf B103} (1981), 21-24.

\vskip 1em
\noindent
We can now see the Ponzano-Regge formula as a purely combinatorial
formula for the partition function of 3d $BF$ theory with gauge group
$\SU(2)$.   Much later, Witten gave a similar formula in the
2-dimensional case:

\vskip 1em
\noindent
E.\ Witten, On quantum gauge theories in two
dimensions, {\sl Comm.\ Math.\ Phys.\ }{\bf 141} (1991) 153-209.

\vskip 1em
\noindent
and Ooguri gave a similar formula in the 4-dimensional case:

\vskip 1em
\noindent
H.\ Ooguri, Topological lattice models in four dimensions, {\sl Mod.\
Phys.\ Lett.\ }{\bf A7 }(1992) 2799-2810.

\vskip 1em
\noindent
For the Dijkgraaf-Witten model see:

\vskip 1em
\noindent
R.\ Dijkgraaf and E.\ Witten, Topological gauge
theories and group cohomology, {\sl Commun.\ Math.\ Phys.\ }{\bf
129} (1990) 393-429.

\noindent
D.\ Freed and F.\ Quinn, Chern-Simons theory with
finite gauge group, {\sl Commun.\ Math.\ Phys.\ }{\bf 156}
(1993), 435-472.

\vskip 1em
\noindent
Ponzano and Regge's original argument relating the asymptotics of
$6j$ symbols to their discretized action for 3-dimensional Riemannian
general relativity turned out to be surprisingly hard to make precise.
A rigorous proof was recently given by Roberts:

\vskip 1em 
\noindent 
J.\ Roberts, Classical $6j$-symbols and the tetrahedron,
{\sl Geometry and Topology} {\bf 3} (1999), 21-66.

\subsection*{7 Spin Foams}

The idea that transition amplitudes in 4d quantum gravity should be 
expressed as a sum over surfaces was proposed in the following paper:

\vskip 1em
\noindent
J.\ Baez, Strings, loops, knots and gauge fields,
in {\sl Knots and Quantum Gravity}, ed.\ J.\ Baez, Oxford U.\ Press,
Oxford, 1994.

\vskip 1em
\noindent
This idea was developed by Iwasaki and Reisenberger, who stressed the
importance of summing over 2-dimensional complexes, as opposed to
2-manifolds:

\vskip 1em
\noindent
J.\ Iwasaki, A definition of the Ponzano-Regge quantum
gravity model in terms of surfaces, {\sl Jour.\ Math.\ Phys.\ }{\bf 36}
(1995), 6288-6298.

\noindent
M.\ Reisenberger, Worldsheet formulations of gauge theories
and gravity, preprint available as gr-qc/9412035.  

\vskip 1em
\noindent
Later, Reisenberger and Rovelli showed how to derive such a `sum over
surfaces' formulation from a formula for the Hamiltonian constraint in
quantum gravity:

\vskip 1em
\noindent
M.\ Reisenberger and C.\ Rovelli, ``Sum over surfaces''
form of loop quantum gravity, {\sl Phys.\ Rev.\ }{\bf D56} (1997), 
3490-3508. 

\vskip 1em
\noindent
The relation between spin network evolution and triangulated spacetime
manifolds was clarified by Markopoulou:

\vskip 1em
\noindent
F.\ Markopoulou, Dual formulation of spin 
network evolution, preprint available as gr-qc/9704013. 

\vskip 1em
\noindent
The general notion of a spin foam was defined in the following paper:

\vskip 1em
\noindent
J.\ Baez, Spin foam models, {\sl Class.\ Quant.\ Grav.}
{\bf 15} (1998) 1827-1858.   

\vskip 1em
\noindent
For an attempt to systematically derive spin foam models from 
the Lagrangians for $BF$ theory and related theories, see:

\vskip 1em
\noindent
L.\ Freidel and K.\ Krasnov, Spin foam models and
the classical action principle, {\sl Adv.\ Theor.\ Phys.\ }{\bf 2} (1998),
1221-1285.

\vskip 1em
\noindent
For a discussion of the mathematical and philosophical 
underpinnings of the spin foam approach, see:

\vskip 1em
\noindent
J.\ Baez, Higher-dimensional algebra and Planck-scale physics, 
to appear in {\sl Physics Meets Philosophy at the Planck Scale,}
eds.\ C.\ Callender and N.\ Huggett, Cambridge U.\ Press, preprint
available as gr-qc/9902017.   

\vskip 1em
\noindent
For a study of volume in 3-dimensional quantum gravity, see:

\vskip 1em
\noindent
L.\ Freidel and K.\ Krasnov, Discrete space-time volume for
3-dimensional $BF$ theory and quantum gravity, {\sl Class.\ Quant.\
Grav.\ }{\bf 16} (1999), 351-362.

\subsection*{8  $q$-Deformation and the Cosmological Constant}

The relation between Chern-Simons theory and the Jones polynomial
was first glimpsed in Witten's seminal paper:

\vskip 1em
\noindent
E.\ Witten, Quantum field theory and the Jones polynomial, {\sl Comm.\
Math.\ Phys.\ }{\bf 121} (1989), 351-399.

\vskip 1em
\noindent
The relation to quantum groups was clarified by Reshetikhin and Turaev:

\vskip 1em
\noindent
N.\ Reshetikhin and V.\ Turaev, Invariants of 3-manifolds via link
polynomials and quantum groups, {\sl Invent.\ Math.\ }{\bf 103} (1991),
547-597.

\vskip 1em
\noindent
By now the subject has grown to enormous proportions, and we can
scarcely begin to list all the relevant refereces here.  Instead,
we merely direct the reader to the following textbooks: 

\vskip 1em
\noindent
M.\ Atiyah, {\sl The Geometry and Physics of Knots,}
Cambridge U.\ Press, Cambridge, 1990.   

\noindent
V.\ Chari and A.\ Pressley, {\sl A Guide to Quantum Groups,} 
Cambridge U.\ Press, Cambridge, 1994.

\noindent
J.\ Fuchs, {\sl Affine Lie Algebra and Quantum Groups,} Cambridge
U.\ Press, Cambridge, 1992.

\noindent
C.\ Kassel, {\sl Quantum Groups,} Springer-Verlag, New York, 1995.

\noindent
L.\ Kauffman, {\sl Knots and Physics}, World Scientific Press, Singapore,
1993.

\noindent
L.\ Kauffman and S.\ Lins, {\sl Temperley-Lieb Recoupling
Theory and Invariants of 3-Manifolds}, Princeton U.\ Press, Princeton,
New Jersey, 1994.

\noindent
V.\ Turaev, {\sl Quantum Invariants of Knots and 3-Manifolds,} de
Gruyter, New York, 1994.

\vskip 1em
\noindent
The book by Kauffman and Lins is especially handy whenever one needs
a compendium of skein relations for $U_q\su(2)$.   For an overview
of the relations between $BF$ theory and Chern-Simons theory, see:

\vskip 1em
\noindent
A.\ Cattaneo, P.\ Cotta-Ramusino, J.\ Fr\"ohlich and M.\
Martellini, Topological $BF$ theories in 3 and 4 dimensions,
{\sl Jour.\ Math.\ Phys.} {\bf 36} (1995), 6137-6160.

\vskip 1em
\noindent
The $q$-deformed version of the Ponzano-Regge model was discovered by
Turaev and Viro:

\vskip 1em
\noindent
V.\ Turaev and O.\ Viro, State sum invariants of 3-manifolds
and quantum $6j$ symbols, {\sl Topology} {\bf 31} (1992), 865-902.

\vskip 1em
\noindent
They formulated the theory both in terms of a triangulation of the
3-manifold and, dually, in terms of a 2-dimensional complex embedded in
the manifold.  We may now see their theory as a spin foam model for 3d
Riemannian quantum gravity with nonzero cosmological constant.   Their
construction was soon generalized by isolating the properties of the
6$j$ symbols that make it work, and tracing these back to the properties
of certain categories of representations.  One can read about these
generalizations in the book by Turaev, and also in the following papers:

\vskip 1em
\noindent
B.\ Durhuus, H.\ Jakobsen and R.\ Nest, Topological quantum field
theories from generalized 6$j$-symbols,  {\sl Rev.\ Math.\ Phys.\ }{\bf
5} (1993), 1-67.

\noindent
J.\ Barrett and B.\ Westbury, Invariants of piecewise-linear 3-manifolds,
{\sl Trans.\ Amer.\ Math.\ Soc.\ }{\bf 348} (1996), 3997-4022.

\noindent
D.\ Yetter, State-sum invariants of 3-manifolds associated to Artinian
semisimple tortile categories, {\sl Topology and its Applications} {\bf 58}
(1994), 47-80.

\vskip 1em
\noindent
Turaev also described a related model in 4 dimensions, formulated in 
terms of a 2-dimensional complex embedded in the manifold:

\vskip 1em
\noindent
V.\ Turaev, Quantum invariants of 3-manifolds and a glimpse of shadow
topology, in {\sl Quantum Groups}, Springer Lecture Notes in Mathematics 1510,
Springer-Verlag, New York, 1992, pp.\ 363-366.

\vskip 1em
\noindent
This model is also discussed in Turaev's book.   Crane and Yetter
developed an isomorphic theory, formulated in terms of a triangulation,
by $q$-deforming Ooguri's formula for the partition function of 4d $BF$
theory with gauge group $\SU(2)$:

\vskip 1em
\noindent
L.\ Crane and D.\ Yetter, A categorical construction
of 4d TQFTs, in {\sl Quantum Topology,} eds.\ L.\ Kauffman and
R.\ Baadhio, World Scientific, Singapore, 1993, pp.\ 120-130.

\vskip 1em
\noindent
The isomorphism between Turaev's theory and the Crane-Yetter model
was worked out by Roberts: 

\vskip 1em
\noindent
J.\ Roberts, Skein theory and Turaev-Viro invariants, {\sl Topology}
{\bf 34} (1995), 771-787.

\vskip 1em
\noindent
The generalization of this theory to other quantum groups was later
worked out by Turaev (see his book above) and in the following paper:

\vskip 1em
\noindent
L.\ Crane, L.\ Kauffman and D.\ Yetter,
State-sum invariants of 4-manifolds, {\sl J.\ Knot Theory \& Ramifications}
{\bf 6} (1997), 177-234.

\vskip 1em
\noindent
For an argument that this theory is really a spin foam model of 
$BF$ theory with cosmological term, see:

\vskip 1em
\noindent
J.\ Baez, Four-dimensional $BF$ theory as a topological
quantum field theory, {\sl Lett.\ Math.\ Phys.\ }{\bf 38} (1996),
129-143.

\vskip 1em
\noindent
Along closely related lines, there is also some interesting work 
on the canonical quantization of Chern-Simons theory and 3d $BF$
theory in the piecewise-linear context:

\vskip 1em
\noindent
A.\ Alekseev, H.\ Grosse and V.\ Schomerus, 
Combinatorial quantization of the Hamiltonian Chern-Simons theory I,
{\sl Commun.\ Math.\ Phys.\ }{\bf 172} (1995), 317-358.

\noindent
A.\ Alekseev, H.\ Grosse and V.\ Schomerus,
Combinatorial quantization of the Hamiltonian Chern-Simons theory II,
{\sl Commun.\ Math.\ Phys.\ }{\bf 174} (1995), 561-604.

\noindent
D.\ Bullock, C.\ Frohman, and J.\ Kania-Bartoszynska,
Topological interpretations of lattice gauge field theory,
{\sl Commun.\ Math.\ Phys.\ }{\bf 198} (1998), 47-81.

\noindent
D.\ Bullock, C.\ Frohman, and J.\ Kania-Bartoszynska, Skein modules and
lattice gauge field theory, preprint available as math.GT/9802023.

\subsection*{9  4-Dimensional Quantum Gravity}

For a tour of various formulations of Einstein's equation, see:

\vskip 1em 
\noindent
P.\ Peldan, Actions for gravity, with generalizations: a review,
{\sl Class.\ Quant.\ Grav.\ }{\bf 11} (1994), 1087-1132.

\vskip 1em 
\noindent
For an introduction to canonical quantum gravity, try the following
books:

\vskip 1em 
\noindent
A.\ Ashtekar and invited contributors, {\sl New
Perspectives in Canonical Gravity,} Bibliopolis, Napoli, Italy, 1988.
(Available through the American Institute of Physics; errata available
from the Center for Gravitational Physics and Geometry at Pennsylvania
State University.)
 
\noindent
A.\ Ashtekar, {\sl Lectures on Non-perturbative
Canonical Quantum Gravity,} World Scientific, Singapore, 1991.

\vskip 1em 
\noindent
The spin foam model of 4-dimensional Riemannian quantum gravity which we
discuss here was invented by Barrett and Crane:

\vskip 1em 
\noindent
J.\ Barrett and L.\ Crane, Relativistic spin networks
and quantum gravity, {\sl Jour.\ Math.\ Phys.\ }{\bf 39} (1998),
3296-3302.

\vskip 1em 
\noindent 
A detailed discussion of their model appears in my first paper on spin
foam models (see the Notes for Section 7).   A more detailed treatment
of general relativity as a constrained $\Spin(4)$ $BF$ theory can be found
in the following papers:

\vskip 1em
\noindent
M.\ Reisenberger, Classical Euclidean general relativity from `lefthanded
area = righthanded area', preprint available as gr-qc/9804061.

\noindent
R.\ De Pietri and L.\ Freidel, $\so(4)$ Plebanski action and relativistic
spin foam model, preprint available as gr-qc/9804071.

\vskip 1em 
\noindent
A heuristic argument for the uniqueness of the Barrett-Crane intertwiner
was given by Barbieri:

\vskip 1em 
\noindent
A.\ Barbieri, Space of the vertices of relativistic
spin networks, preprint available as gr-qc/9709076.

\vskip 1em 
\noindent
Later, Reisenberger gave a rigorous proof:

\vskip 1em 
\noindent
M.\ Reisenberger, On relativistic spin network 
vertices, preprint available as gr-qc/9809067.

\vskip 1em 
\noindent
An explanation of the uniqueness of the Barrett-Crane intertwiner in
terms of geometric quantization was given in my paper with Barrett on
the quantum tetrahedron (see the Notes for Section 5.)   Similar
intertwiners for vertices of higher valence have been constructed by
Yetter:

\vskip 1em 
\noindent
D.\ Yetter, Generalized Barrett-Crane vertices and invariants
of embedded graphs, preprint available as math.QA/9801131.

\vskip 1em 
\noindent
Barrett found an integral formula for the Barrett-Crane intertwiner:

\vskip 1em 
\noindent
J.\ Barrett, the classical evaluation of relativistic spin networks, 
preprint available as math.QA/9803063.

\vskip 1em 
\noindent
Later, he and Williams used this to give a heuristic argument relating
the asymptotics of the amplitudes in the Barrett-Crane model to 
the Regge action:

\vskip 1em 
\noindent
J.\ Barrett and R.\ Williams, 
The asymptotics of an amplitude for the 4-simplex,
preprint available as gr-qc/9809032.

\vskip 1em 
\noindent
For the Regge action, see:

\vskip 1em 
\noindent
T.\ Regge, General relativity without coordinates,
{\sl Nuovo Cimento} {\bf 19} (1961), 558-571.

\vskip 1em 
\noindent
Reisenberger and Iwasaki have proposed alternative spin foam models of
4-dimensional Riemannian quantum gravity.  As with the Barrett-Crane
model, the basic idea behind these models is to treat general relativity
as a constrained $BF$ theory.  However, the models of Reisenberger
and Iwasaki involve only the left-handed part of the spin connection, 
so the gauge group is $\SU(2)$:

\vskip 1em 
\noindent
M.\ Reisenberger, A lattice worldsheet sum for 4-d Euclidean general
relativity, preprint available as gr-qc/9711052.

\noindent
J.\ Iwasaki, A surface theoretic model of quantum
gravity, preprint available as gr-qc/9903112.

\vskip 1em 
\noindent
Freidel and Krasnov have constructed spin foam models of Riemannian
quantum gravity in higher dimensions by treating the theory as
a constrained $BF$ theory with gauge group $\SO(n)$:

\vskip 1em 
\noindent
L.\ Freidel, K.\ Krasnov, and R.\ Puzio, $BF$ description
of higher-dimensional gravity theories, preprint available as 
hep-th/9901069.

\vskip 1em 
\noindent
Barrett and Crane have also begun work on a Lorentzian version of their
theory, but so far their formula for the amplitude of a spin foam vertex
remains formal, because the evaluation of spin networks typically diverges 
when the gauge group is noncompact, apparently even after $q$-deformation:

\vskip 1em 
\noindent
J.\ Barrett and L.\ Crane, A Lorentzian signature model for quantum
general relativity, preprint available as gr-qc/9904025.

\vskip 1em 
\noindent
In a different but related line of development, Markopoulou and Smolin
have considered a class of local, causal rules for the time evolution of
spin networks.   Rules in this class are the same as spin foam models.

\vskip 1em 
\noindent
F.\ Markopoulou and L.\ Smolin, Quantum geometry with
intrinsic local causality, {\sl Phys.\ Rev.\ }{\bf D58}:084032 (1998).

\vskip 1em 
\noindent
Smolin has suggested a relationship between these models and string
theory, and proposed a specific model of this type as a candidate for
a background-free formulation of $M$-theory.  Ling and Smolin have begun
to develop the supersymmetric analogue of the theory of spin networks:

\vskip 1em 
\noindent
L.\ Smolin, Strings as perturbations of evolving spin networks, 
preprint available as hep-th/9801022.

\noindent 
L.\ Smolin, Towards a background-independent approach to 
$M$ theory, preprint available as hep-th/9808192.

\noindent
Y.\ Ling, L.\ Smolin, Supersymmetric spin networks and quantum
supergravity, preprint available as hep-th/9904016. 

\subsection*{Appendix}

For more details on piecewise linear cell complexes, try:

\vskip 1em
\noindent
C.\ Rourke and B.\ Sanderson, {\sl Introduction to
Piecewise-Linear Topology}, Springer Verlag, Berlin, 1972.

\vskip1em
\noindent 
The 2-dimensional case is explored more deeply in:

\vskip 1em
\noindent
C.\ Hog-Angeloni, W.\ Metzler, and A.\ Sieradski, 
{\sl Two-dimensional Homotopy and Combinatorial Group Theory}, 
London Mathematical Society Lecture Note Series 197, Cambridge U.\ 
Press, Cambridge, 1993.

\end{document}